# Can Sequentially Linked Gamma-Ray Bursts Nullify Randomness?

*Charles Fleischer*

*2012*

ABSTRACT

In order to nullify the property of randomness perceived in the dispersion of gamma-ray bursts (GRB's) we introduce two new procedures. 1. Create a segmented group of sequentially linked GRB's and quantify the resultant angles. 2. Create segmented groups of sequentially linked GRB's in order to identify the location of GRB's that are positioned at equidistance, by using the selected GRB as the origin for a paired point circle, where the circumference of said circle intercepts the location of other GRB's in the same group.

## 1. Introduction

The study of GRB's requires orbiting satellites with scintillating sensor systems that respond when activated. The engineers and scientists who designed and launched them into orbit are the giants on whose shoulders all who study GRB's stand.
The spacecraft:   Vela -1962
Compton Gamma-Ray Observatory with
The Burst and Transient Source Experiment (BATSE)- April 1991
Wind-Konus -November 1994
BeppoSAX -April 1996
The High Energy Transient Explorer (HETE-2)- October, 2000
The International Gamma-Ray Astrophysics Laboratory (INTEGRAL) - October 17, 2002
Swift- November 20, 2004
Suzaku (Astro-E2)- July 2005
AGILE -April 23, 2007
Fermi- June 2008.

GRB's are the most luminous explosions in the universe. Following the initial release of Gamma radiation is an afterglow, a subsequent emission of diminished electromagnetic energies, X-rays, ultra-violet, visible light, infra-red, microwaves and radio waves.

Since their serendipitous discovery in 1967 and the first unclassified scientific paper in 1973 (Klebesadel et al.) much has been discovered about their behavior yet the true nature of this cosmic phenomena still remains an elusive mystery without a single conclusive explanation regarding the nature of their origin. In 1993 Nemiroff wrote a paper that listed more than 100 models to explain the origins of GRB's. They included a comet falling into a white dwarf (Schlovskii 1974), an asteroid falling into a neutron star (Newman 1980) and the evaporation of a primordial black hole (Cline et al. 1992).

GRB's were initially divided into two fundamental categories based on the duration of each burst. Those under 2 seconds were referred to as short GRB's, (SGRB's) those over two seconds were called long GRB's (LGRB's) (Kouveliotou et al. 1993). In 1998 Horvath suggested there might also be a third grouping with a duration between long and short (Mukherjee et al. 1998).

Some scientists conclude that the LGRB's are created by the collapse of massive stars (Woosley & Bloom 2006; Hjorth et al.2003; Stanek et al. 2003). Others conclude LGRB's are driven by the merger of compact objects (Kluzniak & Ruderman 1998; Rosswog et al. 2003).

Some say the origin of SGRB's are due to the merger of compact binary objects, neutron stars or black holes (Eichler et al. 1989; Narayan et al. 1992). Some say the compact binaries were formed in primordial binaries (Belcynskiet al. 2002), some say they were formed dynamically in dense cluster cores (Davies 1995; Grindlay et al.2006).

Some say the afterglow is explained by the synchrotron emission of accelerated electrons interacting with the surrounding medium (Piran 2005; Meszaros2006; Zhang 2007). Some say the burst emission area is penetrated by a globally structured magnetic field (Spruit et al.2001; Zhang and Meszaros 2002; Lyutikov et al. 2003), or possibly by Compton drag of ambient soft photons (Shaviv and Dar 1995; Lazzati et al. 2004), or the combination of a thermal component from the photosphere and a non thermal component (Ioka et al. 2007). Some question if the outflow jet is collimated (Zhang et al. 2004; Toma et al. 2005).

It is difficult to find any realm of scientific research that has generated so many different hypotheses. What is needed is a new understanding regarding the origin of GRB's, one that encompasses the totality of all GRB's into a unified system of related events.

## 2. Distribution and Duration

To effectively determine the presence of randomness it is necessary to observe the celestial location for the position of every GRB. Evaluating GRB sky distribution patterns played an important part in determining that GRB's originate outside of our galaxy. (Zhang & Meszaros 2004; Fox et al. 2005; Meszaros 2006). If they were emanating from our galaxy we would have expected to see a concentration of GRB's along the galactic plane of the Milky Way. But we did not the GRB's were distributed randomly in an isotropic manner all across the universe sky map (Meegan et al. 1992). For the first 30 years most scientists agreed on the idea that GRB's were distributed randomly in an isotropic manner (Paczynski 1991a; Dermer 1992; Mao & Paczynski 1992a; Piran 1992; Fenimore et al.1993 Woods & Loeb 1994; Paczynski & Xu 1994).

In 1999 Balazs et al. found that there were differences in the level of randomness that correlated to the duration of each GRB, whereby the short and intermediate bursts showed a higher level of non randomness than the long bursts. The same findings were validated by Meszaros et al.2000 and Litvin et al.2001. In 2003 Meszaros & Stocek reported that even the long bursts might display patterns indicating anisotropic distribution.

The differentiation between long and short GRB's became a bigger issue with the appearance of GRB 060505 and GRB 060614 because they have characteristics associated with both long and short GRB's (Fynbo et al.2006; Della Valle et al. 2006; Gal-Yam et al. 2006).

In 2008 Vavrek et al. presented a paper "Testing the Randomness in the Sky-Distribution of Gamma-Ray Bursts". For their study they divided the GRB's detected by BATSE into 5 groups: short1, short2, intermediate, long 1 and long2. Using Monte Carlo simulations they employed three different methodologies to test for randomness: Voronoi tessellation, minimal spanning tree and multi-fractal spectra. They determined that the short and intermediate groups deviated significantly from being fully random and that the long groups did not.

In 2010 O.V.Verkhodanov et al. presented the paper "GRB Sky Distribution Puzzles" They utilized GRB information from BATSE and BeppoSAx with data from the WMAP (Wilkson Microwave Anisotropy Probe) to correlate GRB distribution with peaks in the Cosmic Microwave Background (CMB). They found correlations between GRB positions and the CMB equatorial coordinate system but could not understand the mechanism of the correlation.

## 3. Sequence and Time

For the present study the author utilized the group of GRB's beginning with GRB 040827A (six GRB's before the SWIFT era) and ranging up to present day. The information regarding the occurrence and location of the GRB's for this study was made possible by Sonoma State University and is presented at http://grb.sonoma.edu/.

This information including the galactic co-ordinates and the time and date for each burst is presented on an interactive sky map where the location of each burst is indicated as a circular blue dot. The sky map provides a clear visual understanding of the position for each GRB.

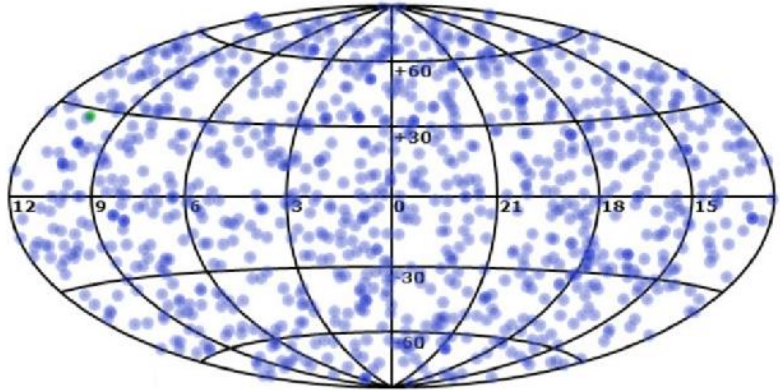

**Figure 1.** Sky map showing GRB distribution.

Creating sub-sets based on commonalities is an effective way to analyze lots of data. A technique exemplified by Vavrek et al. 2008, dividing over 2,000 BATSE GRB's into 5 groups based on the time duration of each GRB.

Time was also a factor in sub-dividing the over 1000 recorded GRB's used in this study. However in this case the element of time was applied to occurrence rather than duration. What differentiates this study from others is the choice to incorporate information regarding the sequence in which the events unfolded, literally combining time and space.

To facilitate this procedure the author divided the chronological progression of GRB's into octaves, creating close to 150 segmented groups, so that each group contained 8 sequential GRB's.

In Fig.2 the 8 GRB's that occurred from 10/8/2011 to 11/22/2011 are connected sequentially by straight lines. To assist the reader in the process of determining the chronological order for each sequence of 8 GRB's the author utilized a color coding that follows the progressive order of spectral vibration; 1- red, 2-orange, 3-yellow, 4-green, 5-cyan, 6-blue,7- indigo, 8-violet.

For the purpose of this study each group of eight sequentially linked GRB's will be referred to as an Ogg.

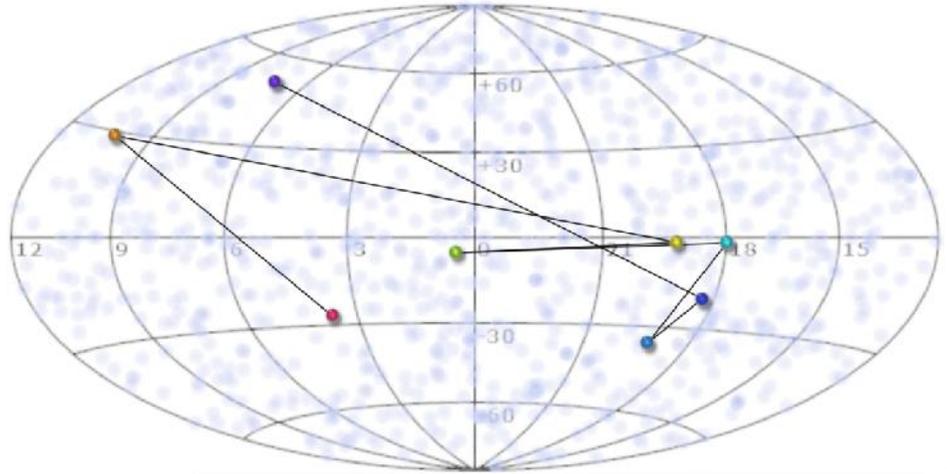

**Figure 2** An Ogg starting with GRB 111008A and ending on GRB 111022B. 10/8/2011 to 11/22/2011

The 1-red GRB is linked to the 2-orange GRB. The 2-orange GRB is linked to the 3-yellow GRB. The 3-yellow GRB is linked to the 4-green GRB. The 4-green GRB is linked to the 5-cyan GRB The 5-cyan GRB is linked to the 6-blue GRB The 6-blue GRB is linked to the 7-indigo GRB. The 7-indigo GRB is linked to the 8-violet GRB. This procedure results in the creation of a figure with 8 colored dots linked sequentially by 7 lines.

Like the octave in musical scales the last burst of one octave is also the first burst of the subsequent octave. The ogg of Fig.2 ends with violet GRB 111022B. Fig. 3 Starts with red GRB 111022B and ends with violet GRB111117.

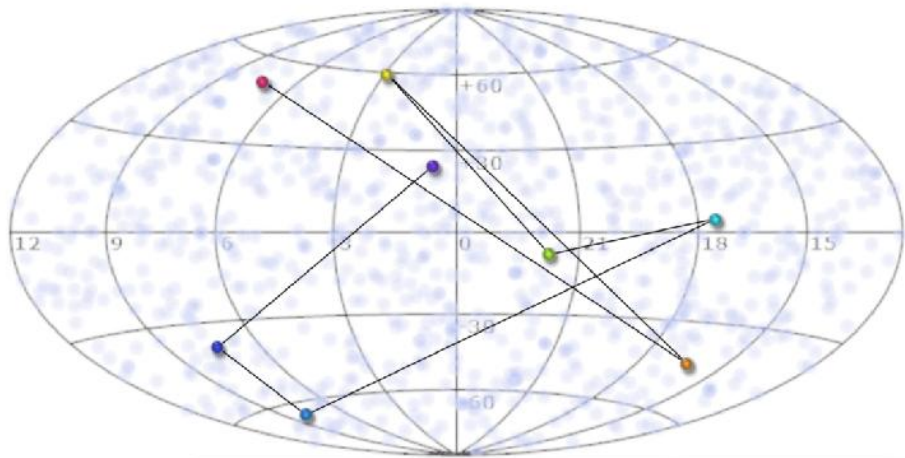

**Figure 3.**

Fig.4 The ogg begins with red GRB 111117A and ends with GRB 111207A.

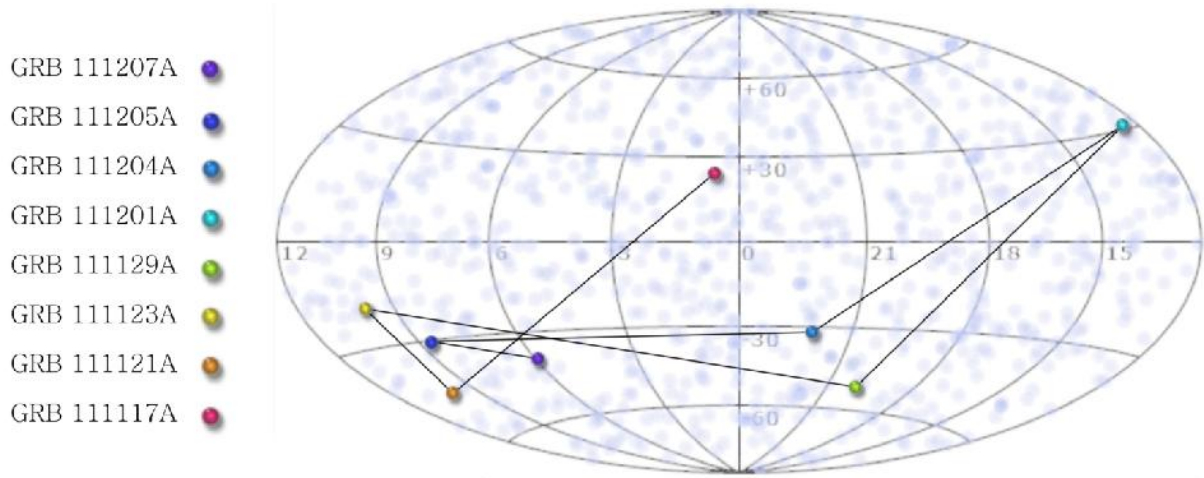

**Figure 4**

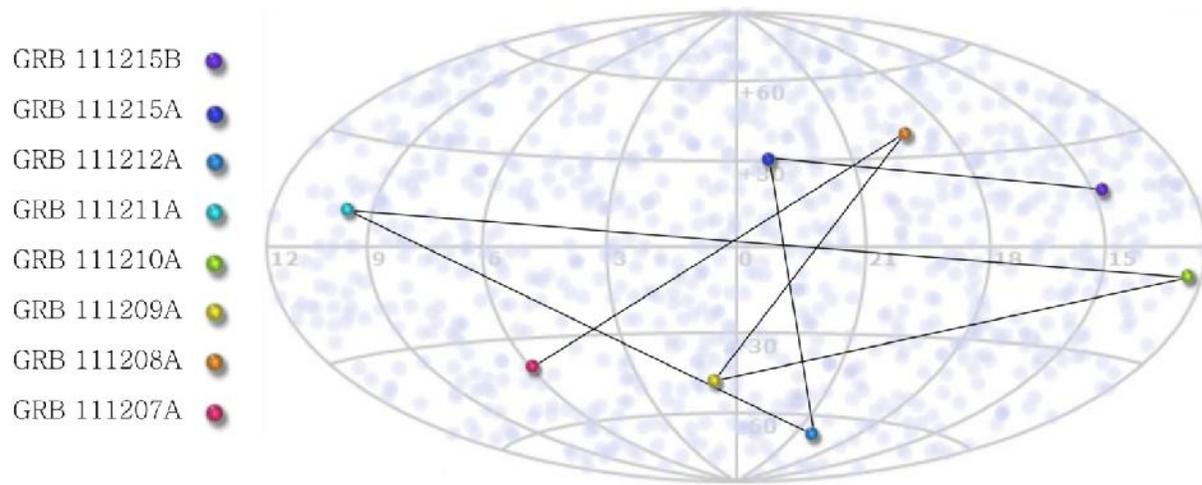

**Figure 5**

Fig.5 The ogg begins with red GRB 111207A and ends with violet GRB 111215B. Fig. 6 The ogg begins with red GRB 111215B and ends with violet GRB 120107A.

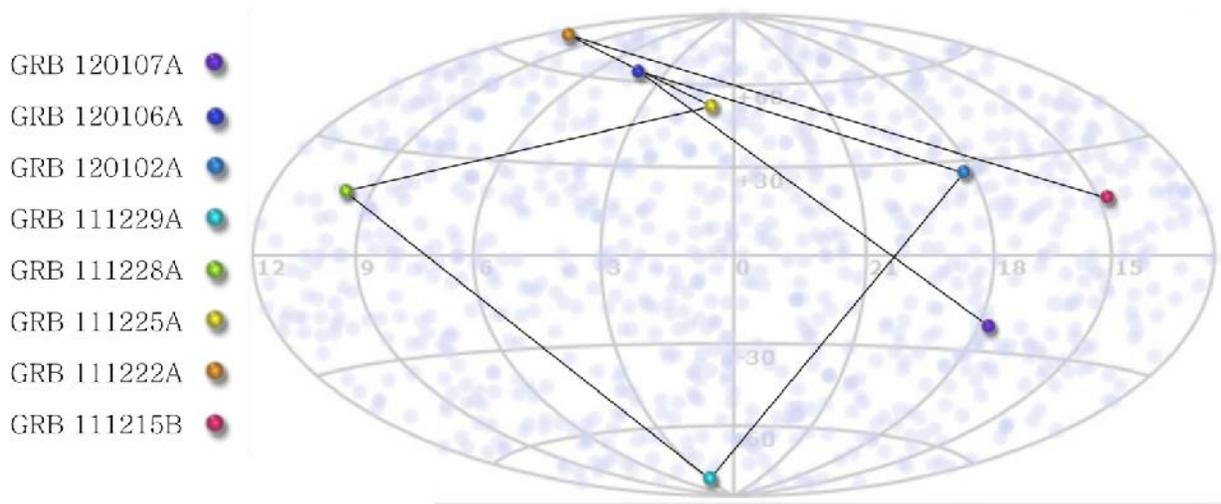

**Figure 6**

The size of the angles formed by linking eight sequential bursts skews to smaller angles. Linking 8 non sequential GRB's creates the pattern in Fig.7. A pattern never seen in sequentially linked GRB's

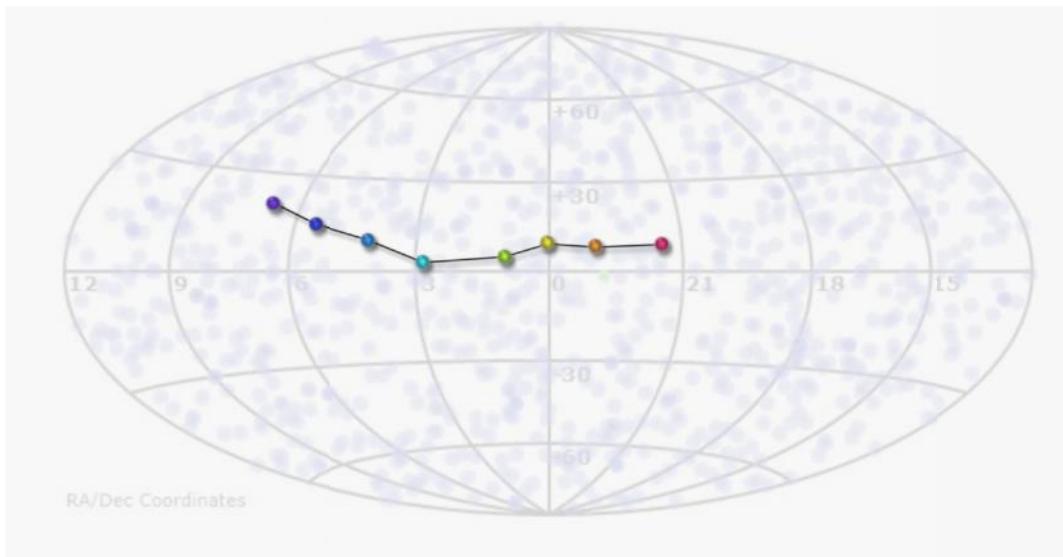

**Figure 7. Non sequential GRB's**

Fig.8. shows another linking of 8 non sequential GRB's this kind of linear pattern that is never seen when linking 8 sequential GRB's.

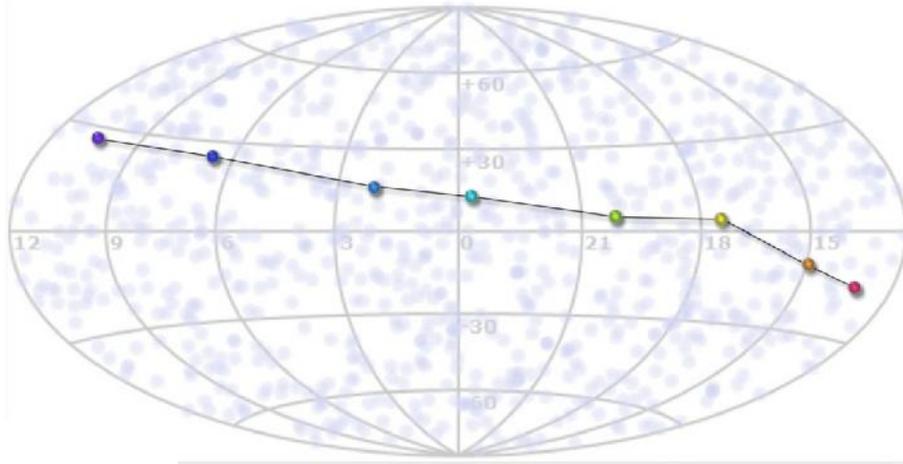

**Figure 8**

Fig.9 shows the 8 sequentially linked GRB's that occurred from 2/1/09 to 2/22/09. Beginning with red GRB 090201A and ending with GRB 090222A.

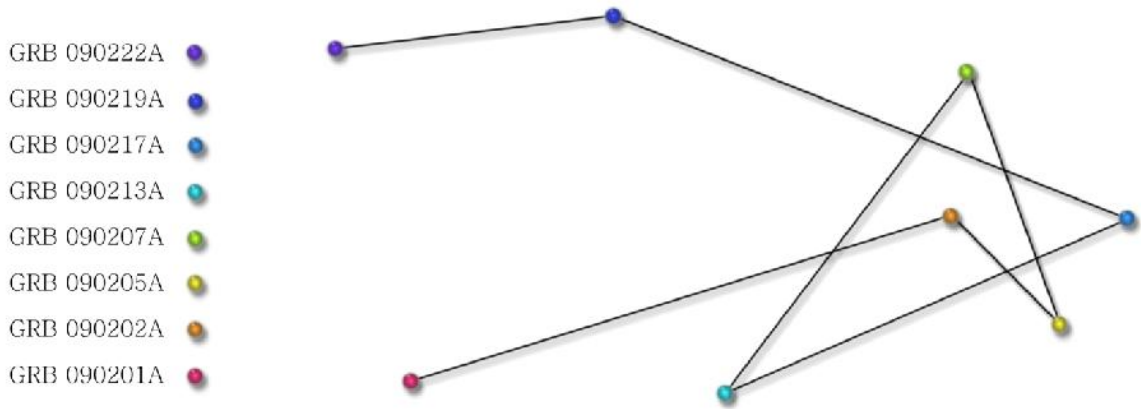

**Figure 9.** An Ogg beginning with GRB 090201A and ending with GRB 090222A

Fig.9 thru Fig. 28 presents some examples of the 140 oggs generated for this study. They all demonstrate the propensity to form smaller angles.

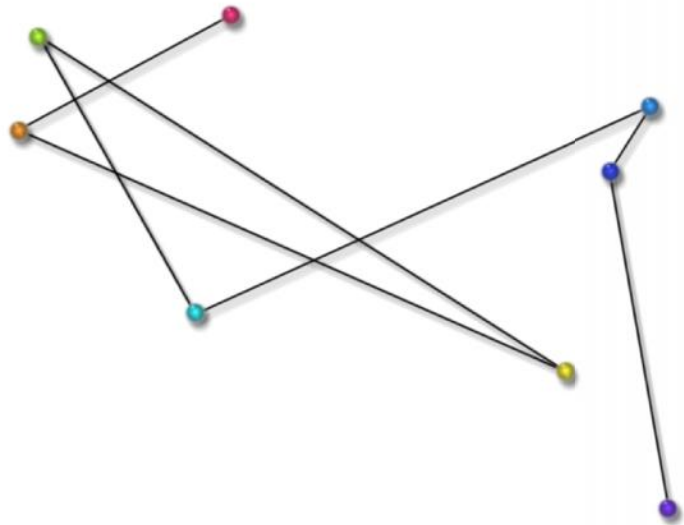

**Figure 10.** An Ogg starting with GRB 090222A and ending with GRB 090304A.

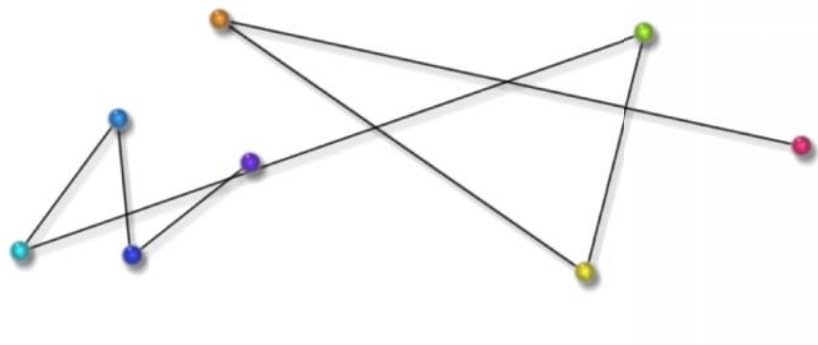

**Figure 11.** An Ogg starting with GRB 060814A and ending with GRB 060912A.

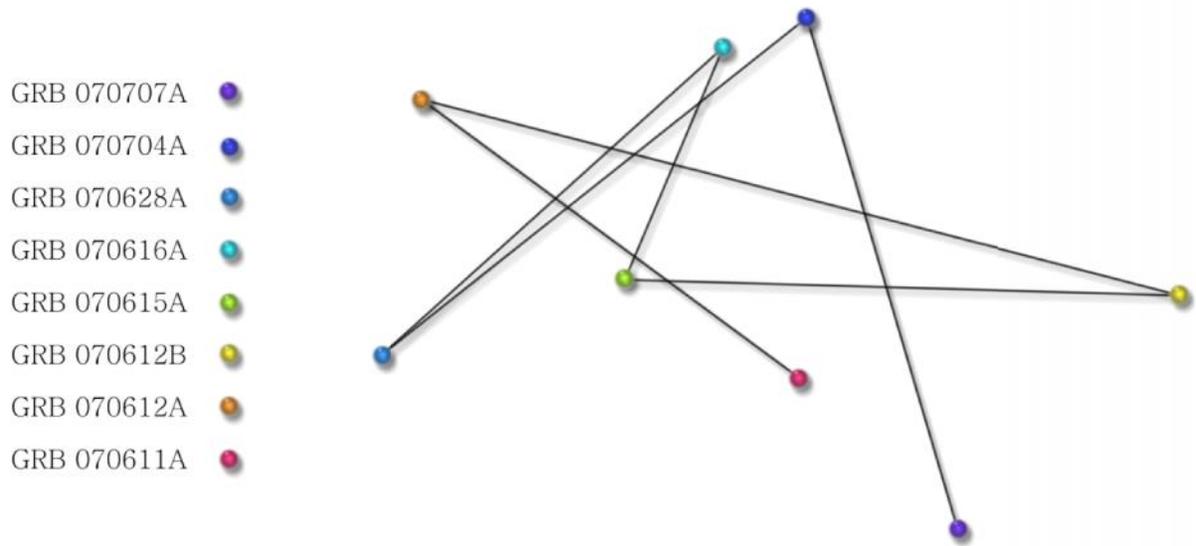

**Figure 12.** An Ogg starting with GRB 070611A and ending with GRB 070707A.

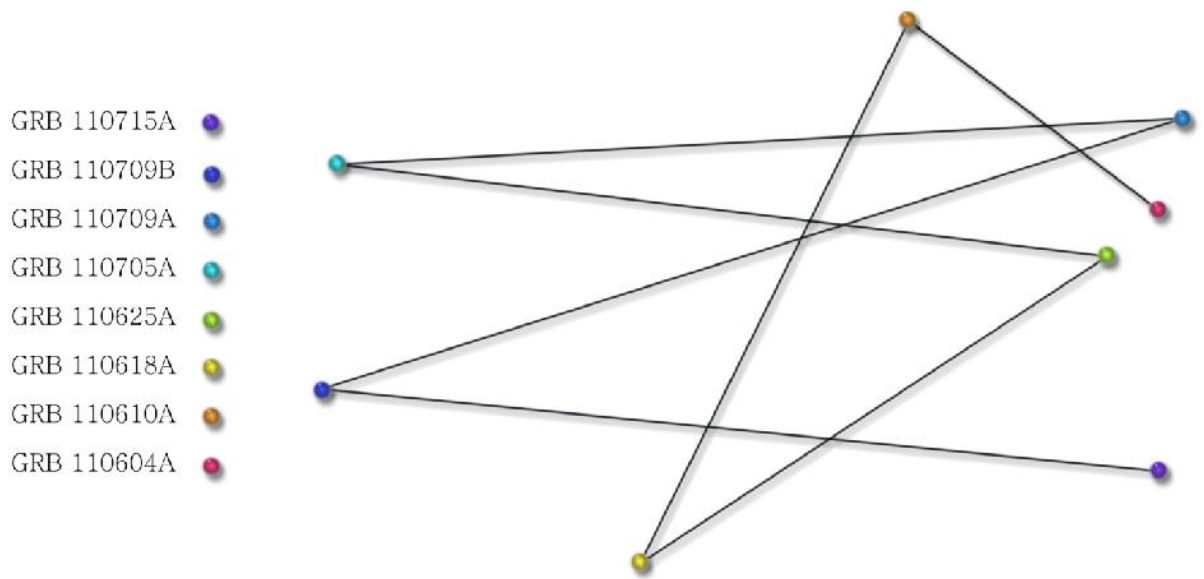

**Figure 13.** An Ogg starting with GRB 110604A and ending with GRB 110715A.

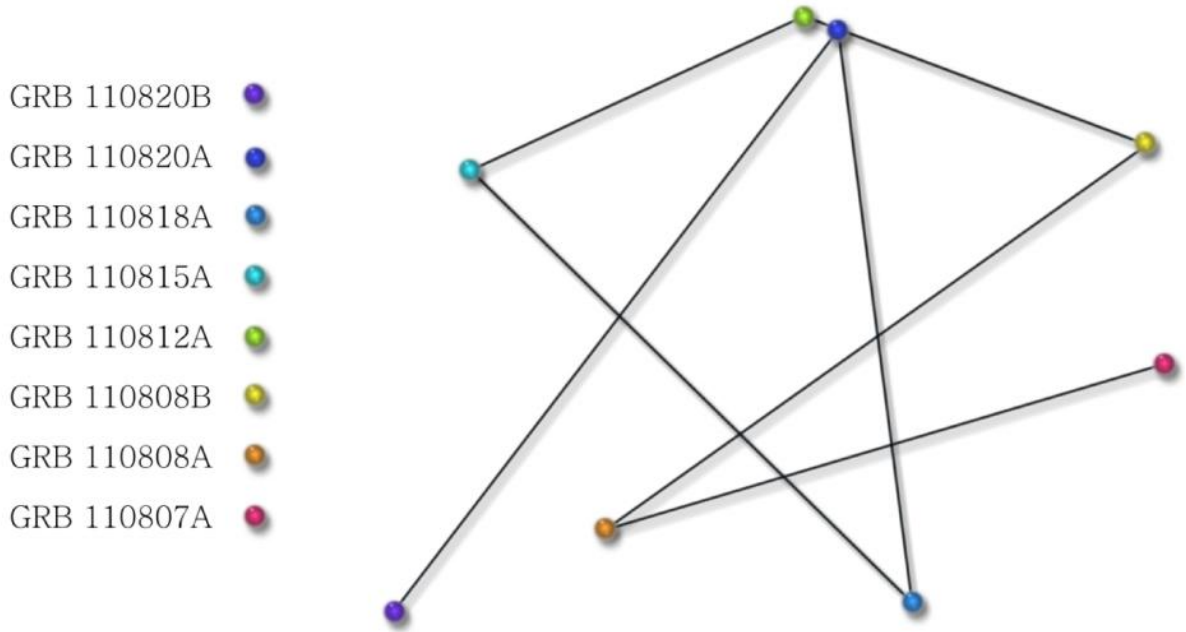

**Figure 14.** An Ogg starting with GRB 110807A and ending with GRB 110820B.

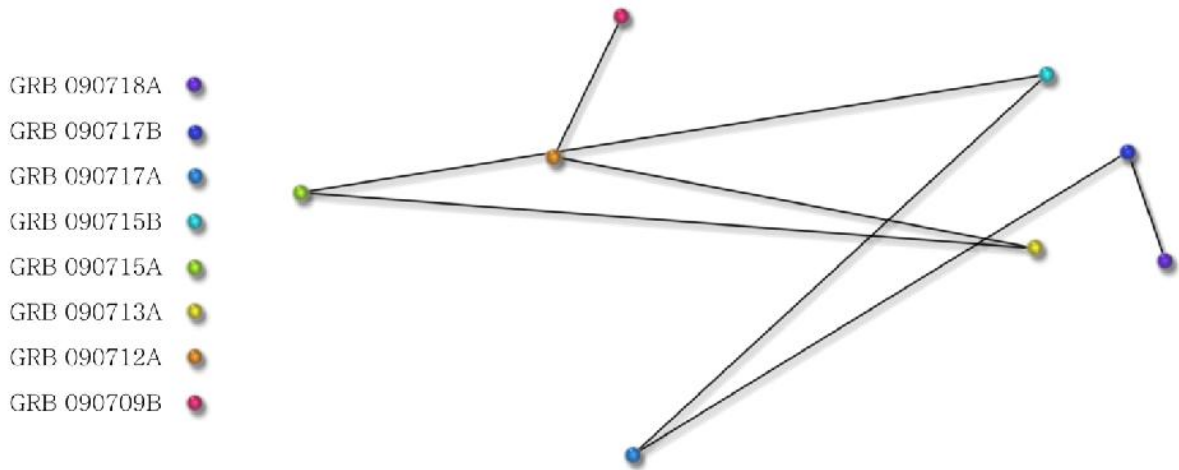

**Figure 15.** An Ogg starting with GRB 090709A and ending with GRB 090718A.

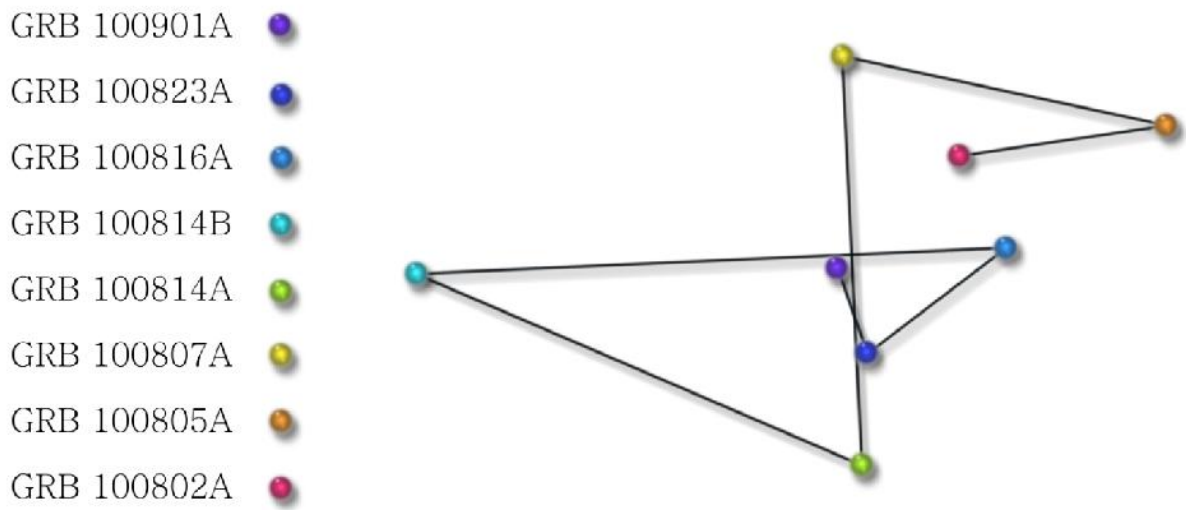

**Figure 16.** An Ogg starting with GRB 100802A and ending with GRB 100901A.

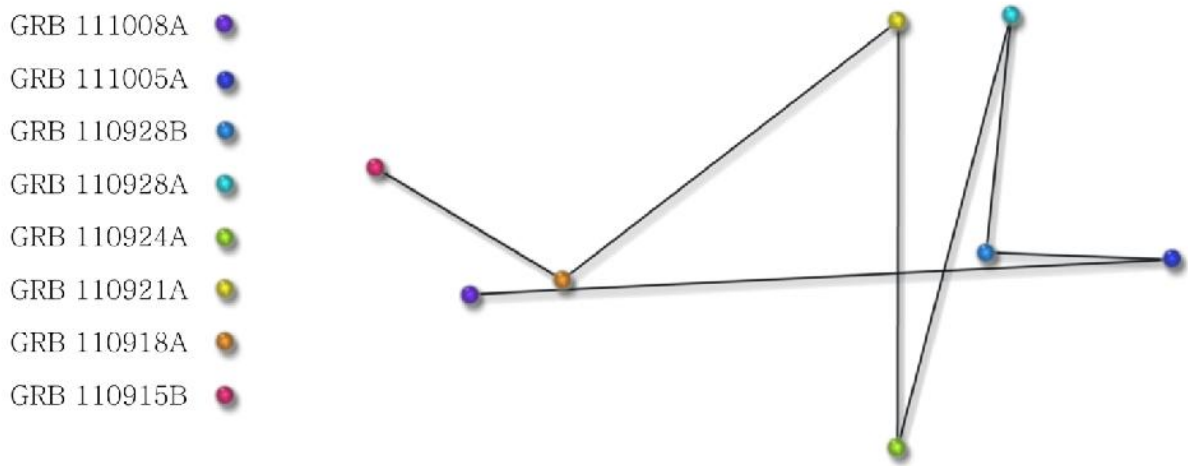

**Figure 17.** An Ogg starting with GRB 110915B and ending with GRB 111008A.

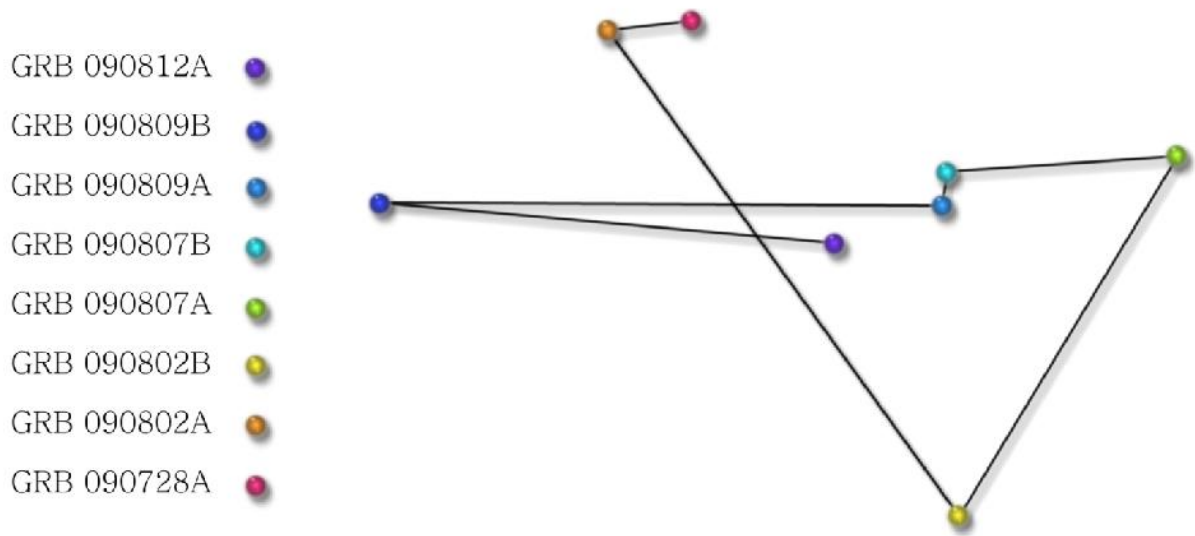

**Figure 18.** An Ogg starting with GRB 090728A and ending with GRB 090812A.

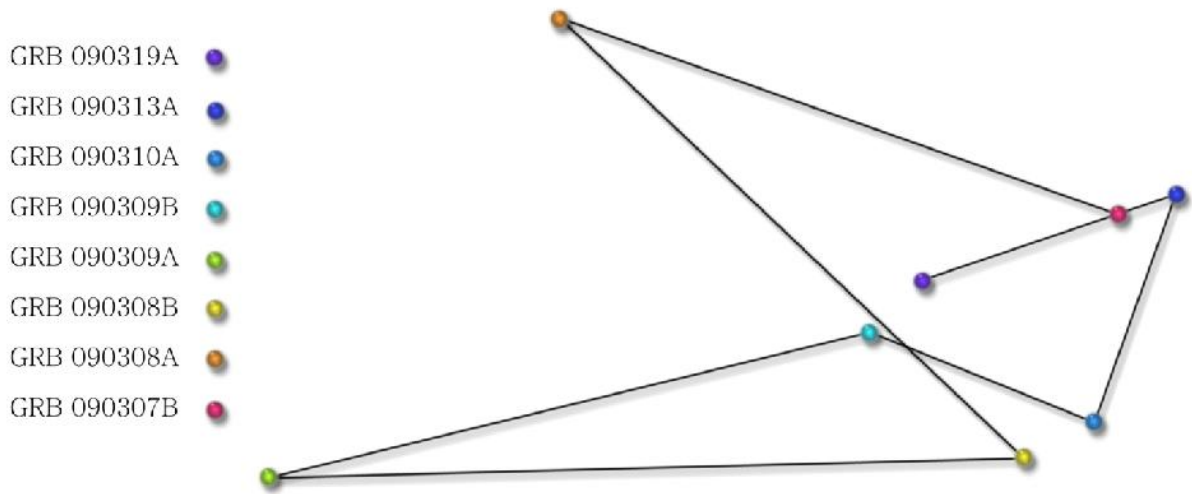

**Figure 19**. An Ogg starting with GRB 090307A and ending with GRB 090319A.

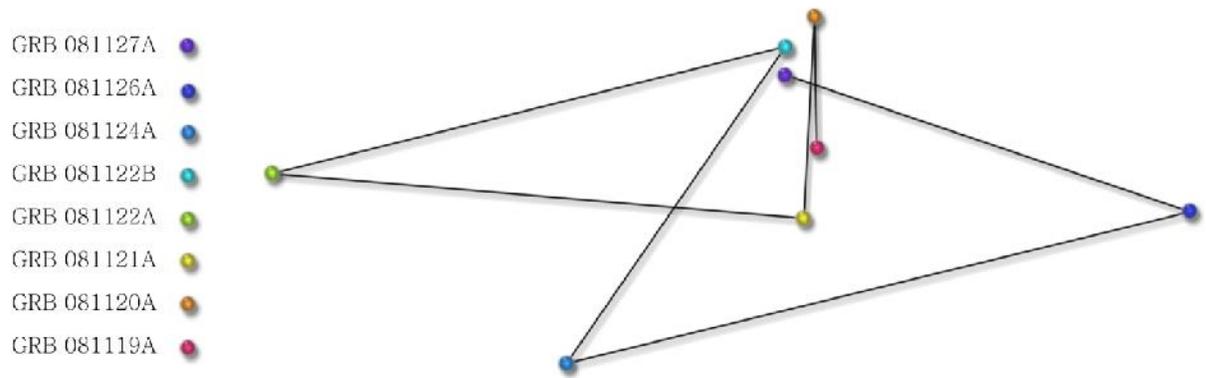

**Figure 20.** An Ogg starting with GRB 081119A and ending with GRB 081127A.

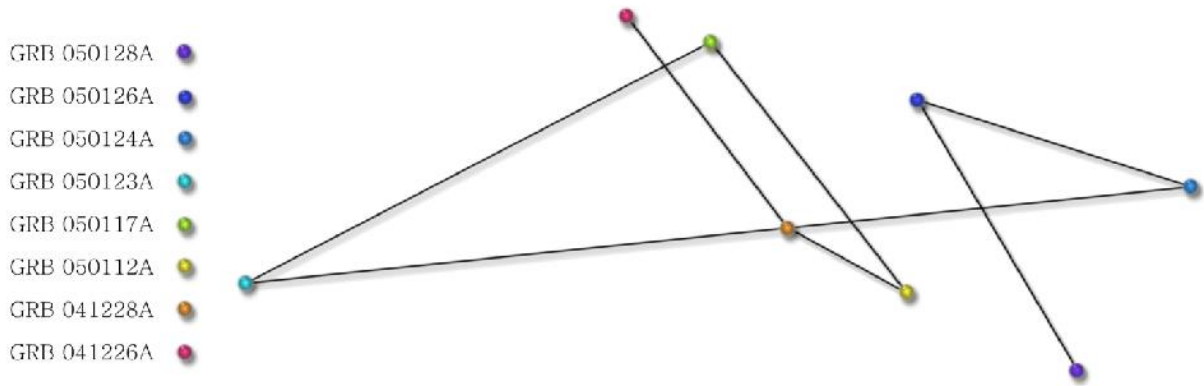

**Figure 21.** An Ogg starting with GRB 041226A and ending with GRB 050128A.

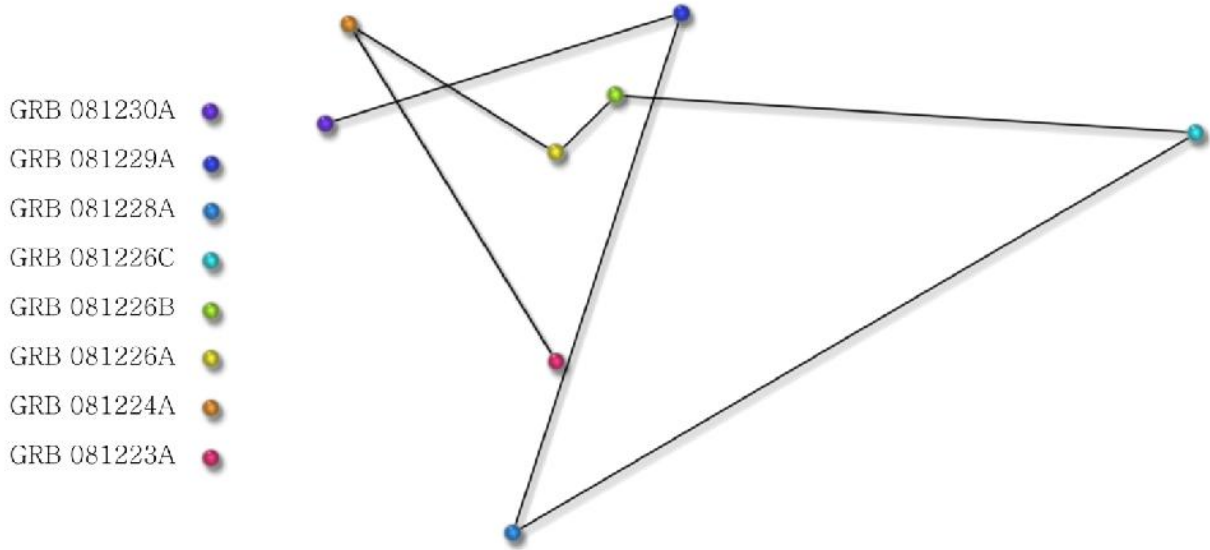

**Figure 22.** An Ogg starting with GRB 081223A and ending with GRB 081230A.

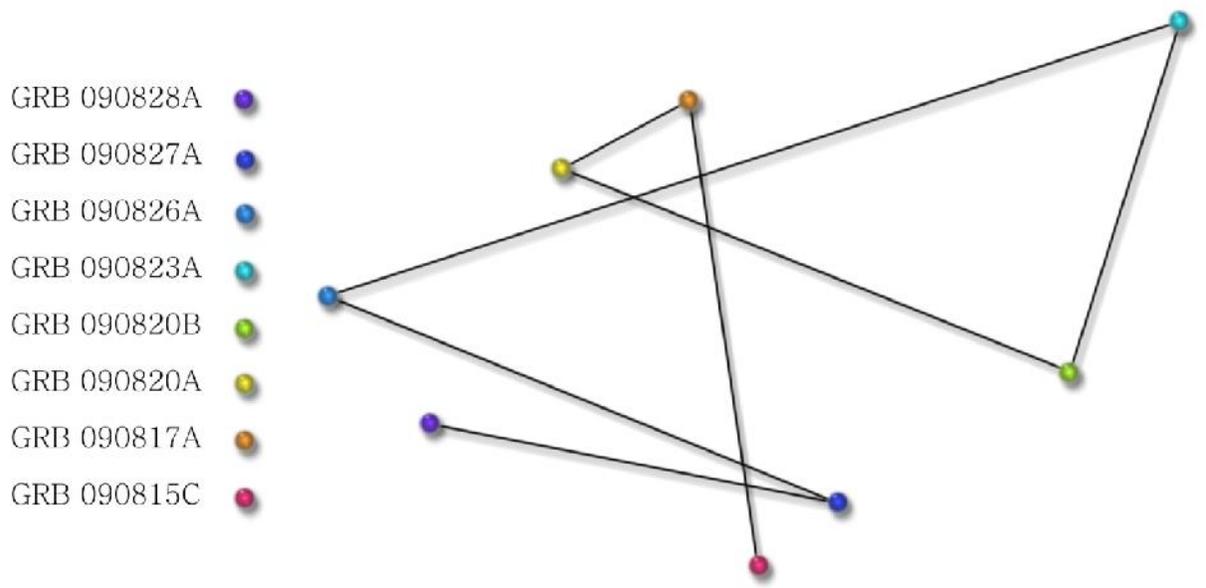

**Figure 23.** An Ogg starting with GRB 090815C and ending with GRB 090828A.

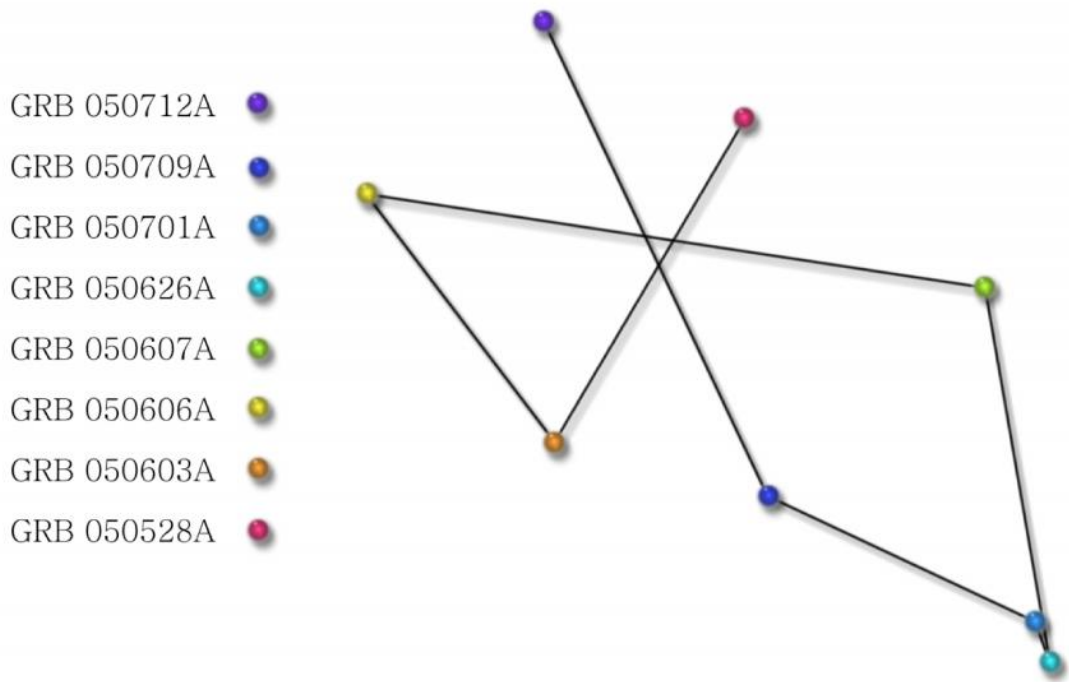

**Figure 24.** An Ogg starting with GRB 050528A and ending with GRB 050712A**.**

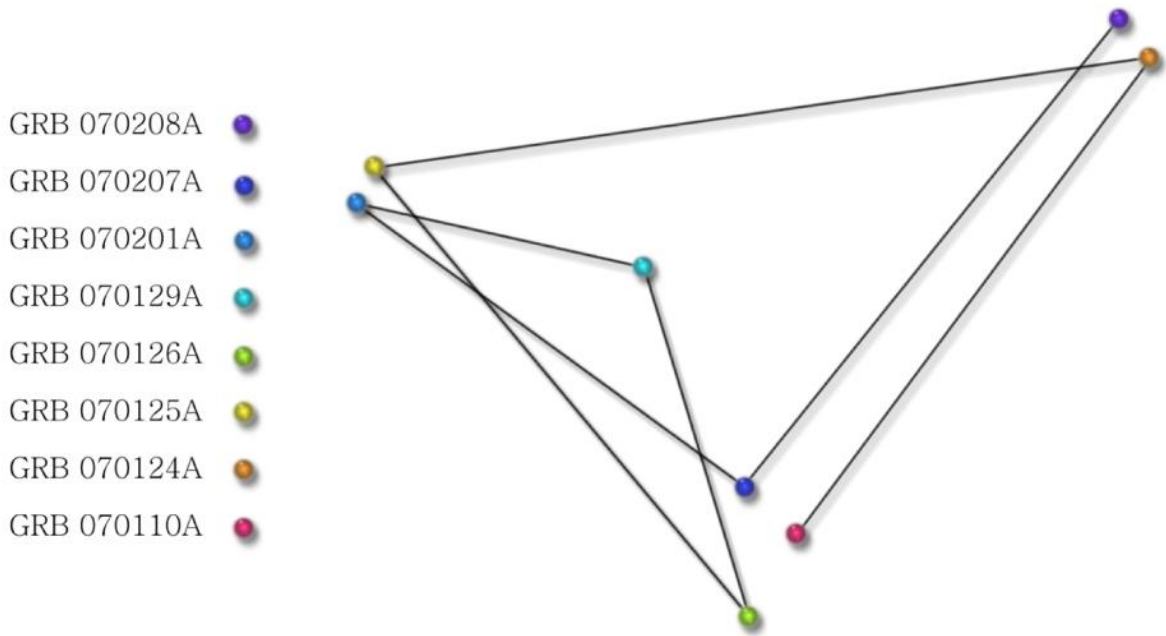

**Figure 25.** An Ogg starting with GRB 070110A and ending with GRB 070208A.

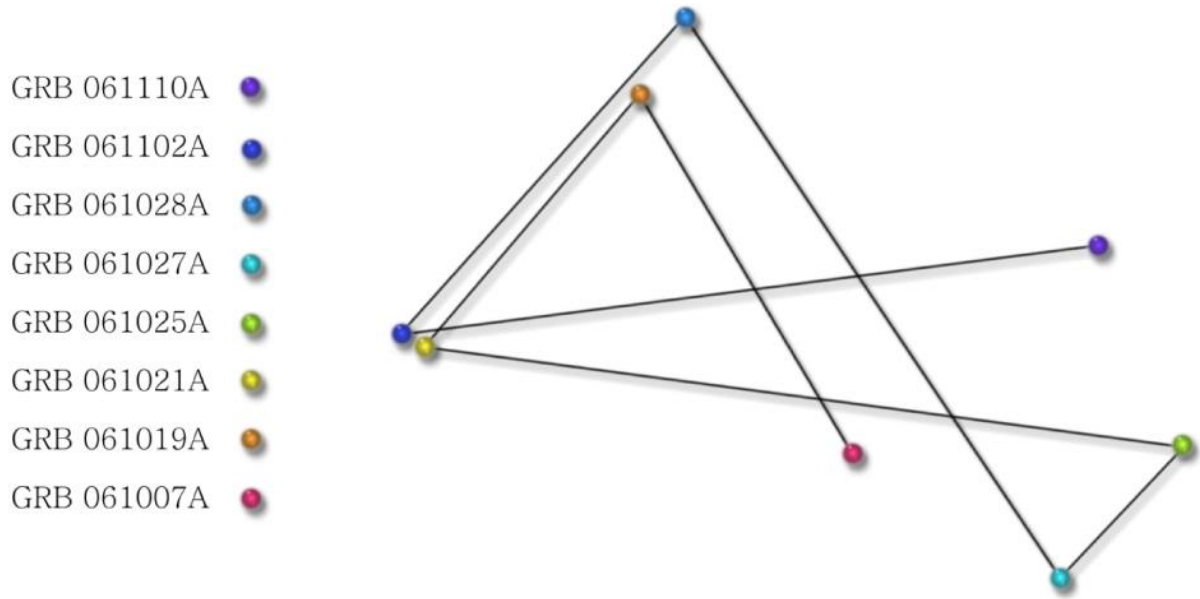

**Figure 26.** An Ogg starting with GRB 061007A and ending with GRB 061110A.

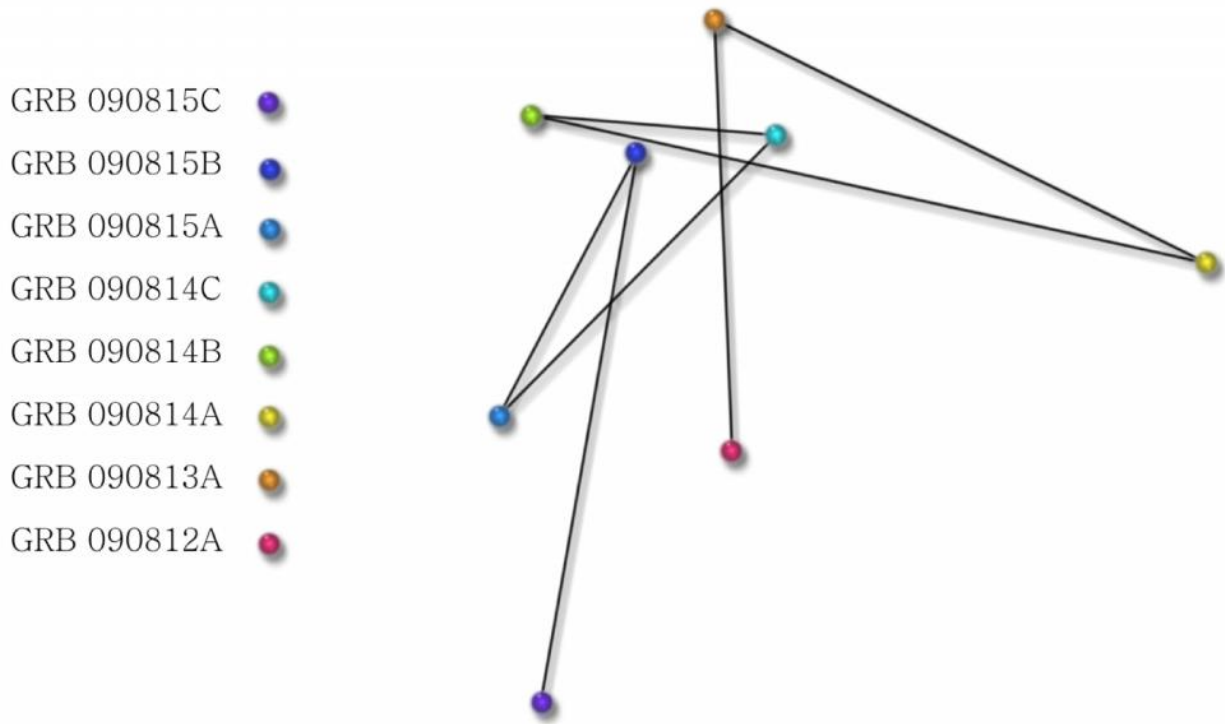

**Figure 27.** An Ogg starting with GRB 090812A and ending with GRB 090815C.

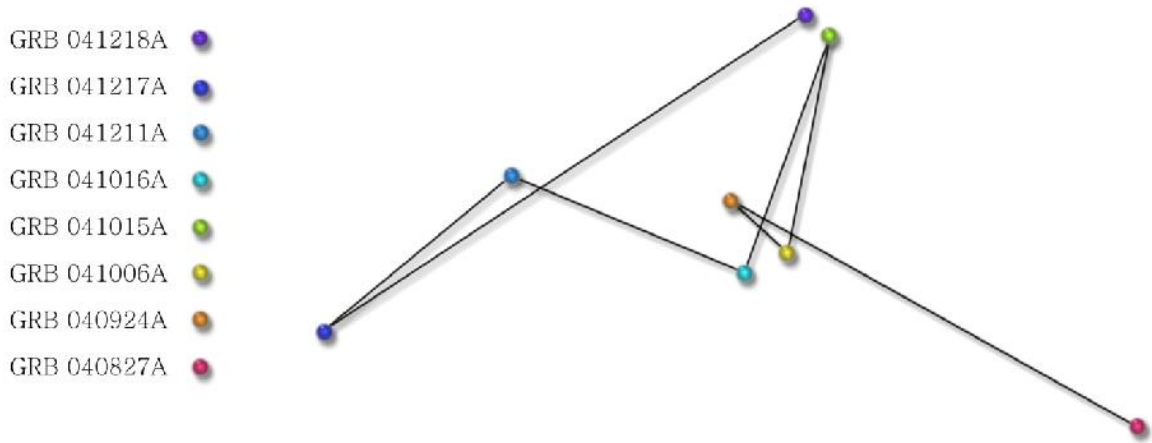

**Figure 28.** An Ogg starting with GRB 040827A and ending with GRB 041218A.

## 4. Equidistant GRB's and CPPC

The following figures feature a GRB that is positioned at a location that is equidistant from two or more bursts in the same series. Using the ogg from Fig. 28 to demonstrate this property Fig.29 uses blue GRB 04112A as the origin of two CPPC (concentric paired point circles). The smaller circle intersects cyan GRB 04106A and indigo GRB 041217A, the larger circle touches green GRB 041015A and violet GRB 041218A.

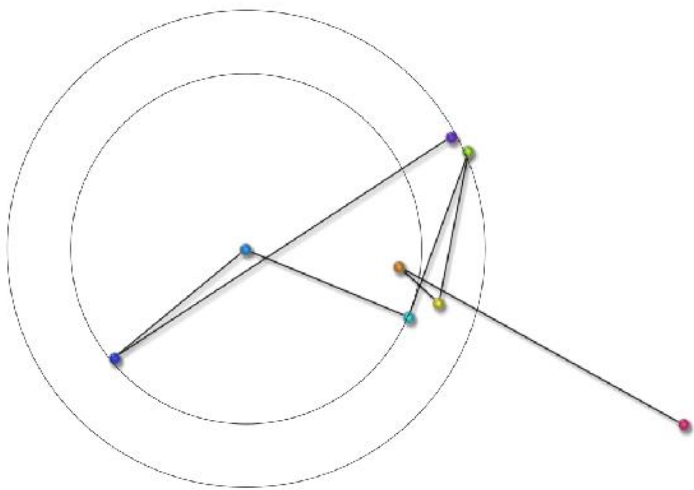

**Figure 29.** The Ogg starting with GRB 040827A and ending with GRB 041218A. Using the blue dot GRB 041211A as a center point for two concentric paired point circles.

Fig.30 Using the same ogg with cyan GRB 041016A as the center for 2 concentric paired point circles, the smaller circle intersects blue GRB 041211A and green GRB 041015A, the larger circle passes through red GRB 040827A and indigo GRB 041217A.

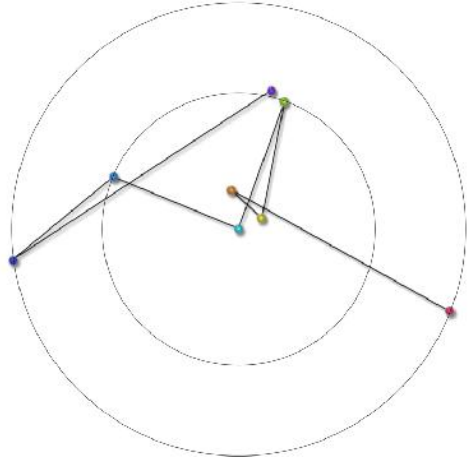

**Figure 30**

In Fig. 31 Orange is the center for 2 concentric paired point circles, yellow GRB 041006A and cyan GRB 041016A (small circle), green GRB 041015A and violet GRB 041218A (large circle). Fig. 32 The indigo GRB 041217A as the origin for 2 CPPC, the smallest intersects orange GRB 040924A and cyan GRB 041016A. The larger circle intersects violet GRB 041218A and green GRB 041015A.

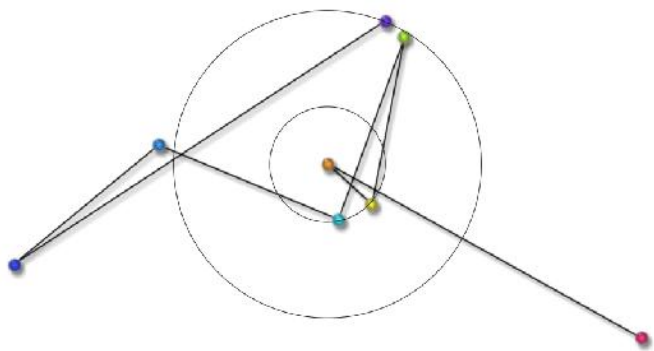

**Figure 31**

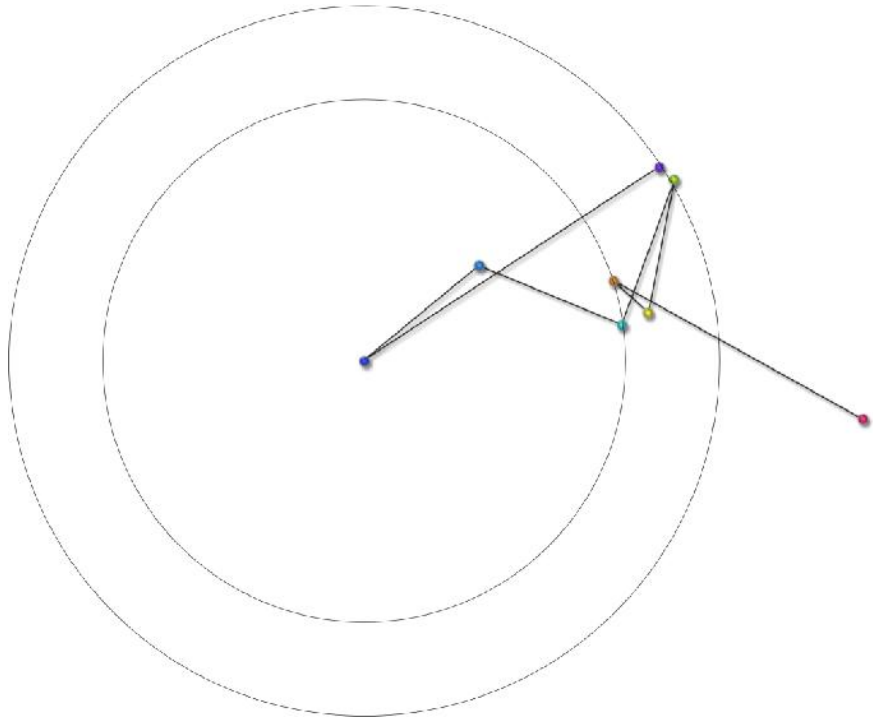

**Figure 32**

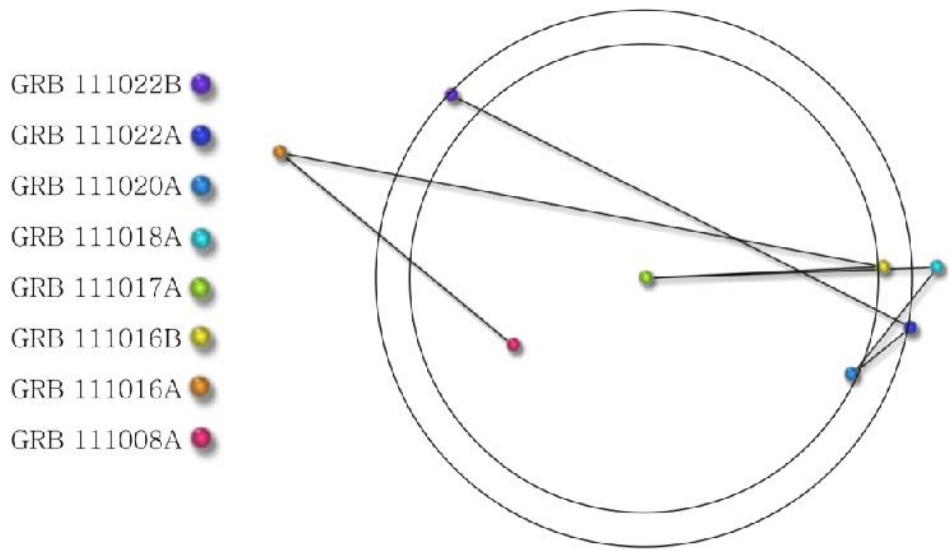

**Figure 33**

Fig.33 The ogg from Fig.2 green GRB 111017A as the center for 2 CPPC blue GRB 111020A and yellow GRB 111016A (small circle), indigo GRB 111022A and violet GRB 111022B (large circle).

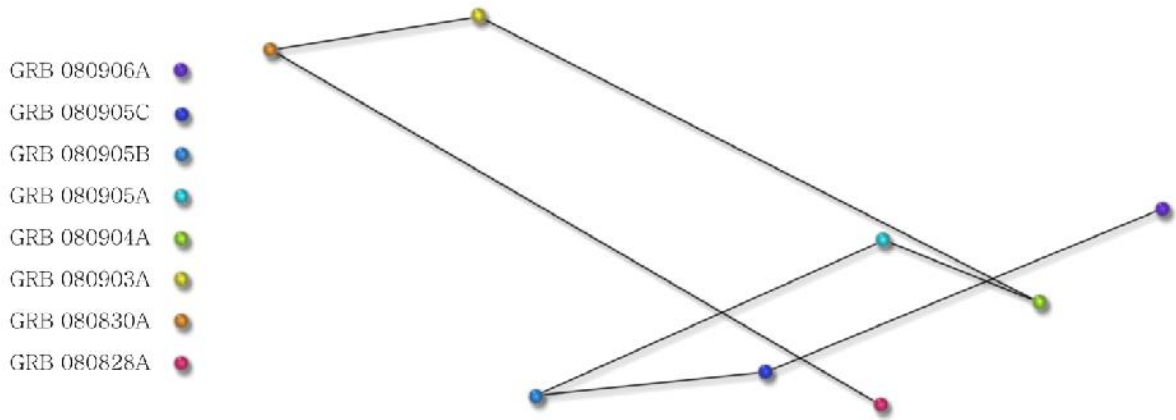

**Figure 34.** An Ogg starting with GRB 090222A and ending with GRB 090304A.

The ogg in Fig.34 has 4 CPPC seen in the following figures.

Fig.35 The cyan GRB 080905A is equidistant from the green GRB 080904A, the red GRB 080828A and the indigo GRB 080905c.

In Fig.36 red GRB 080828A is equidistant from blue GRB 080905B and violet GRB 080906A.

In Fig.37 blue GRB 080905B is equidistant from cyan GRB 080905A and yellow GRB 080830A.

In Fig.38 yellow GRB 080830A is equidistant from indigo GRB 080905c and cyan GRB 080905A.

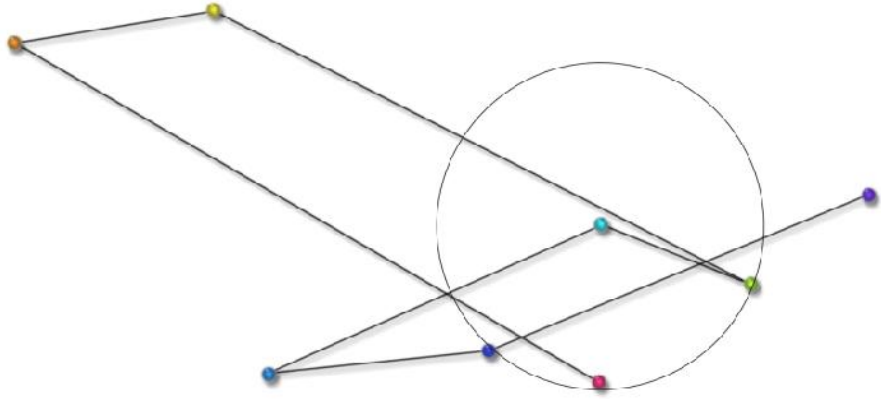

**Figure 35**

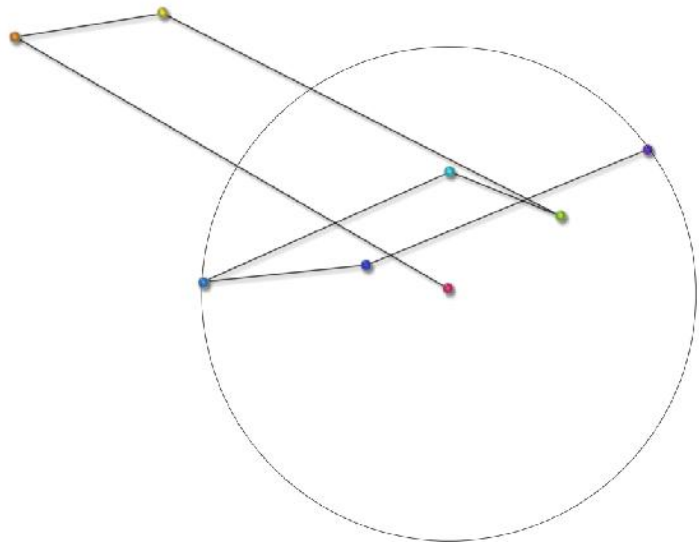

**Figure 36.**

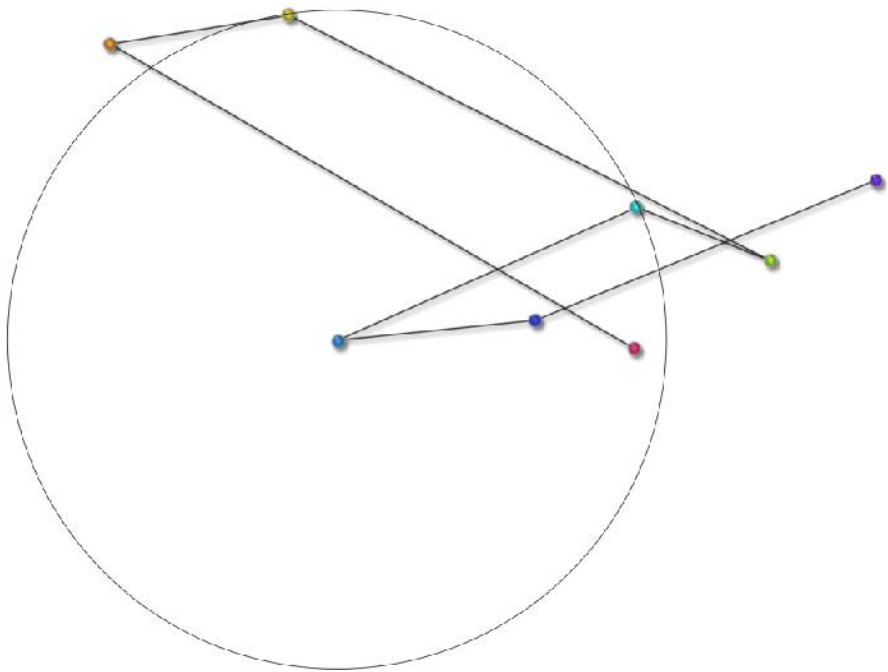

**Figure 37**

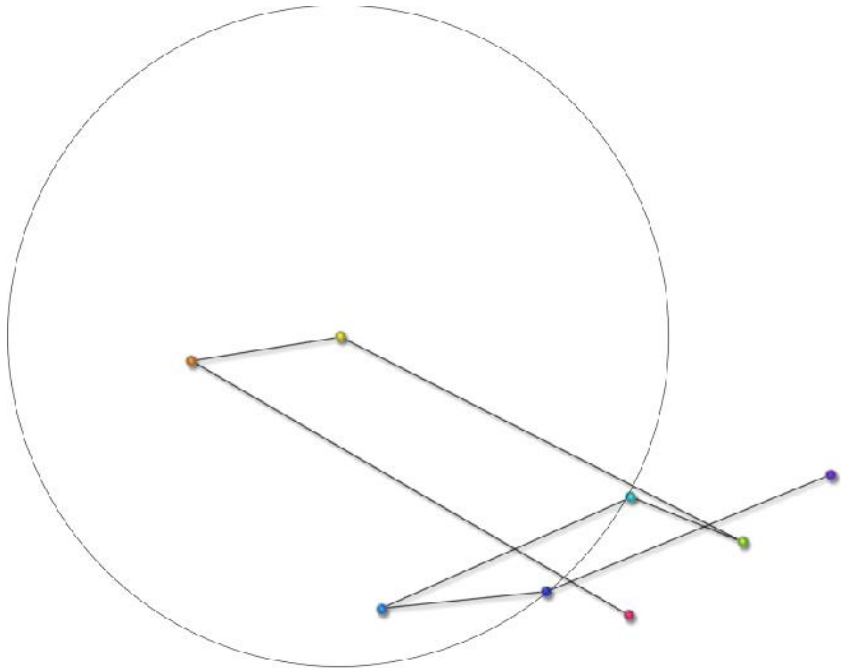

**Figure 38**

Fig.39 has 4 CPPC seen in the following figures.

Fig.40 red GRB 100901A is equidistant to cyan GRB 100906A and blue GRB 100909A.
Fig.41. cyan GRB 100906A is equidistant to yellow GRB 100905A and indigo GRB 100910A.
Fig.42 violet 100906A is equidistant orange GRB 100902A and green GRB 100905A.
Fig.43 indigo 100906A is equidistant red GRB 100901A and green GRB 100905A.

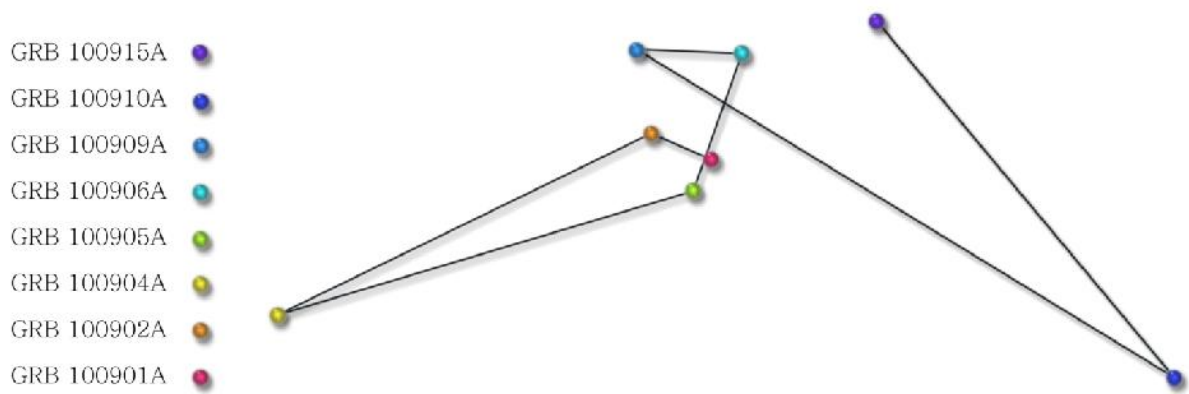

**Figure 39.** An Ogg starting with GRB 100901A and ending with GRB 100915A.

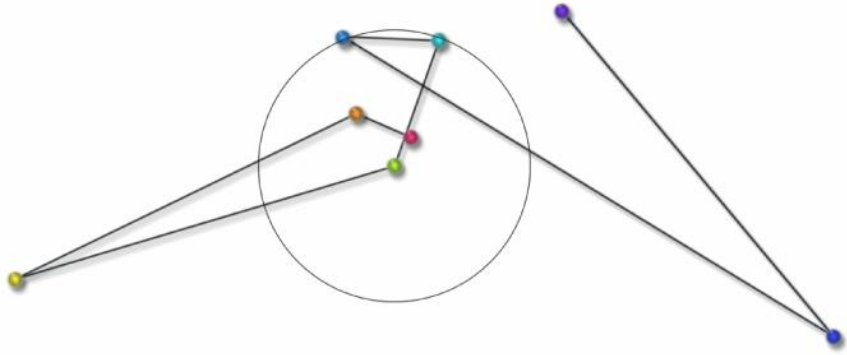

**Figure 40**

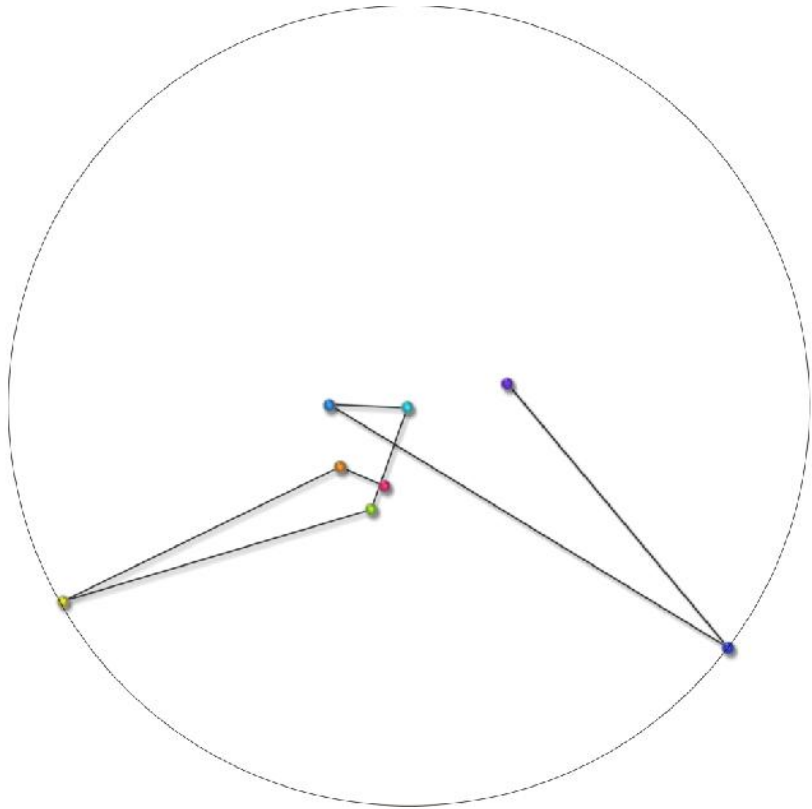

**Figure 41**

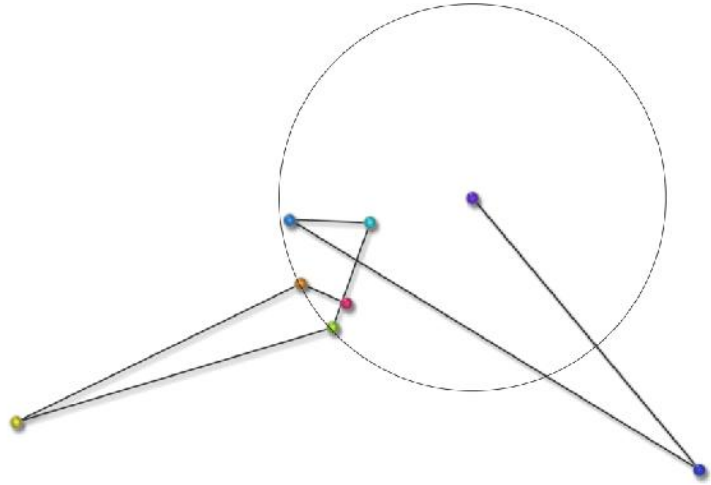

**Figure 42**

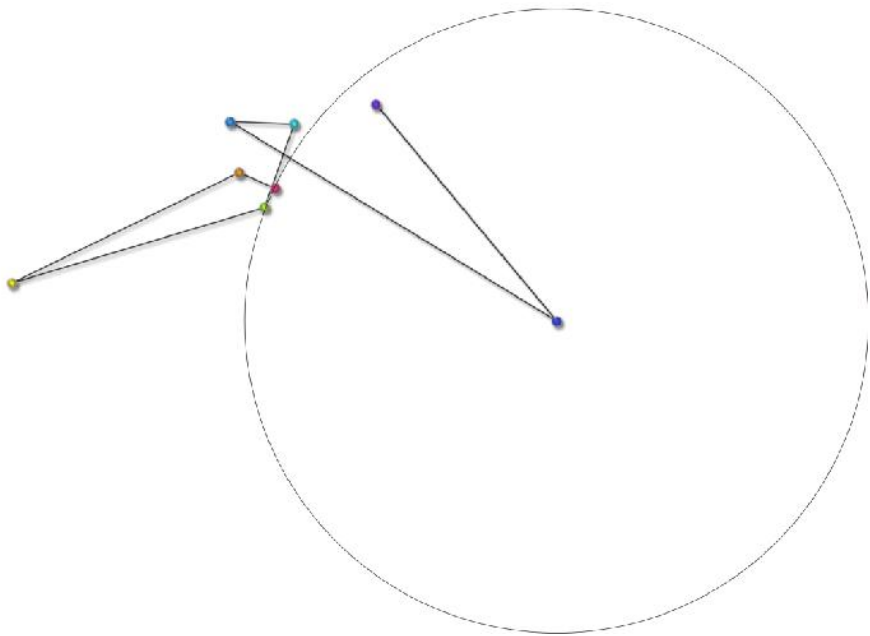

**Figure 43**

## 5. Linking Sequential Oggs

To further explore GRB's positioned at equidistance the author combined three sequential oggs to create a pattern of 22 sequential GRB's. Fig.44, Fig.45, and Fig.46

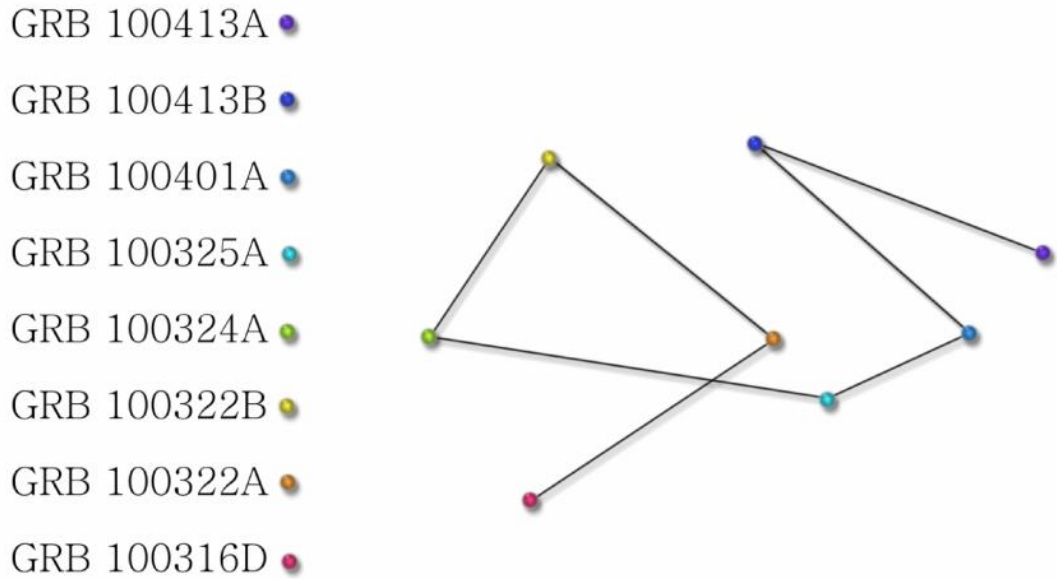

**Figure 44.** This ogg begins with GRB 100316D and ends with GRB100413A.

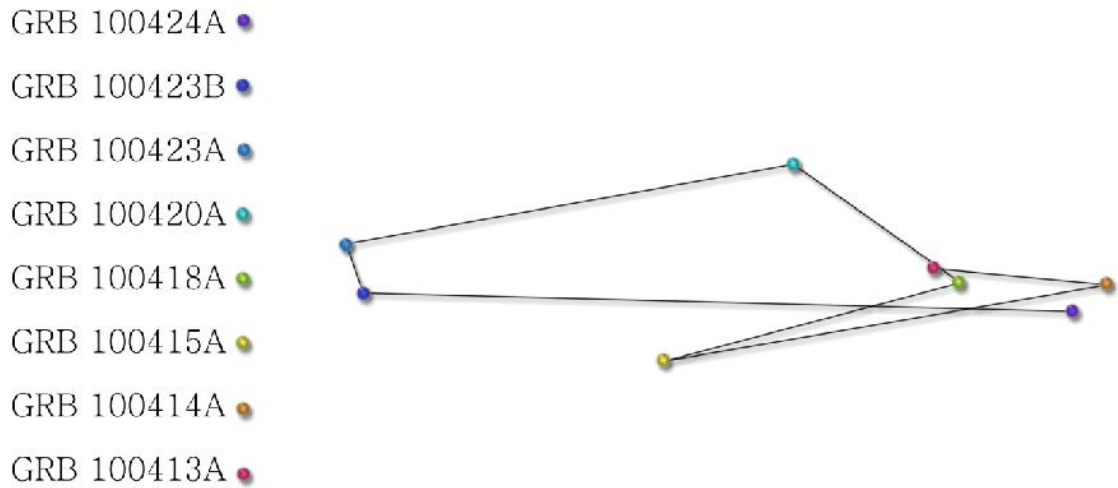

**Figure 45.** This ogg begins with GRB 100413A and ends with GRB 100424A.

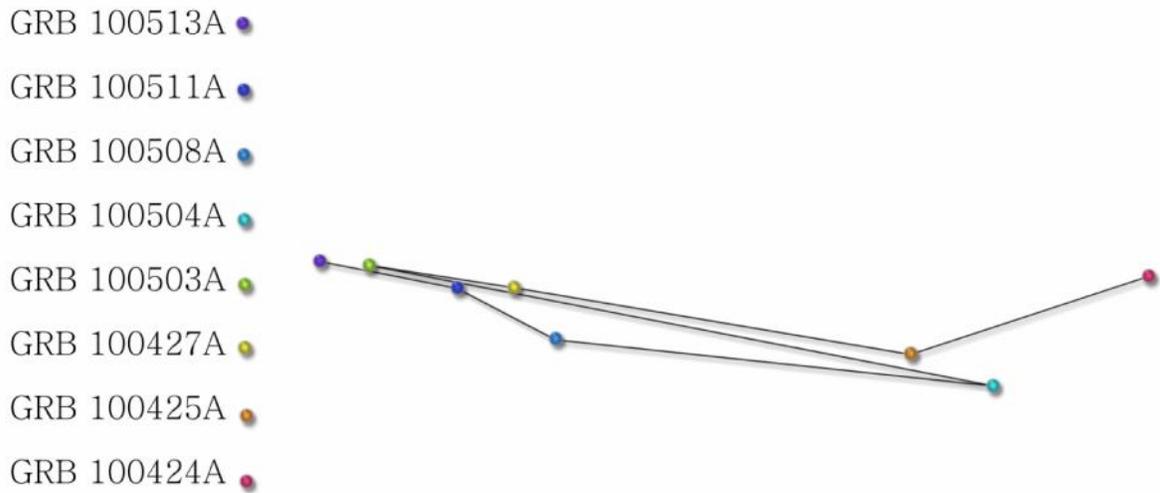

**Figure 46.** This ogg begins with GRB 100424A and ends with GRB 100513A.

Fig.47. The 22 sequential bursts (beginning with 1, GRB 100316D and ending with 22, GRB 100513A) and the 21 lines that link them. The pattern formed by sequentially linking three Oggs will be referred to as a Throgg.

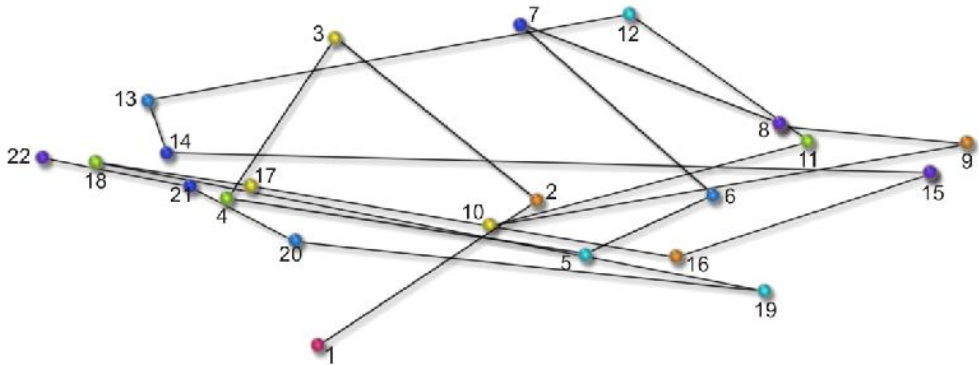

**Figure 47.** A throgg indicating the location of 22 sequential gamma ray bursts numbered from 1-22.

Fig.48 to Fig.69. Every GRB, from 1-22, as the center point for the creation of CPPC. The circles are labeled alphabetically A being the smallest. The box indicates by number which GRB's are intersected by each circle.

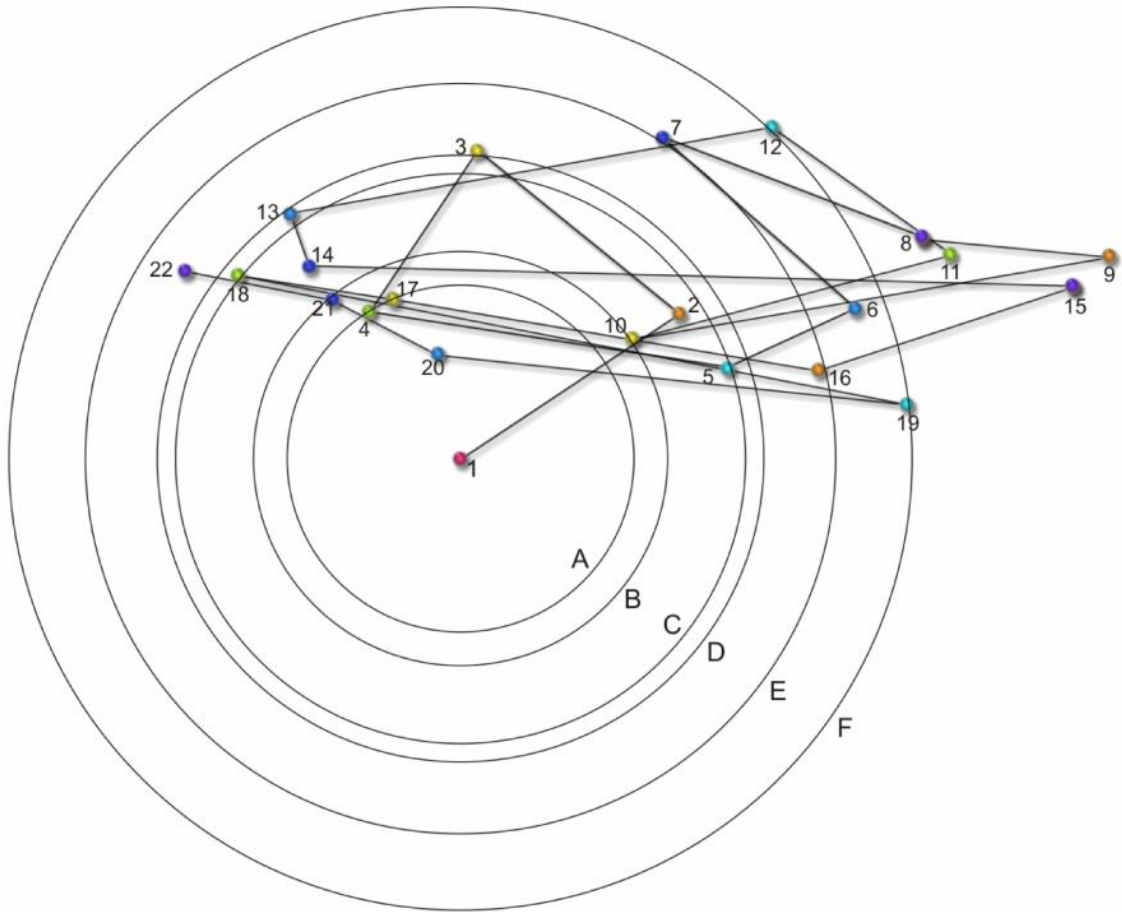

| A | 4  | 17 |  |
|---|----|----|--|
| B | 10 | 21 |  |
| C | 5  | 18 |  |
| D | 3  | 13 |  |
| E | 7  | 16 |  |
| F | 12 | 19 |  |

**Figure 48.** Using 1 (GRB 100316D) as the center of 6 CPPC.

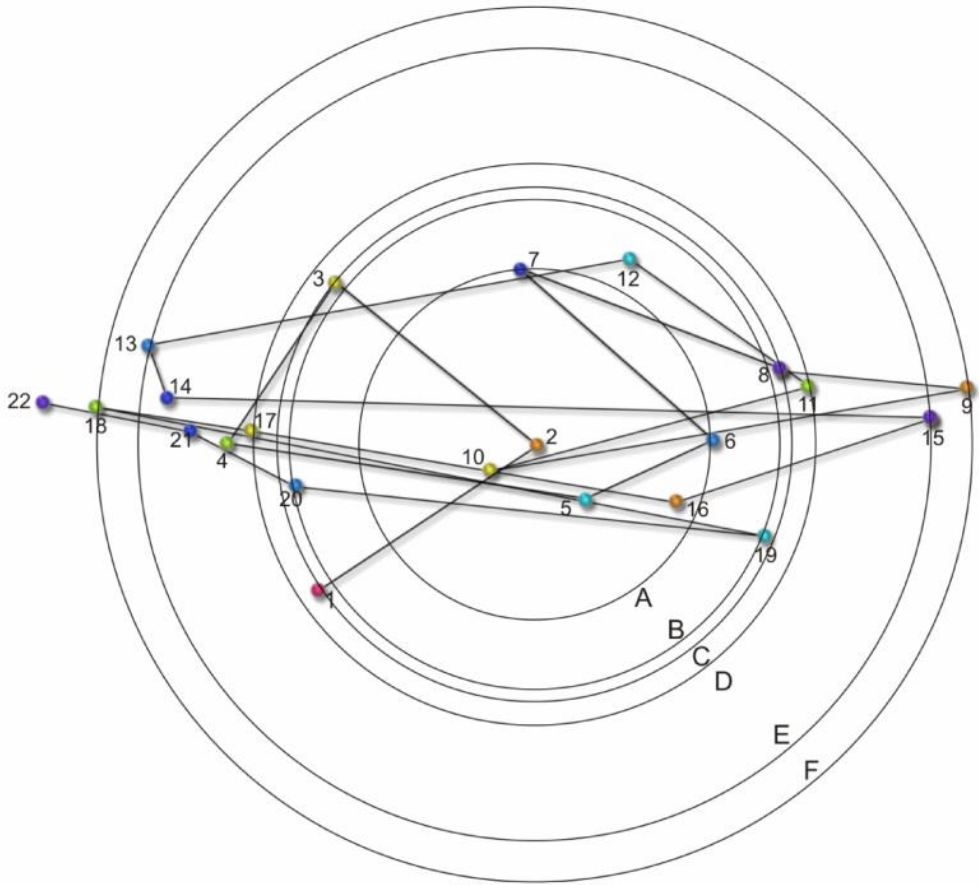

| A | 6  | 7  |   |
|---|----|----|---|
| B | 19 | 20 |   |
| C | 1  | 3  | 8 |
| D | 11 | 17 |   |
| E | 13 | 15 |   |
| F | 9  | 8  |   |

**Figure 49.** Using 2 (GRB 100322A) as the center of 6 CPPC labeled A, B, C, D, E, and F, each circle intersects two or more GRB's that are identified by number from 1-22.

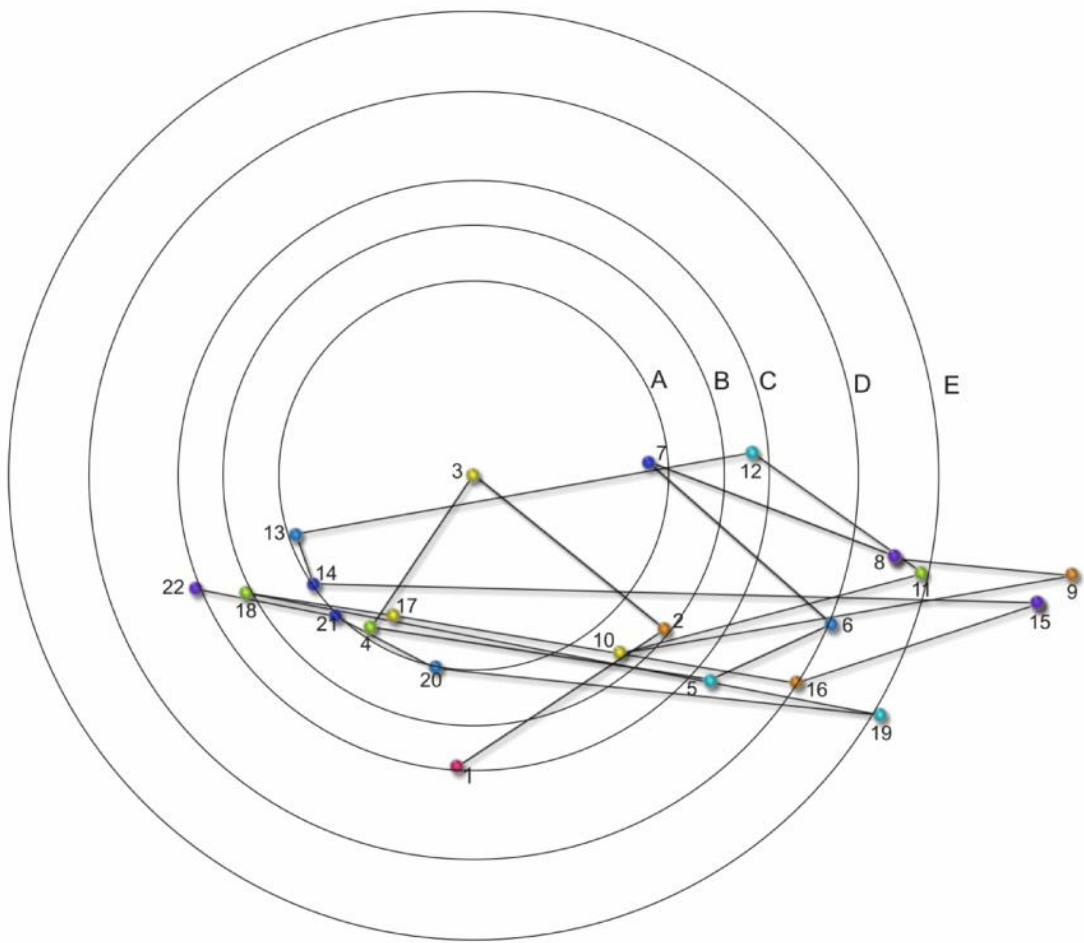

| A | 14 | 21 |  |
|---|----|----|--|
| B | 2  | 18 |  |
| C | 1  | 22 |  |
| D | 6  | 16 |  |
| E | 11 | 19 |  |

**Figure 50.** Using 3 (GRB 100322B) as a center point creates 5 CPPC labeled A, B, C, D and E. Each circle intersects two GRB's that are identified by number from 1-22.

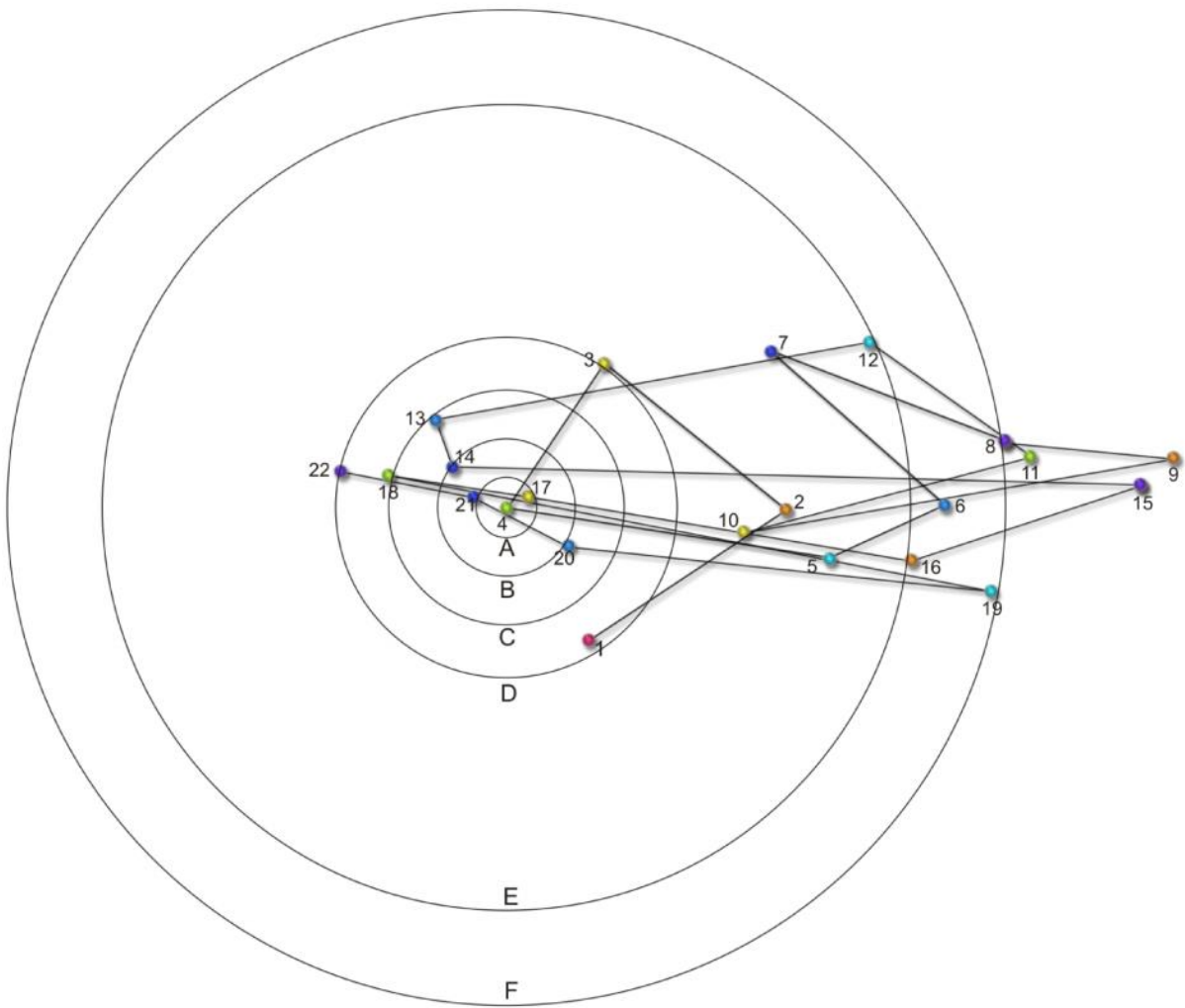

|   |    |    |   |
|---|----|----|---|
| A | 17 | 21 |   |
| B | 14 | 20 |   |
| C | 13 | 18 |   |
| D | 3  | 22 |   |
| E | 12 | 16 |   |
| F | 8  | 19 |   |

**Figure 51.** Using 4 (GRB100324A) as a center point creates 6 CPPC labeled A, B, C, D, E and F. Each circle intersects two GRB's that are identified by number from 1-22.

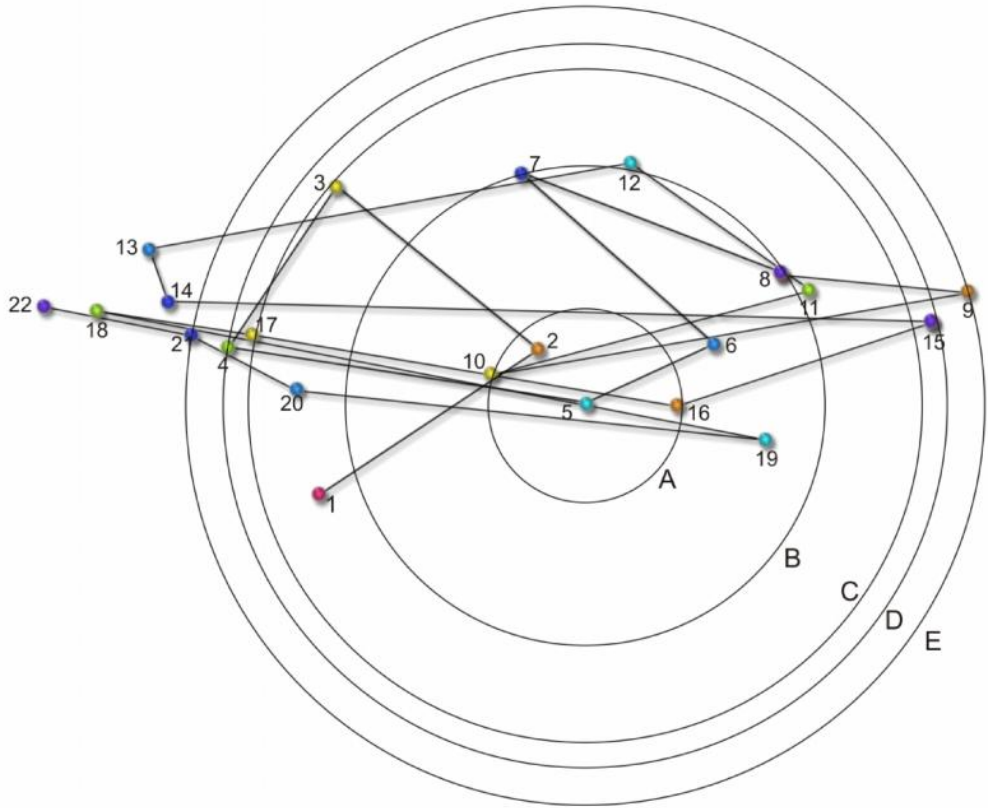

| A | 10 | 16 |    |
|---|----|----|----|
| B | 7  | 8  | 12 |
| C | 3  | 17 |    |
| D | 4  | 15 |    |
| E | 2  | 9  |    |

**Figure 52.** Using 5 (GRB 100325A) as a center point creates 5 CPPC labeled A, B, C, D and E. Each circle intersects two or more GRB's that are identified by number from 1-22.

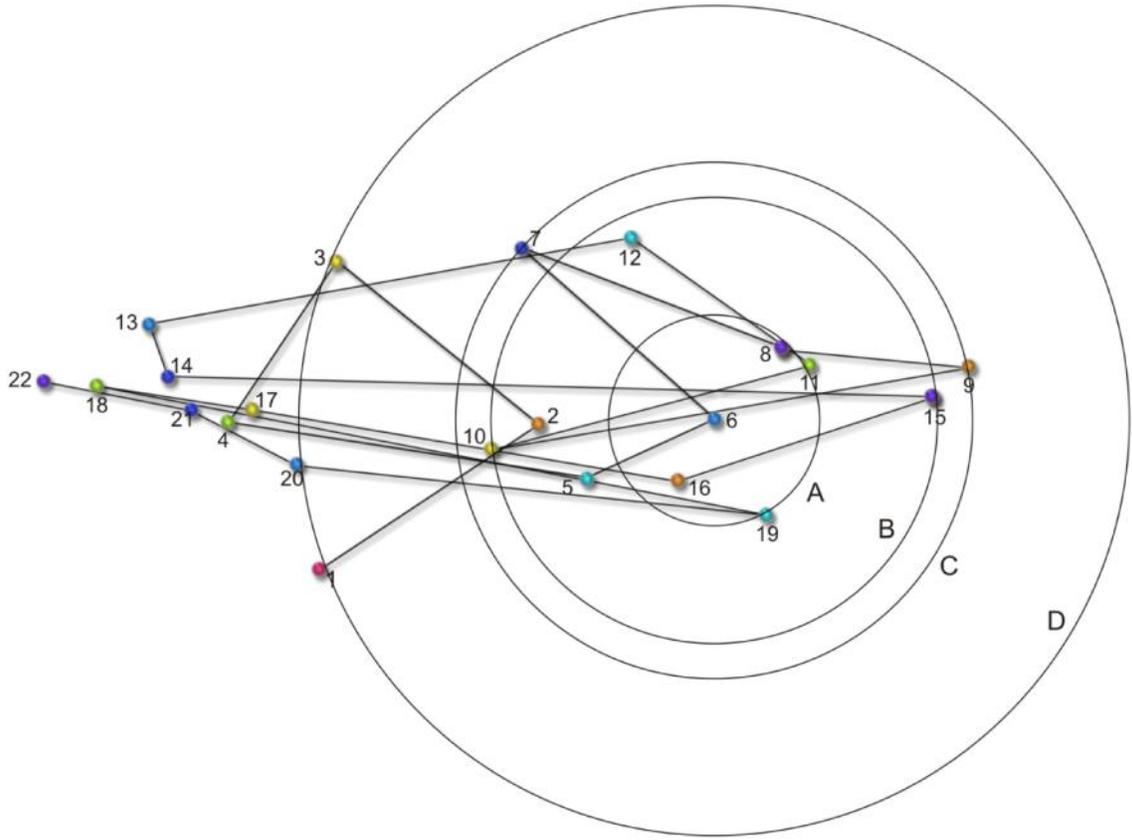

| A | 8  | 11 | 19 |
|---|----|----|----|
| B | 10 | 15 |    |
| C | 7  | 9  |    |
| D | 1  | 3  | 20 |

**Figure 53.** Using 6 (GRB 100401A) as a center point creates 4 CPPC labeled A, B, C, and D. Each circle intersects two GRB's that are identified by number from 1-22.

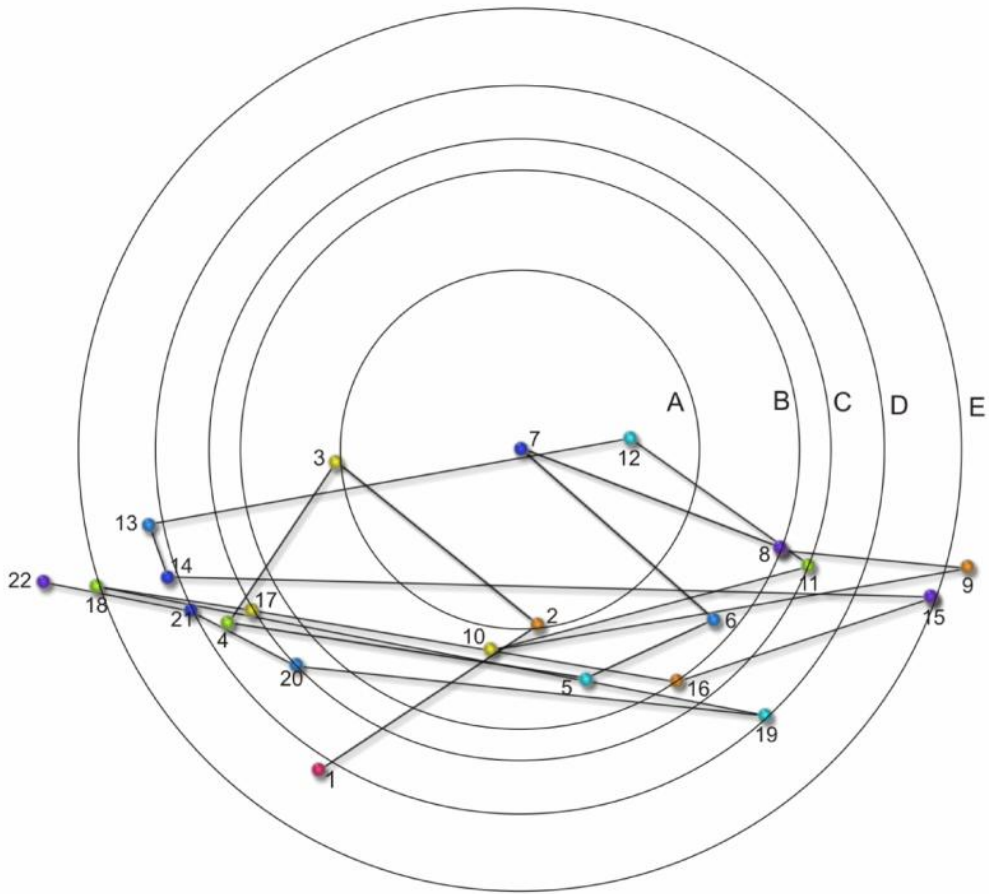

| A | 2 | 3 | |
|---|---|---|---|
| B | 8 | 16 | |
| C | 11 | 17 | 20 |
| D | 19 | 21 | |
| E | 15 | 18 | |

**Figure 54.** Using 7 (GRB 100413B) as a center point creates 5 CPPC A, B, C, D and E, Each circle intersects two or more GRB's that are identified by number from 1-22.

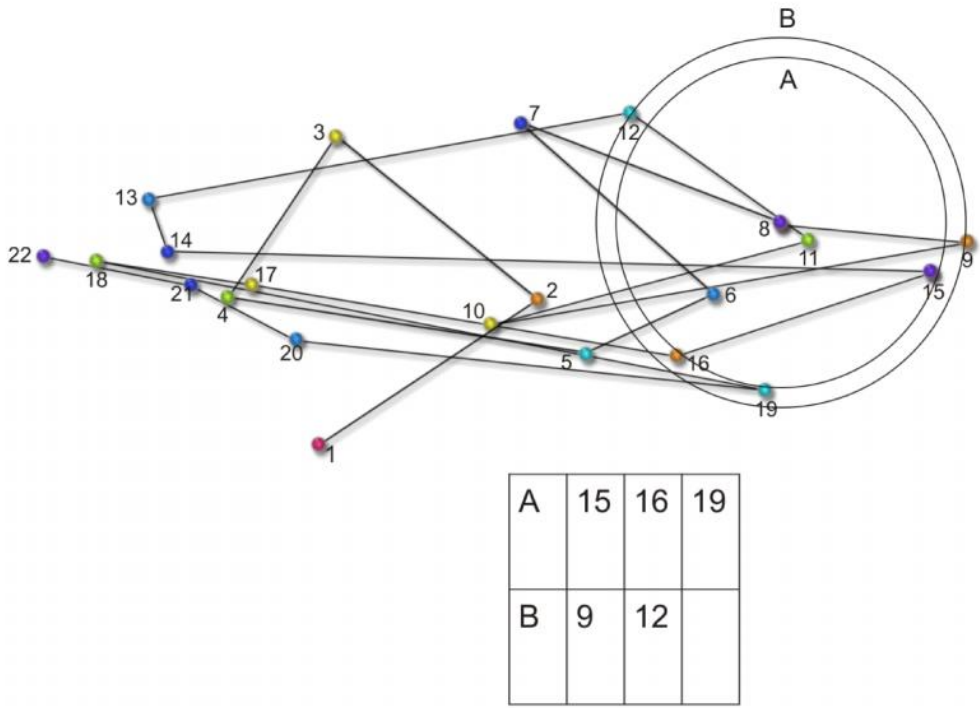

**Figure 55.** Using 8 (GRB 100413A) as a center point creates 2 CPPC labeled A and B. Each circle intersects two or more GRB's that are identified by number from 1-22.

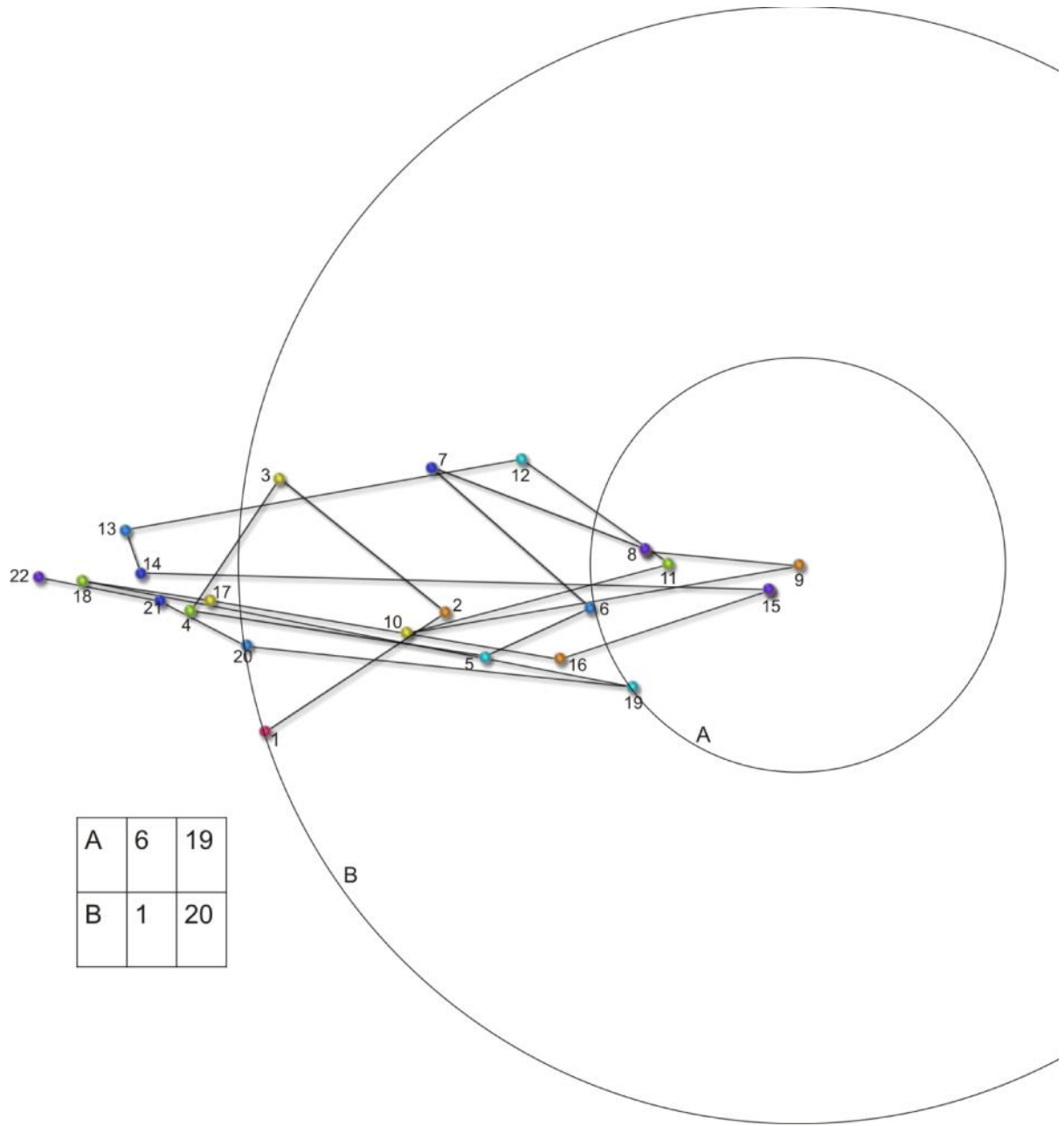

**Figure 56.** Using 9 (GRB 100414A) as a center point creates 2 CPPC labeled A and B. Each circle intersects two GRB's that are identified by number from 1-22.

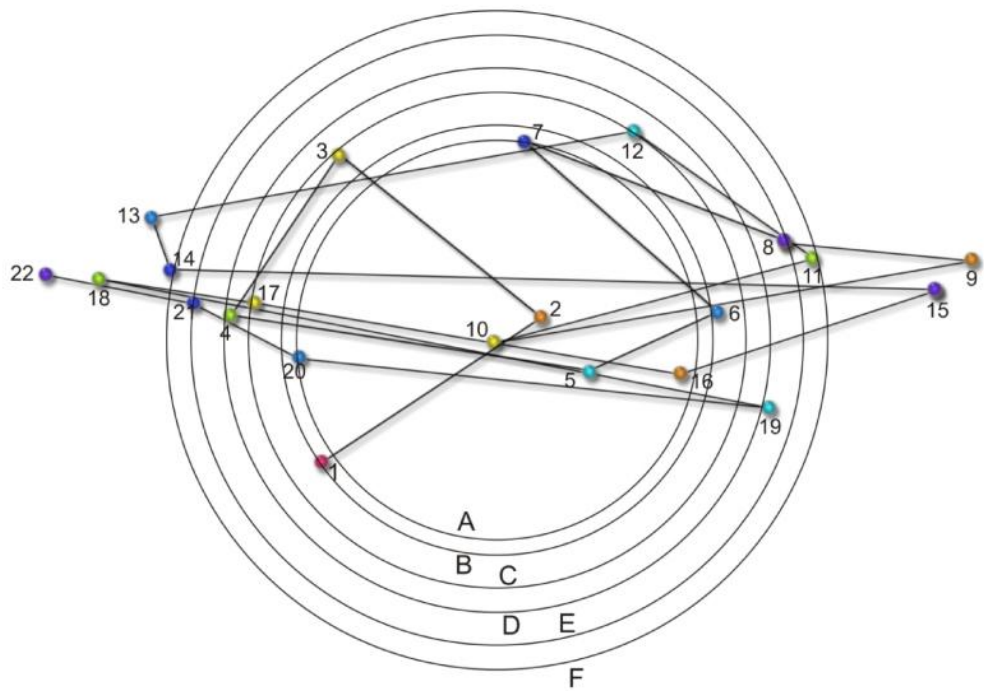

| A | 7  | 20 |    |
|---|----|----|----|
| B | 1  | 6  |    |
| C | 3  | 12 | 17 |
| D | 4  | 19 |    |
| E | 2  | 8  |    |
| F | 11 | 14 |    |

**Figure 57.** Using 10 (GRB 100415A) as a center point creates 6 CPPC labeled A, B, C, D, E, and F. Each circle intersects two or more GRB's that are identified by number from 1-22.

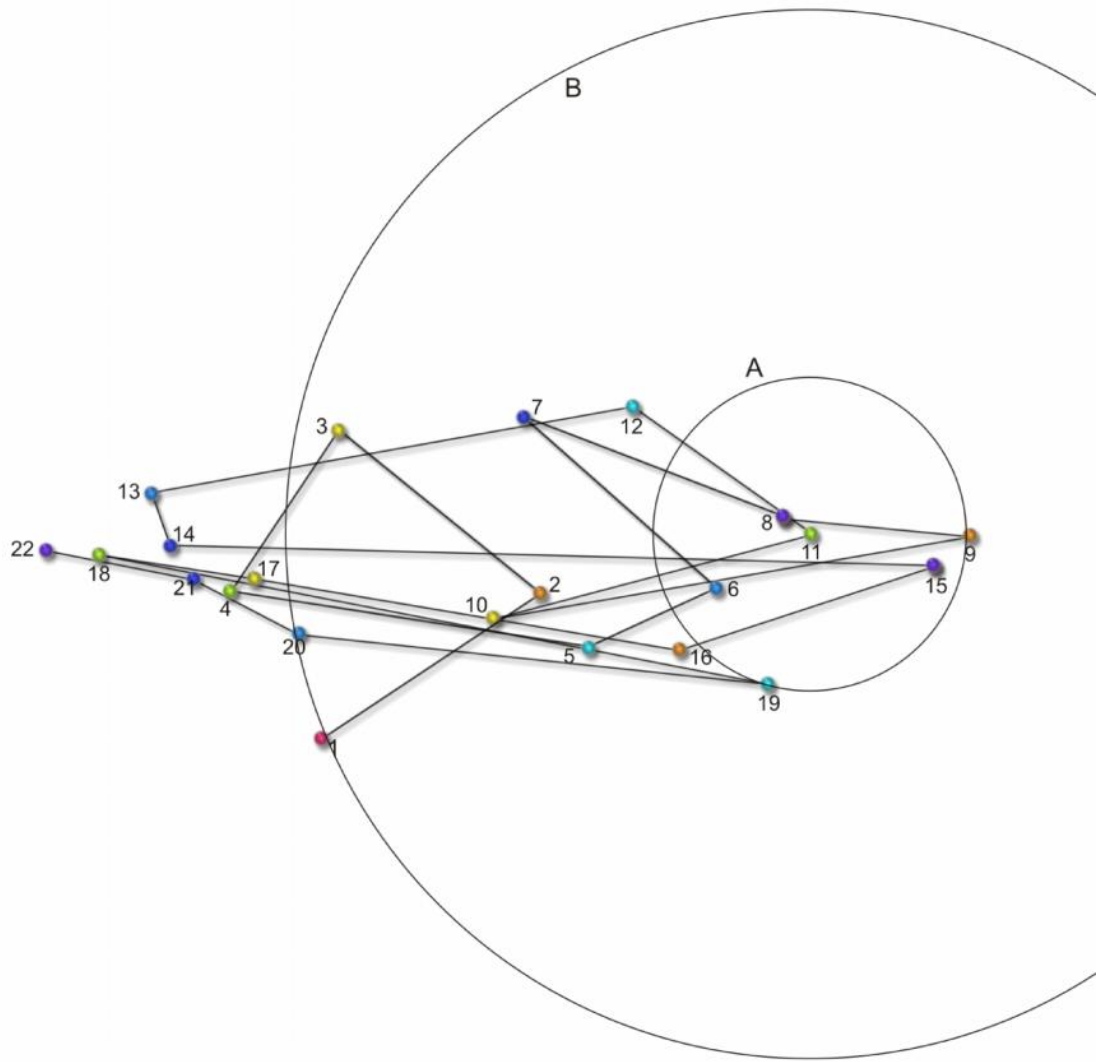

| A | 9 | 19 |
| B | 1 | 20 |

**Figure 58.** Using 11 (GRB 100418A) as a center point creates 2 CPPC labeled A and B. Each circle intersects two GRB's that are identified by number from 1-22.

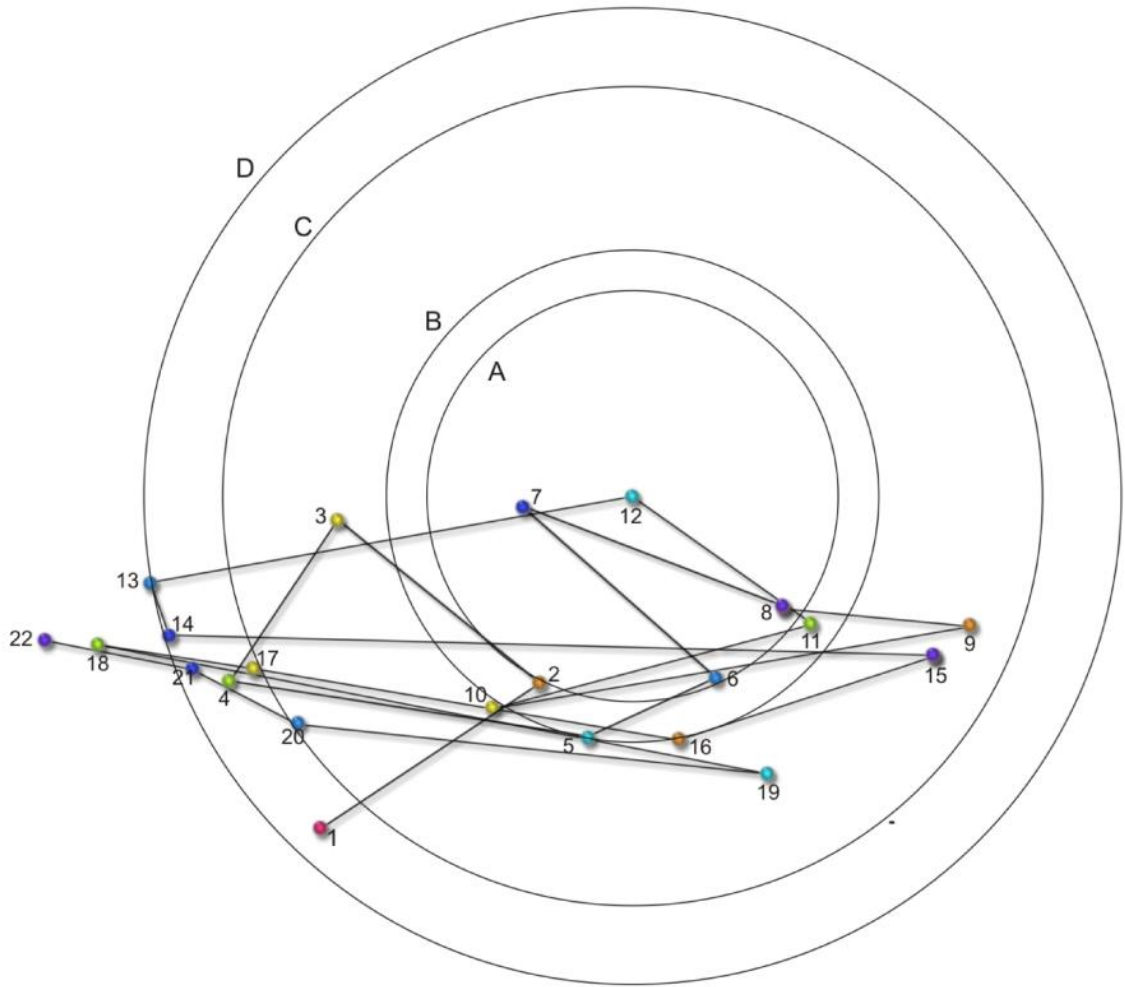

| A | 2  | 6  |    |
|---|----|----|----|
| B | 5  | 10 | 16 |
| C | 17 | 20 |    |
| D | 13 | 14 |    |

**Figure 59.** Using 12 (GRB 100420A) as a center point creates 4 CPPC labeled A, B, C and D. Each circle intersects two or more GRB's that are identified by number from 1-22.

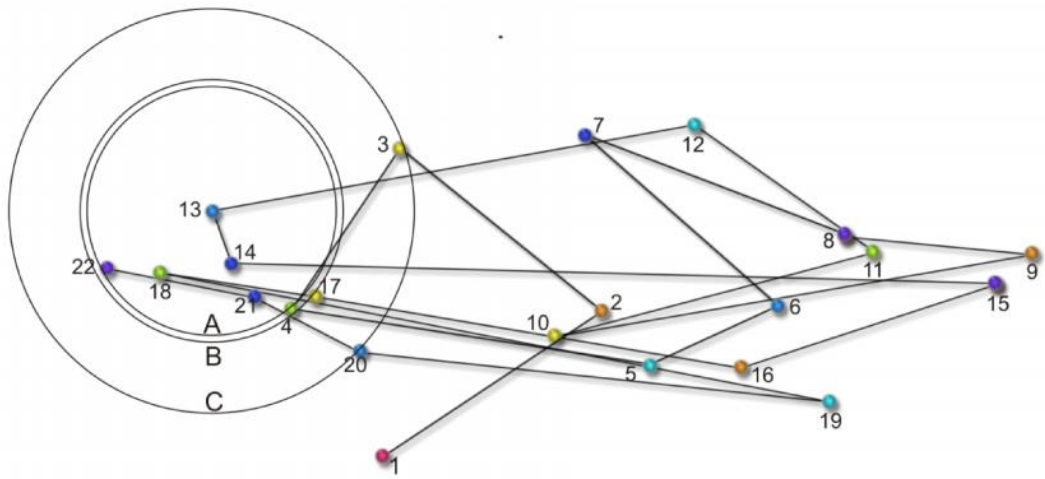

| A | 4 | 22 |
| --- | --- | --- |
| B | 3 | 20 |
| C | 8 | 19 |

**Figure 60.** Using 13 (GRB 100423A) as a center point creates 3 CPPC labeled A, B, C. Each circle intersects two GRB's that are identified by number from 1-22.

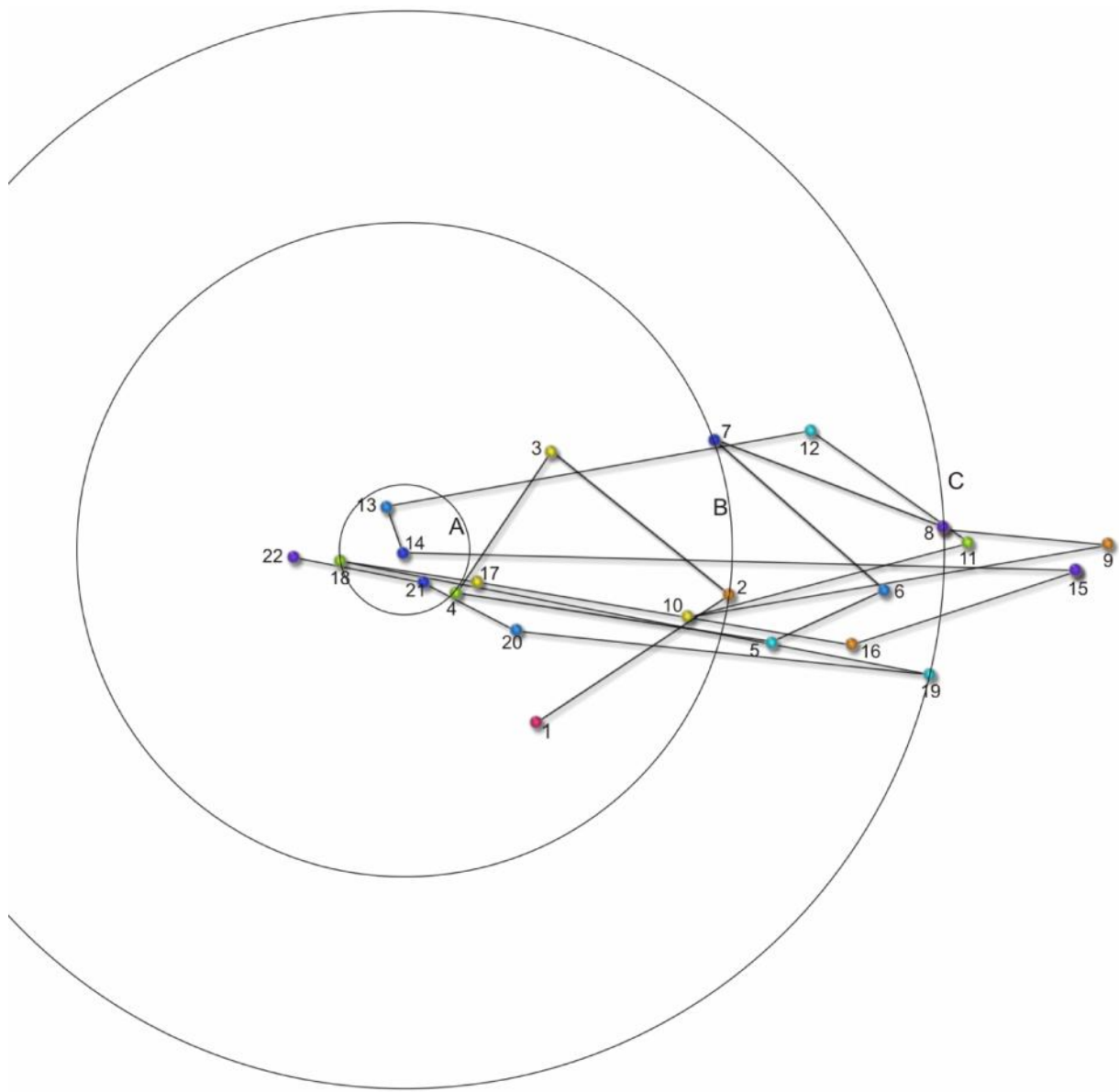

| A | 4 | 18 |
|---|---|----|
| B | 2 | 7  |
| C | 8 | 19 |

**Figure 61.** Using 14 (GRB 100423B) as a center point creates 3 CPPC labeled A, B, C. Each circle intersects two GRB's that are identified by number from 1-22.

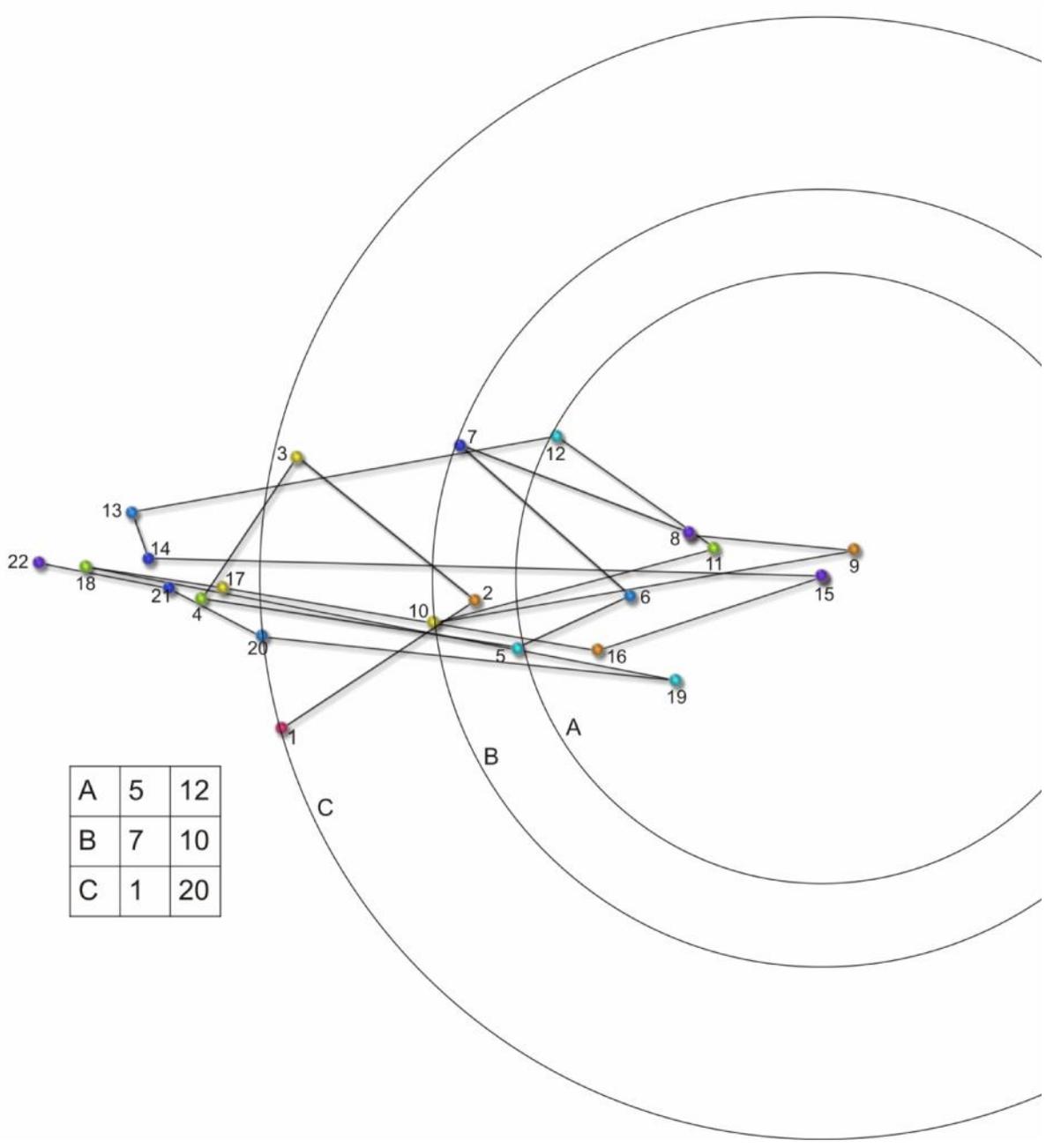

**Figure 62.** Using 15 (GRB 100424A) as a center point creates 3 CPPC labeled A, B, C. Each circle intersects two GRB's that are identified by number from 1-22.

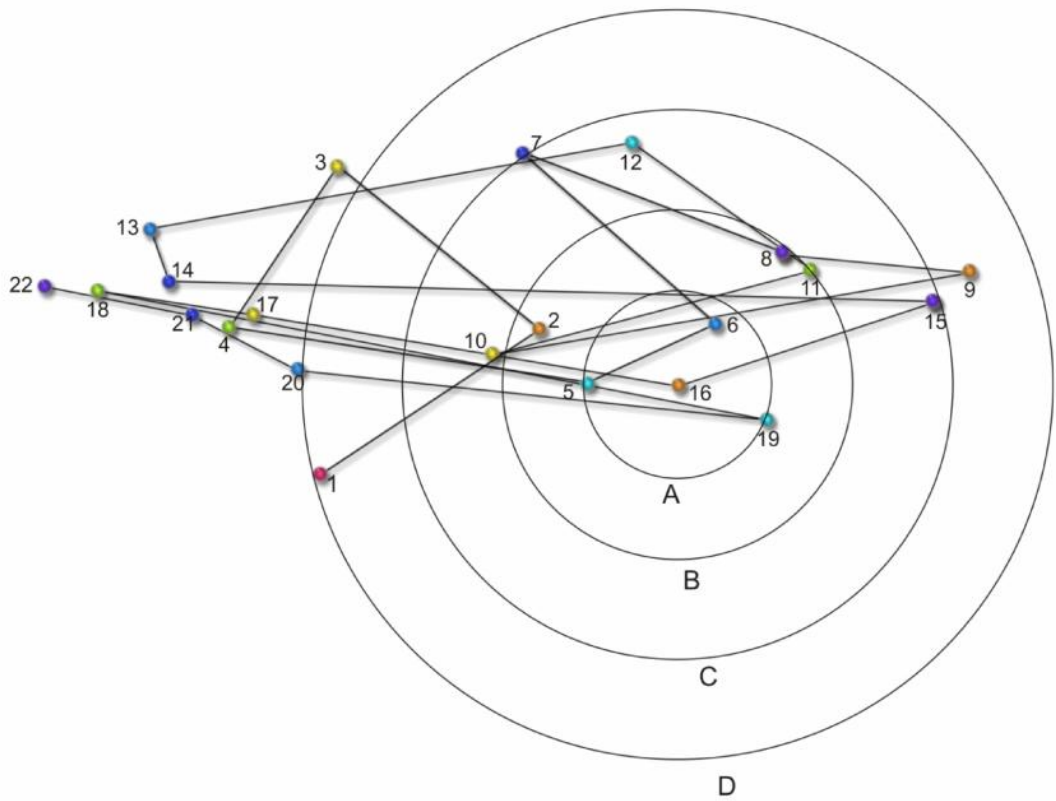

| A | 5 | 19 |  |
|---|---|----|---|
| B | 8 | 11 |  |
| C | 7 | 15 |  |
| D | 1 | 20 |  |

**Figure 63.** Using 16 (GRB 100425A) as a center point creates 4 CPPC labeled A, B, C and D. Each circle intersects two GRB's that are identified by number from 1-22.

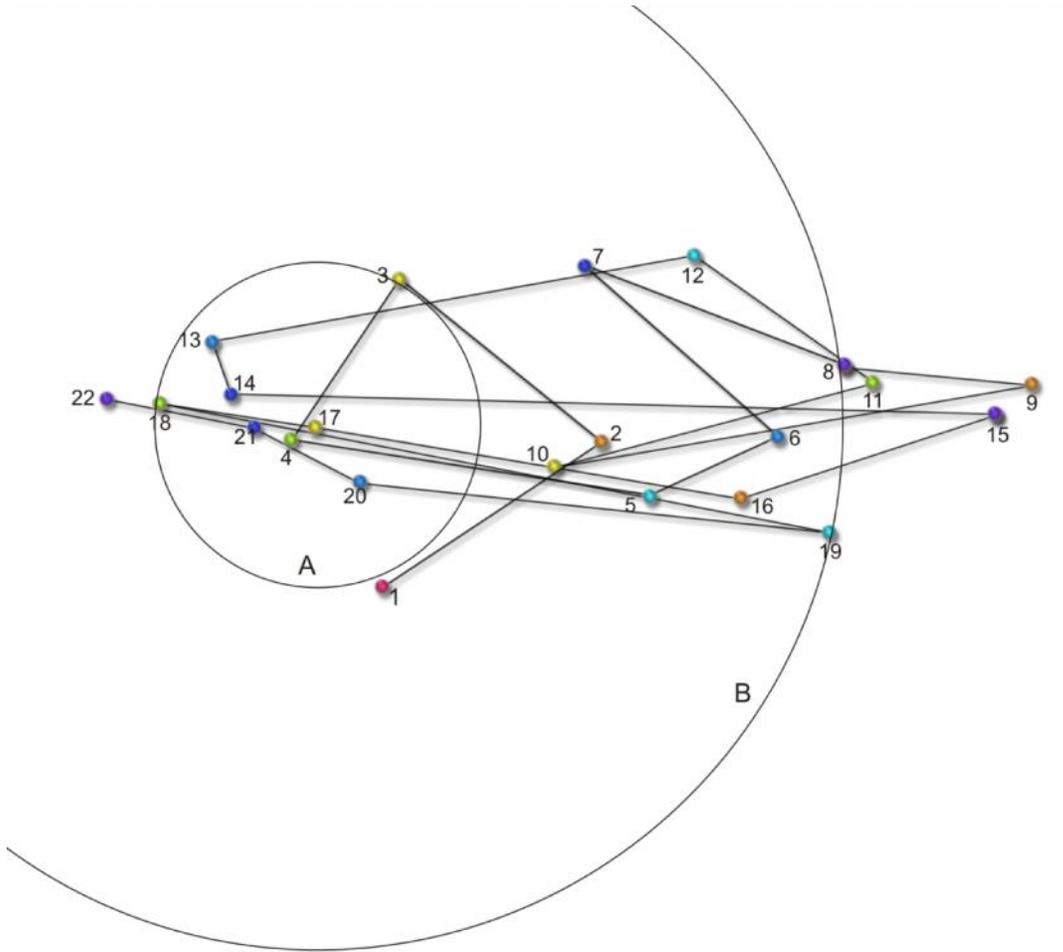

| A | 3 | 18 |
|---|---|----|
| B | 8 | 19 |

**Figure 64.** Using 17 (GRB 100427A) as a center point creates 2 CPPC labeled A and B. Each circle intersects two GRB's that are identified by number from 1-22.

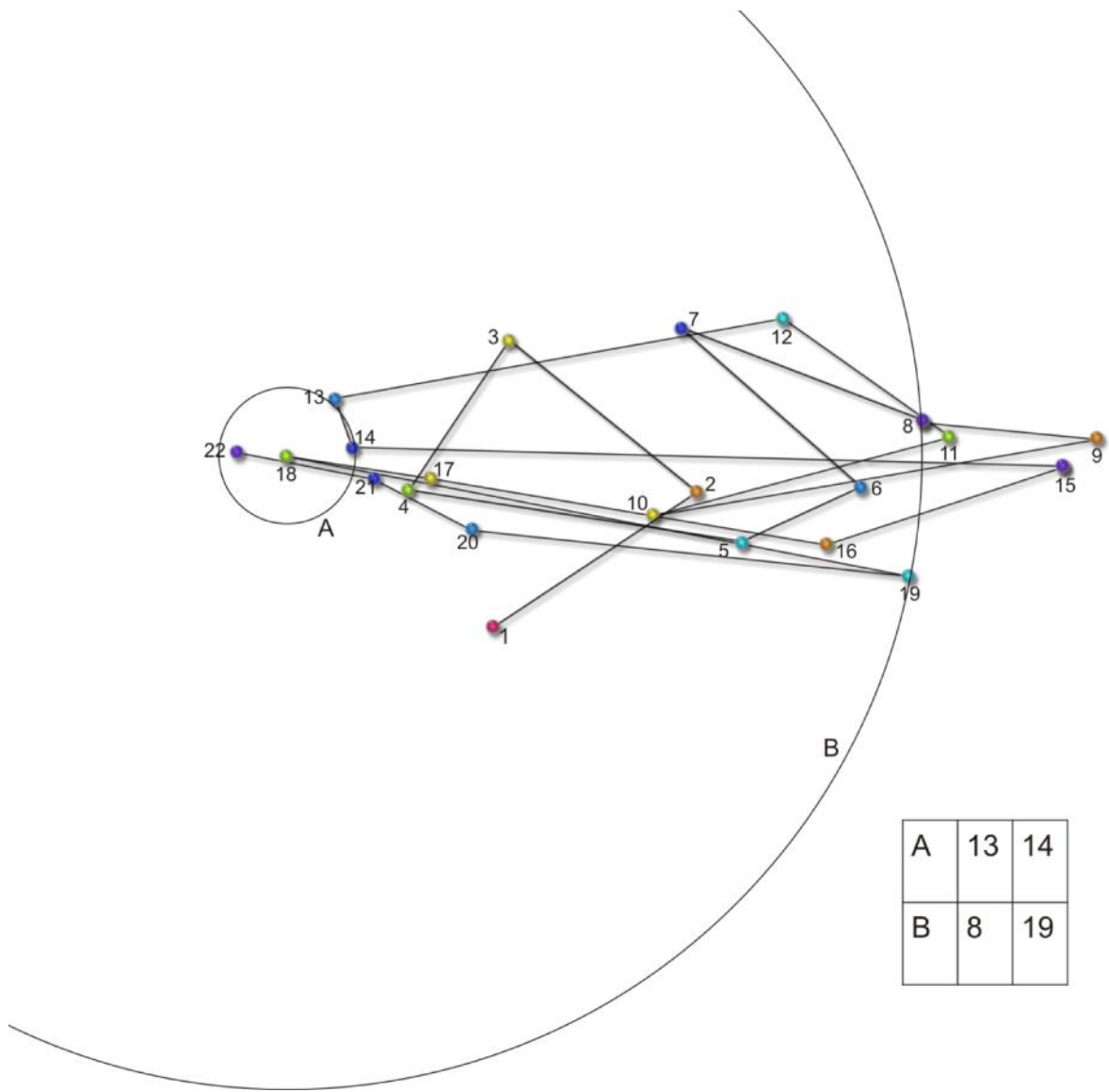

| A | 13 | 14 |
| B | 8  | 19 |

**Figure 65.** Using 18 (GRB 100503A) as a center point creates 2 CPPC labeled A and B. Each circle intersects two GRB's that are identified by number from 1-22.

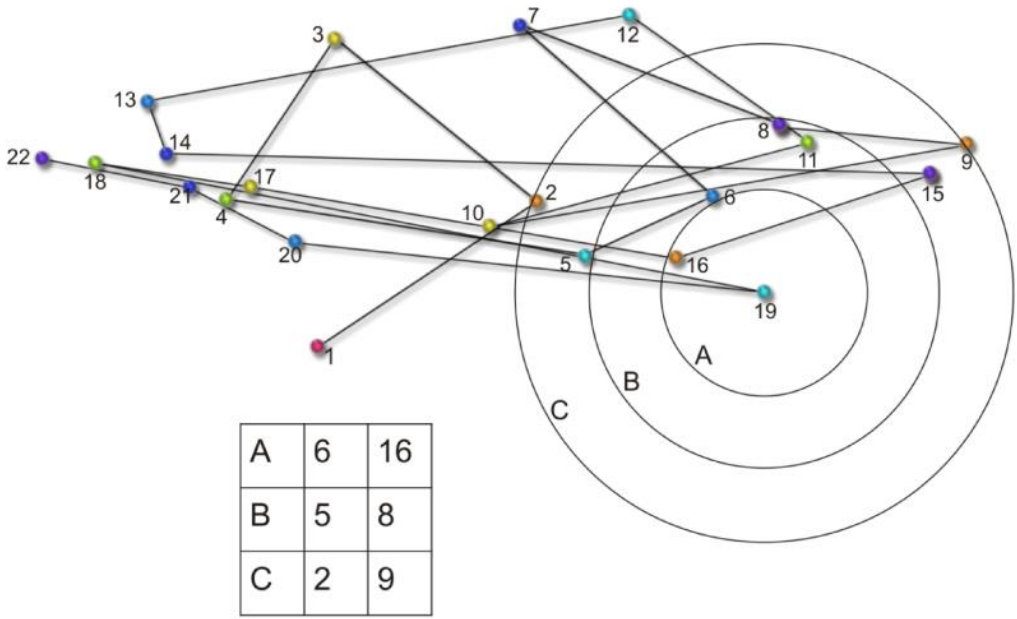

| A | 6 | 16 |
| B | 5 | 8 |
| C | 2 | 9 |

**Figure 66.** Using 19 (GRB 100504A) as a center point creates 3 CPPC labeled A, B, C. Each circle intersects two GRB's that are identified by number from 1-22.

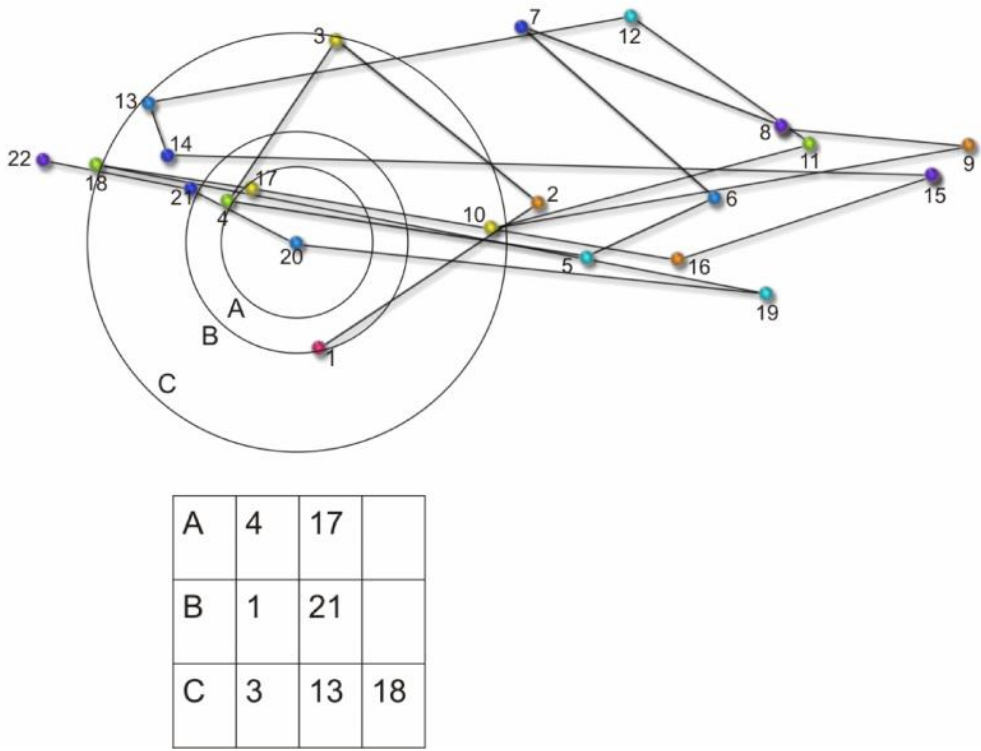

| A | 4 | 17 | |
| B | 1 | 21 | |
| C | 3 | 13 | 18 |

**Figure 67.** Using 20 (GRB 100508A) as a center point creates 3 CPPC labeled A, B, C. Each circle intersects two or more GRB's that are identified by number from 1-22.

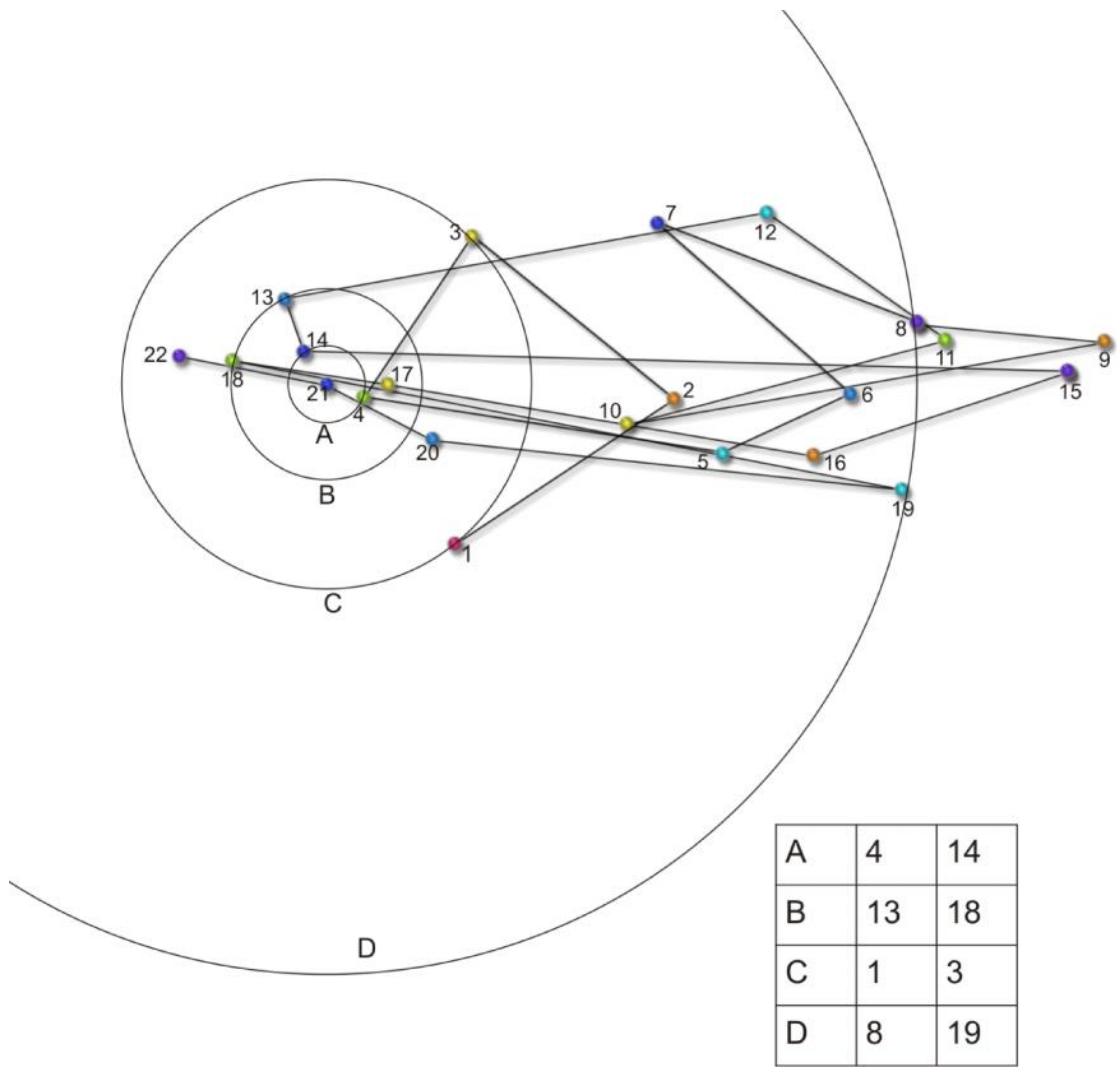

| A | 4 | 14 |
| B | 13 | 18 |
| C | 1 | 3 |
| D | 8 | 19 |

**Figure 68.** Using 21 (GRB 100511A) as a center point creates 4 CPPC labeled A, B, C and D. Each circle intersects two GRB's that are identified by number from 1-22.

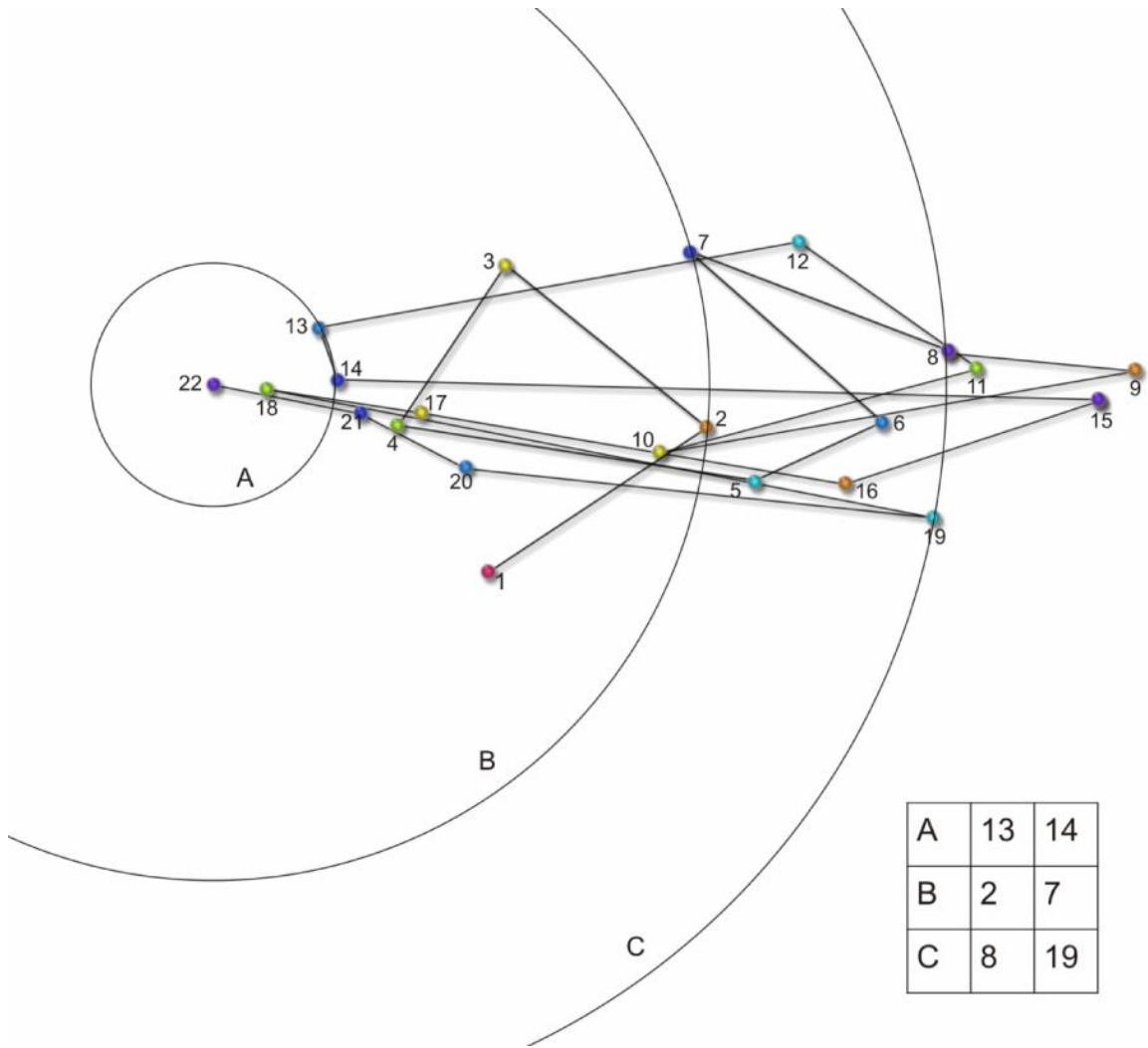

**Figure 69**. Using 22 (GRB 100513A) as a center point creates 3 CPPC labeled A, B, C. Each circle intersects two GRB's that are identified by number from 1-22.

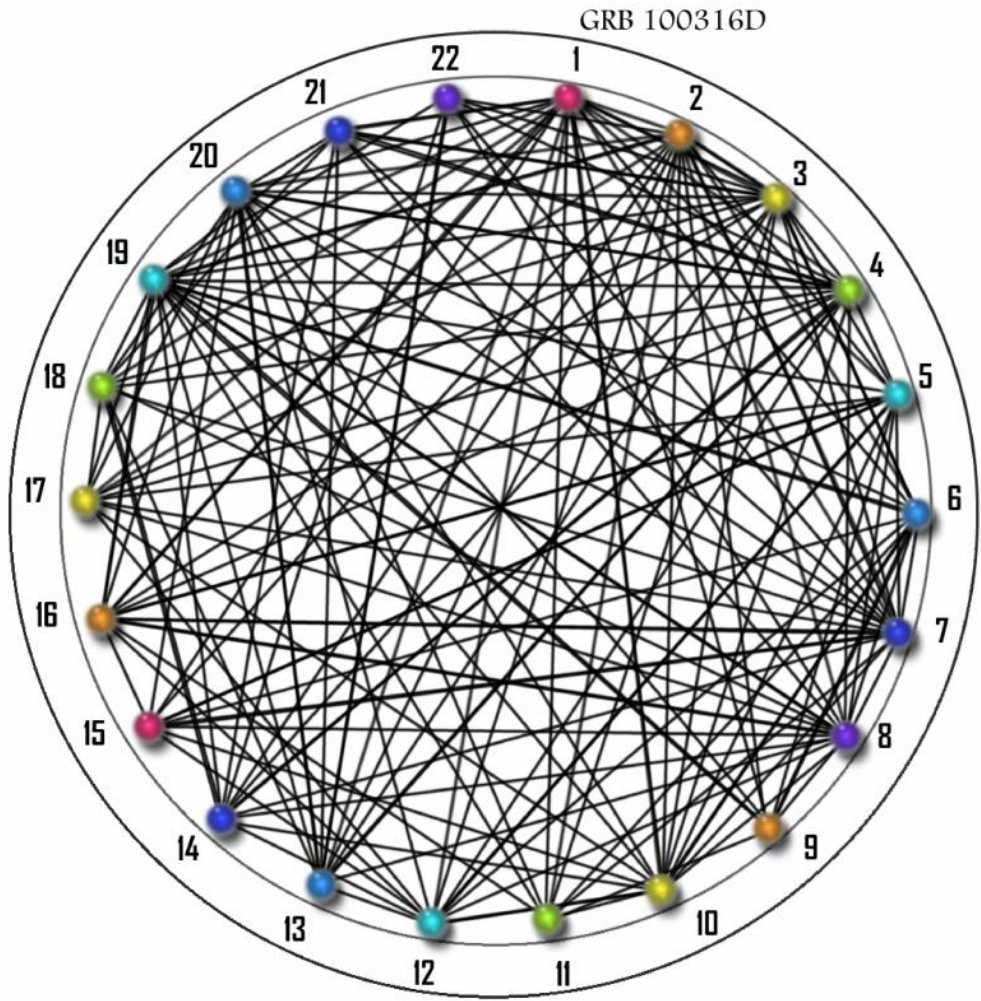

**Figure 70**

Fig.70 is a circle with 22 divisions. Each division represents one of the 22 sequential GRB's in the throgg from Fig.47. The lines from each number radiate to the GRB's that are equidistant from it. There are 175 examples of GRB's at equal distant positions.

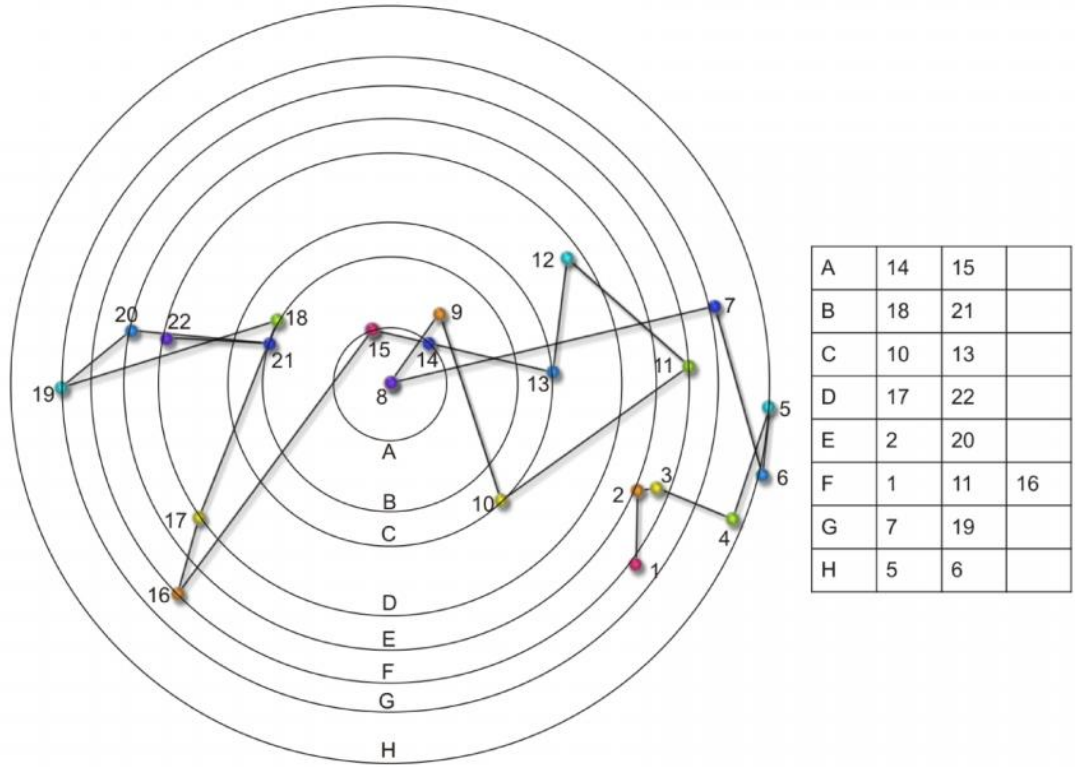

**Figure 71**

8CPPC

Fig.71 features another example of a throgg. Using violet 8 as the origin creates 8 CPPC. The throgg is constructed by linking sequential GRB's that range from GRB 050717A to GRB 050906. July 17, 2005 to September 6, 2005. The origin of the circle is violet 8, GRB 050803A. It is equidistant from (14, 15) A circle, (18, 21) B circle, (10, 13) C circle, (17, 22) D circle, (2, 20) E circle, (1, 11, 16) F circle, (7, 19) G circle, (5, 6) H circle.

# Synthesized Oggs

Linking sequential GRB's creates patterns with angles that skew to smaller. A propensity to create smaller angles doesn't seem random. A random distribution should range from 1 degree 180 degrees, but does it?  To investigate we created an adjustable ogg generator, to randomly create a synthetic ogg (nogg) as a pattern of sequentially linked points in a box.

Fig.71 shows a nogg and the histogram for the angle distribution of sequential points. Points are chosen uniformly random in a box. The histogram is in turn units, (angle in radians),

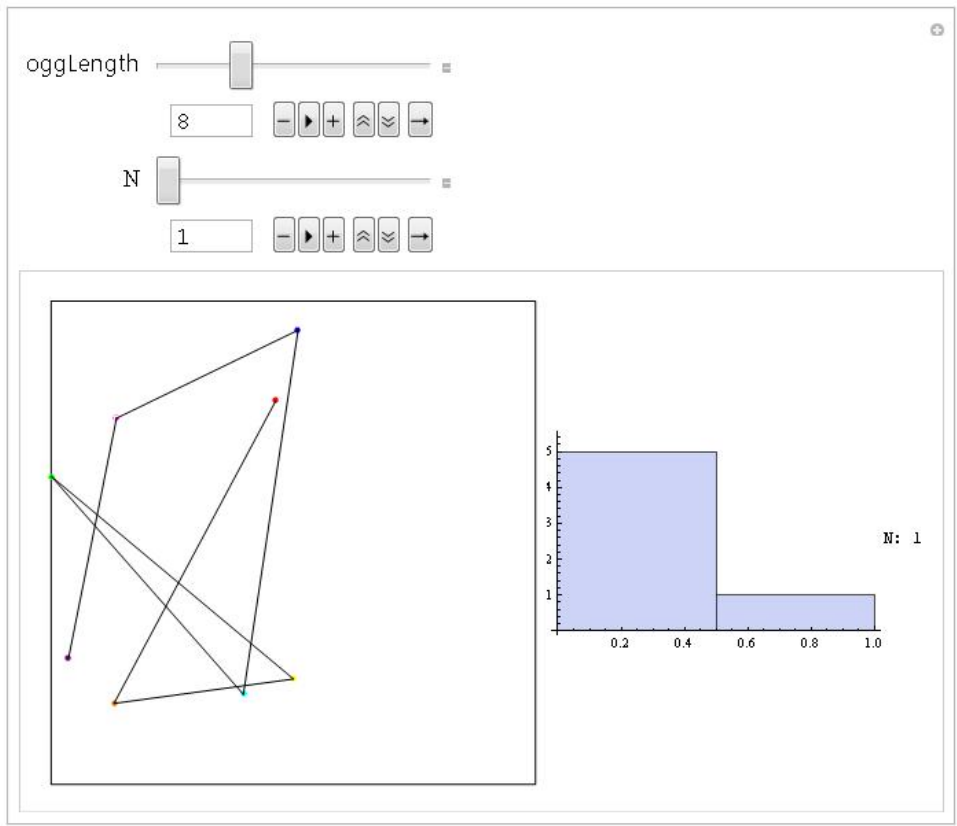

**Figure 71**

Raising the value of N increases the size of the sample, and the number of bins in the histogram. Fig.71 The value of N=1 so the histogram has only two bins. The skew towards small angles is visible.

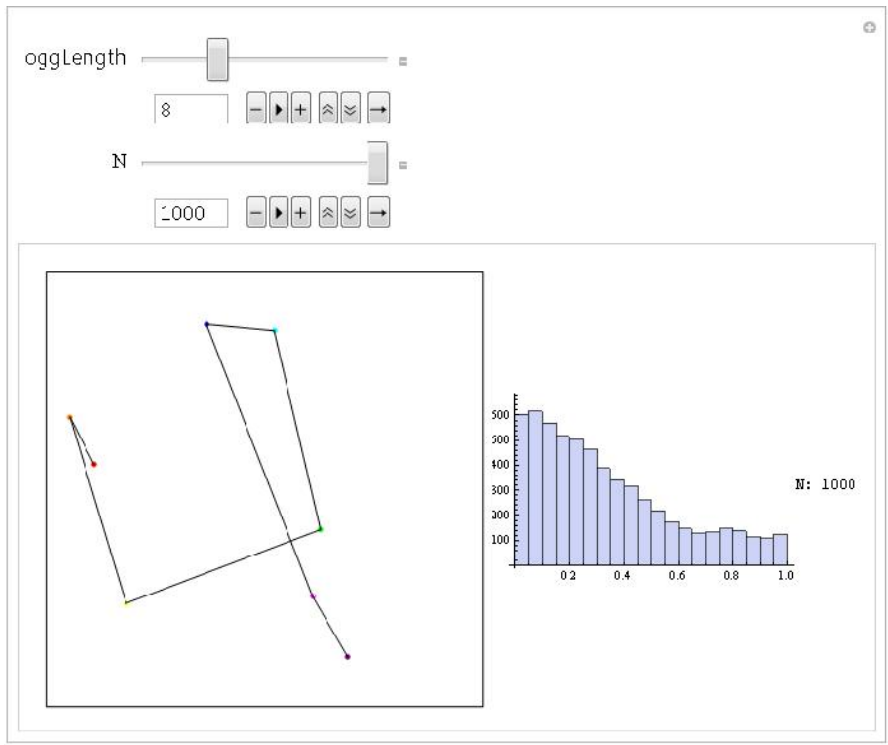

**Figure 72**

In Fig. 72 The value of N=1000, the histogram has 20 bins. The slope is the same regardless of N's value. The skew towards small angles is visible.

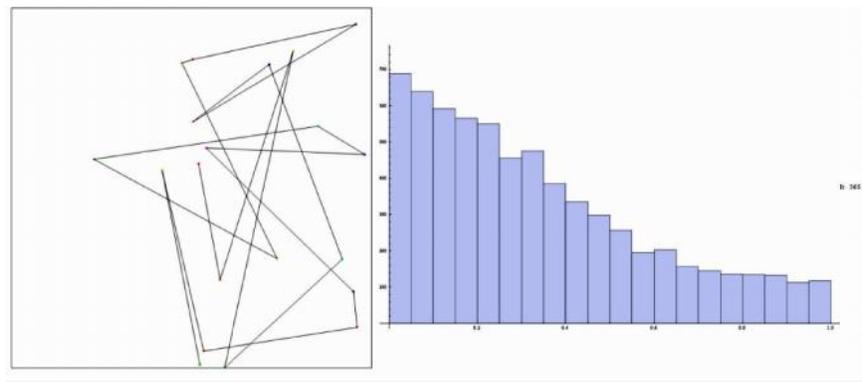

**Figure 73**

Fig.73 Increasing the length of the nogg does not alter the distribution. The skew towards small angles is visible.

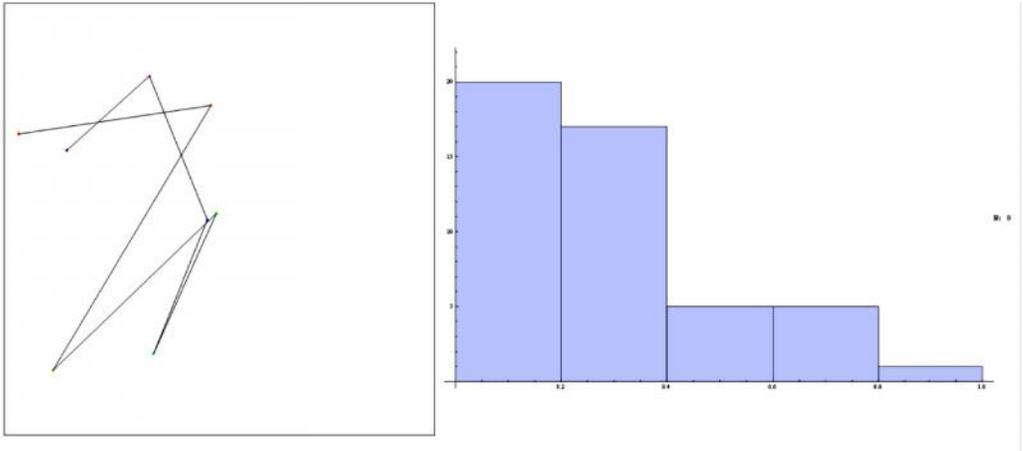

**Figure 74**

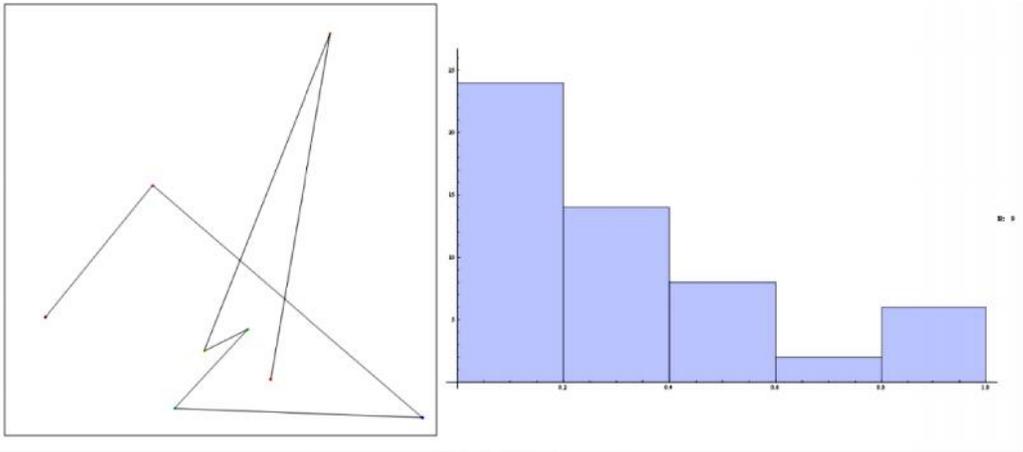

**Figure 75**

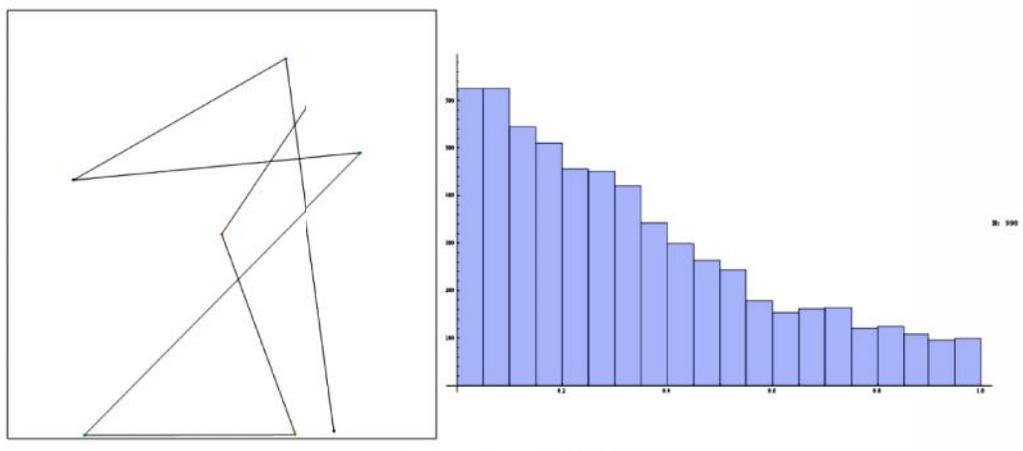

**Figure 76**

Fig.74, 75, 76 Noggs with intersecting lines creates patterns that appear similar to natural oggs.

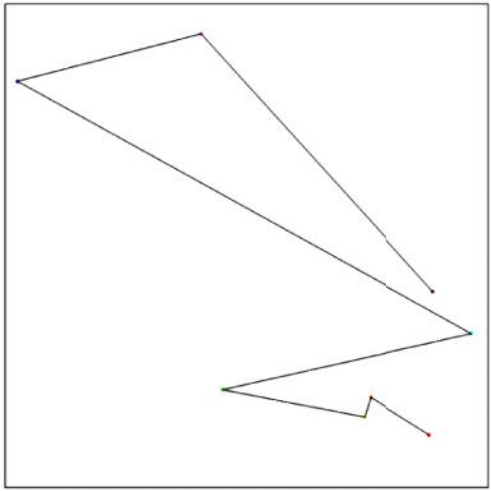

**Figure 77**

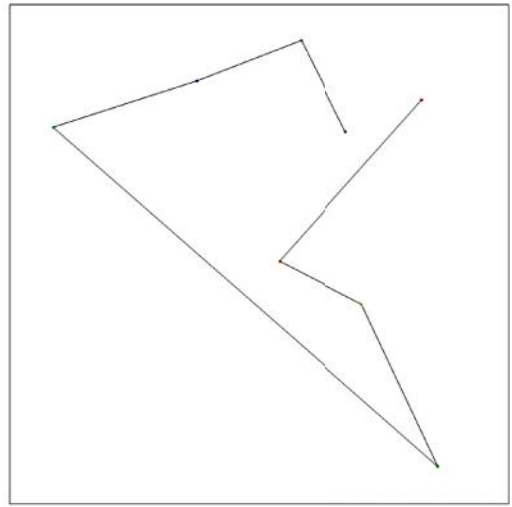

**Figure 78**

Fig. 77, 78, 79, 80 Noggs made from non-intersecting line patterns that do not appear similar to natural oggs.

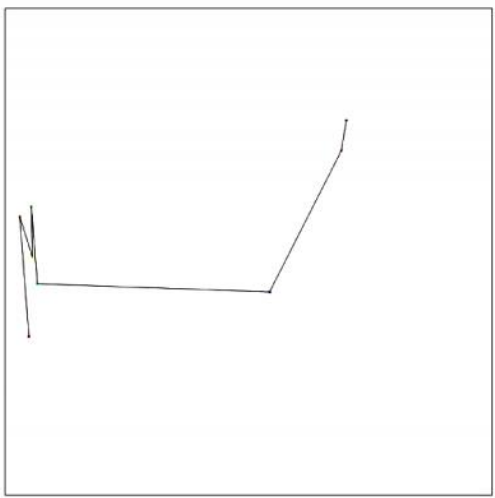

**Figure 79**

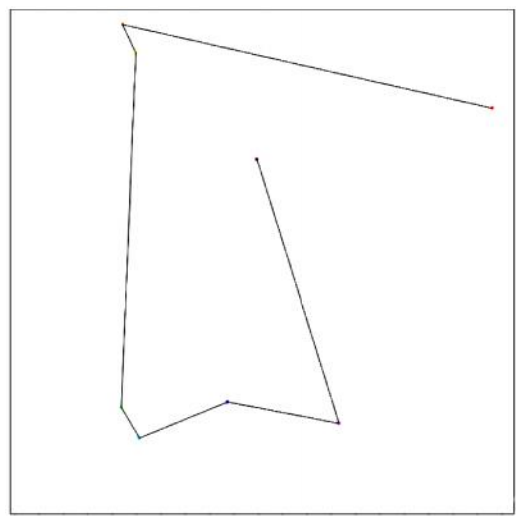

**Figure 80**

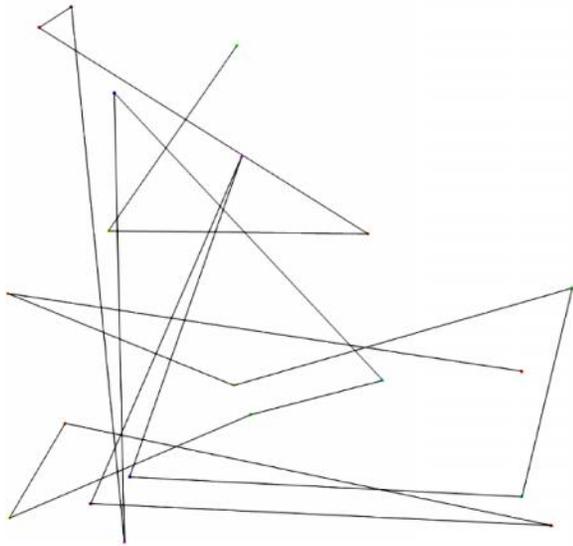
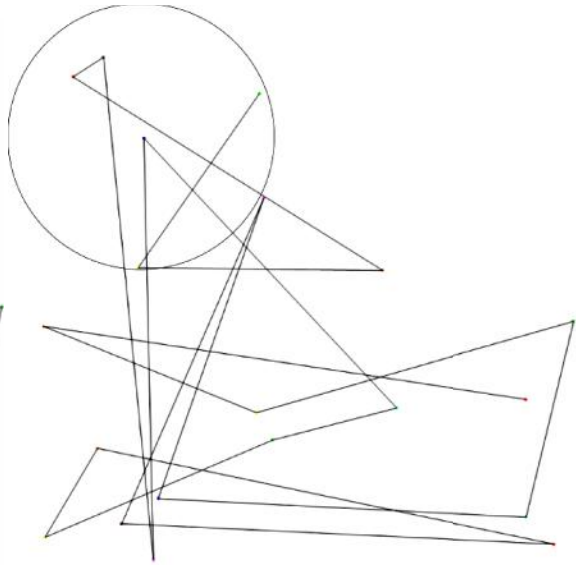

**Figure 82**  **Figure 83**

Fig.82 is a 20 point nogg sequence. Fig.83 is Fig.82 with one example of a PPC (paired point circle). Fig.84 A point and a paired point circle. Fig.85 A point with a larger PPC paired point circle.

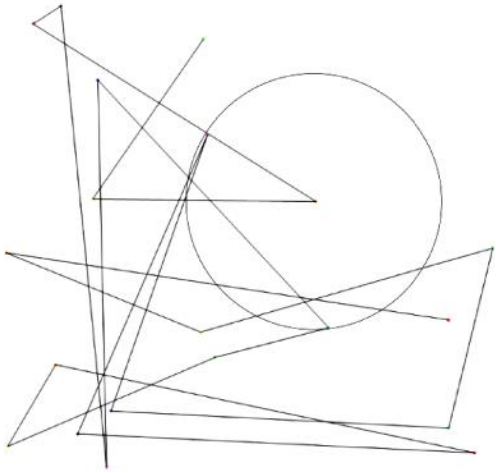
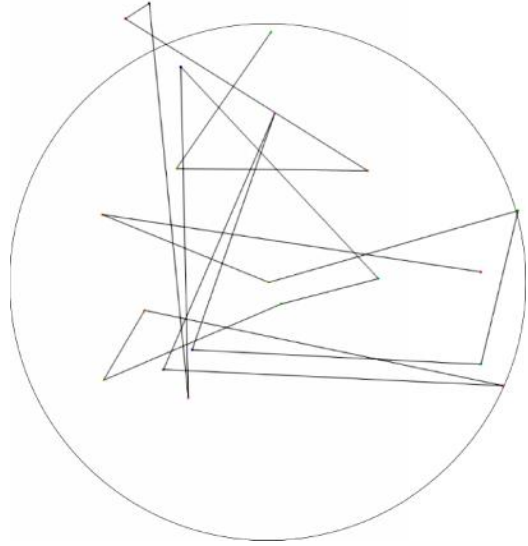

**Figure 84**  **Figure 85**

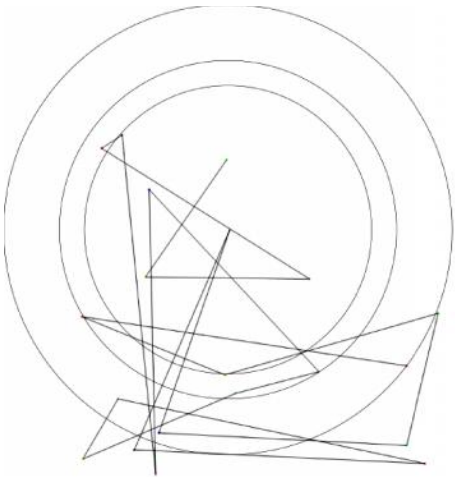
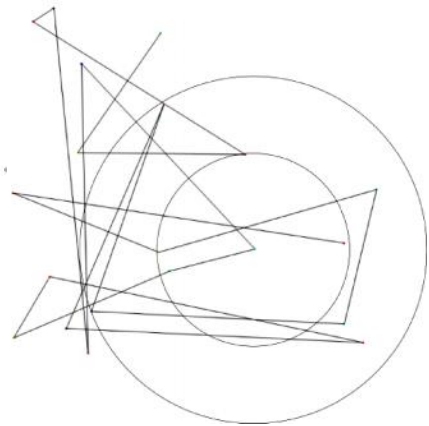

**Figure 86**  **Figure 87**

Fig.86 One point with 3 concentric paired point circles. Fig.87 One point with two concentric paired point circles. A 20 point nogg can produce paired point circles (PPC) and (CPPC) concentric paired point circles.

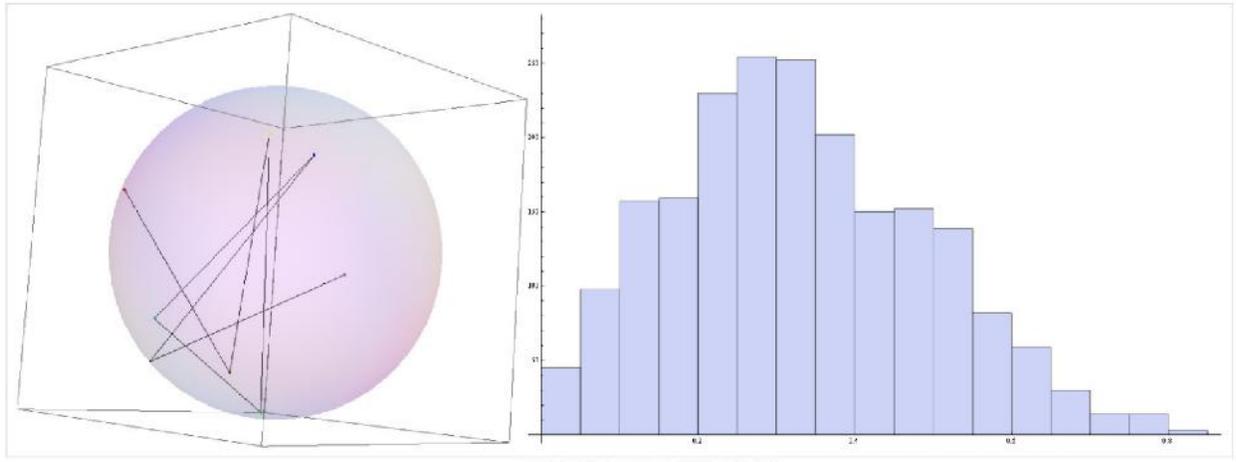

**Figure 88   3D Nogg**

Fig.88 On the left a three dimensional nogg with random points chosen uniformly from the surface of sphere, the histogram on the right is not like the 2D distribution in Figs. 71-76.

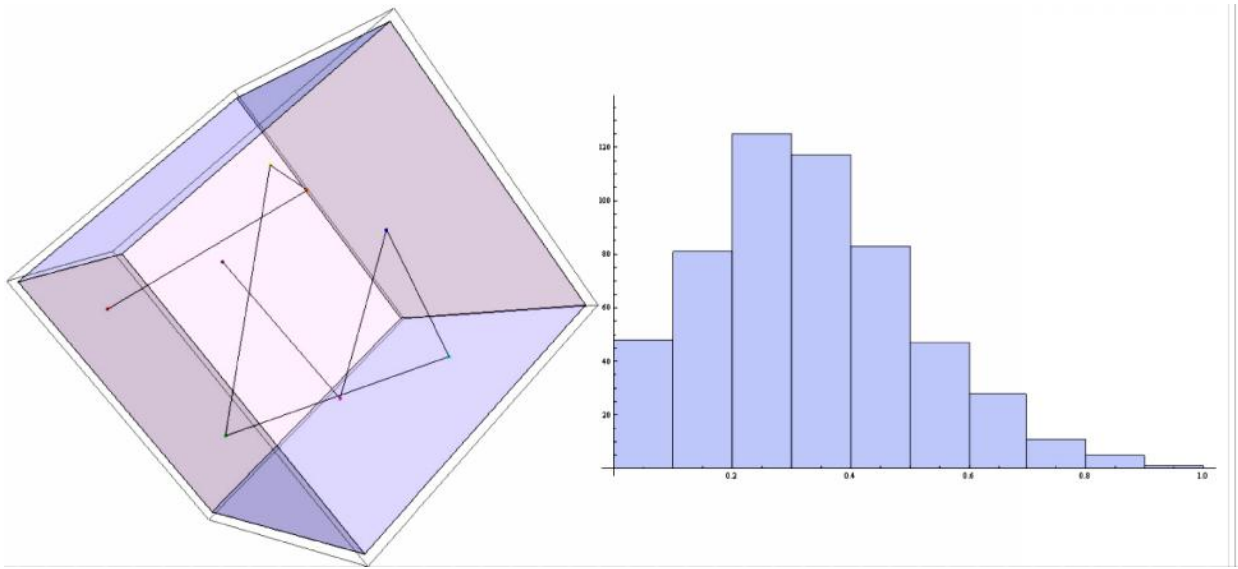

**Figure 89**

Fig.89 Another 3D nogg, 8 random points distributed uniformly in a box, sequentially connected, measured and sorted to reveal angle distribution. The nogg in the 3D cube has the same distribution as points chosen uniformly from the surface of a sphere. Fig.88

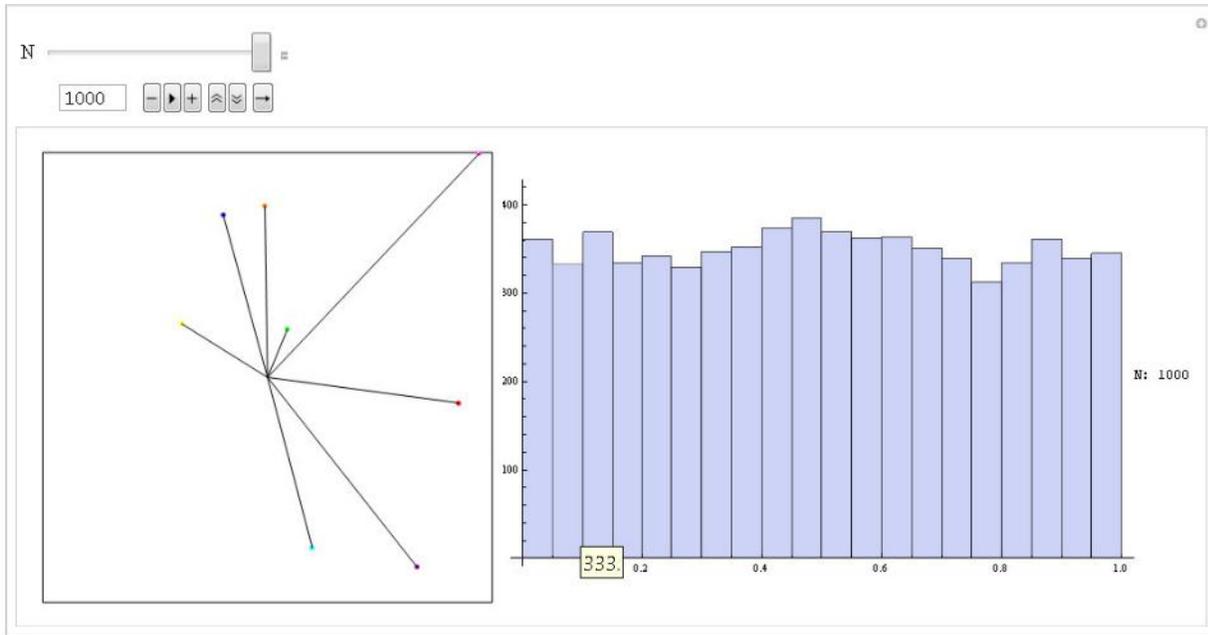

**Figure 90**

Fig.90 Another kind of distribution occurs when the angles are not between sequential points but between location vectors.

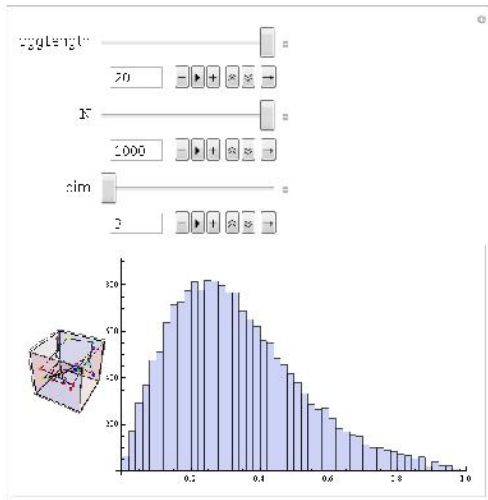

**Figure 91**

Fig. 91 shows a 20 point nogg in three dimensions. The distribution matches fig.88 and fig.89

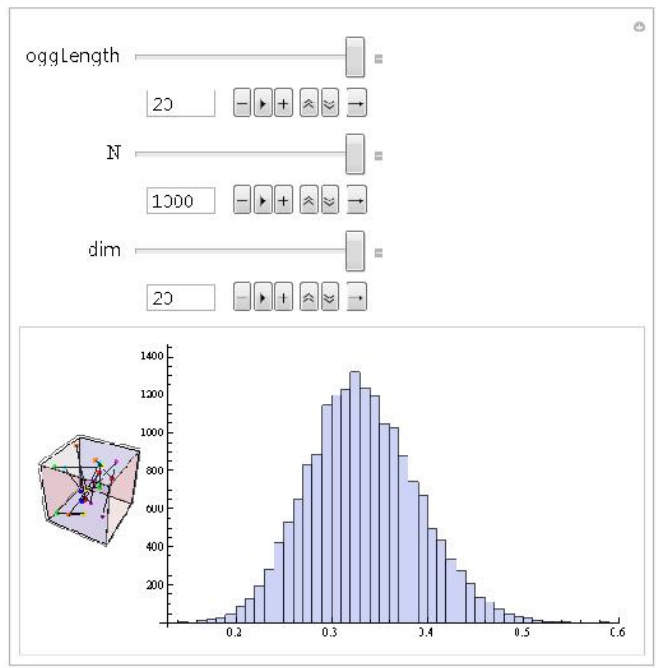

**Figure 92**

Fig.92 shows a 20 point nogg in 20 dimensions. The distribution has shifted towards Gaussian symmetry.

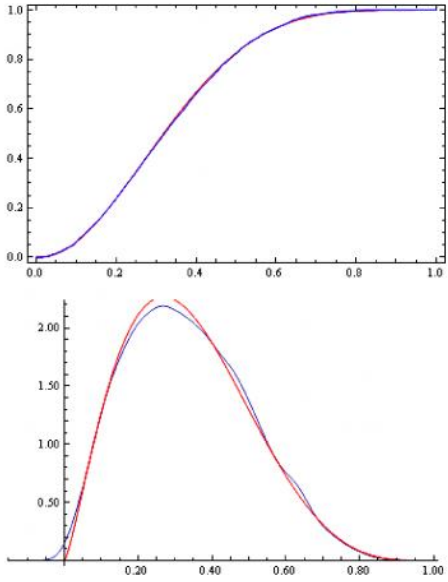
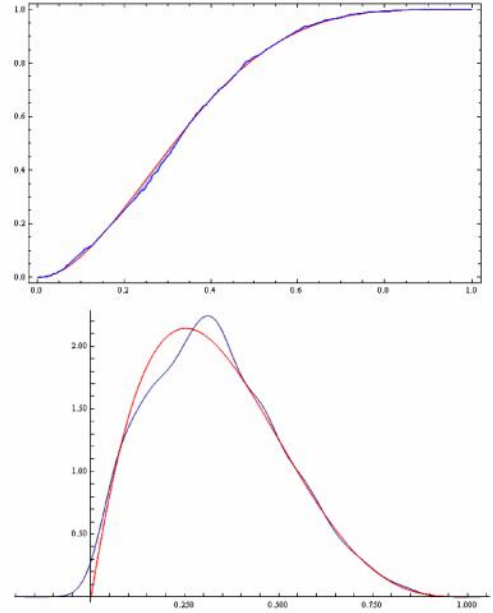

**Figure 93 Random Vectors on a sphere**  **Figure 94 Sonoma, http://grb.sonoma.edu/#**

Fig.93 Selecting random vectors on a sphere creates the blue arc. The null hypothesis that the data is distributed according to the (red arc) Beta Distribution [a,b] is not rejected at the 5. Percent level based on the Cramér-von Mises test. The variance between paths is slight. Random vectors on a sphere are closely aligned with beta distribution.

Fig.94 Using the galactic coordinates from Sonoma, http://grb.sonoma.edu/# (blue).The null hypothesis that the data is distributed according to the Beta Distribution [a,b] is not rejected at the 5. Percent level based on the Cramér-von Mises test. The GRB arc is closely aligned with beta distribution except for the top 1/3 that appears to have been pulled up and pinched to the right.

## Maps Transpose Dimension

All forms of mapping require a method of projection. The Aitoff sky map used in this study (Fig. 1) is neither conformal nor equal area it is a modified azimuth projection map. The center of the map is the only point that is free from distortion. It is positioned to align with the center of our galaxy.

When evaluating the results of this study it is important to consider that the GRB locations are 3D coordinates transposed to Aitoff projection, 2D geometry. How reliable is 3D information when presented in 2D? Is transposed data valid? The structure of DNA was revealed using x-ray diffraction. Watson and Crick determined the shape of the unknown molecule by examining the two-dimensional image that resulted from the directed projection of x-rays towards the crystalline molecule. By measuring the distances between structures they found some points were located at equidistant positions. The information from the 2D image identified the double helix as the 3D structure of DNA.  In the paper "Torsional fluctuations in columnar DNA assemblies" (D.J.Lee and A.Wynveen 2008) the authors describe the usefulness of X-ray diffraction in the process of evaluating the effect of torsion on strands of DNA at varying lengths.

In a paper about the appearance of concentric circles in the WMAP data (Wilkinson microwave anisotropy probe) as evidence of violent pre-Big-Bang activity, the authors present findings regarding CCC (conformal cyclic cosmology) and LCDM (lambda cold dark matter) in the CMB (cosmic microwave background). They conclude the appearances of concentric circles located at different positions on the CMB map (Aitoff projection) are not random. (V.G. Gurzadyan and R. Penrose 2010)

## Transposing Randomness and Probability Density Function

In the study "Random points in a circle and the analysis of chromosome patterns" the authors used a microscope to photograph human chromosomes. The 2D images they observed were evaluated to correlate with the 3D nucleus. (Barton, Fix, 1963)  Herbert Solomon (Geometric Probability, 1978) refers to Barton and Fix and probability density function (PDF) , "To find the probability density function of the distance between the two points.  Derive the distribution of the distance between points randomly dropped on the picture of a cell nucleus."

Bertrand's paradox examines the probability that a random chord will be larger than one side of an equilateral triangle inscribed within the circle.  There are 3 answers $p=1/3$, $p=1/2$, $p=1/4$, they are mutually inconsistent, still they are all correct.  Bertrand's Paradox: a Physical solution" (P. Di Porto et.al 2010) presents a physical solution for the ambiguity of randomness by throwing a long straw into a circle. To derive $p=1/2$ the authors used a straw with a defined length and a circle with a known radius. Amanda Maxham (2010) presented a paper that explored random distribution by throwing poker chips onto a table to determine the probability of two chips overlapping.

## Central Limit Theorem

Following are three different distribution patterns for 8 random points in a box. Fig.95 shows 8 sequential random points in 2D.  Fig.96 shows 8 sequential random points in 3D.  Fig.97 shows 8 sequential random points in 20D.  Increasing the number of dimensions shifts the direction of distribution towards Gaussian, a feature associated with Central Limit Theorem. The shape of randomness conforms to the number of dimensions required to contain it.

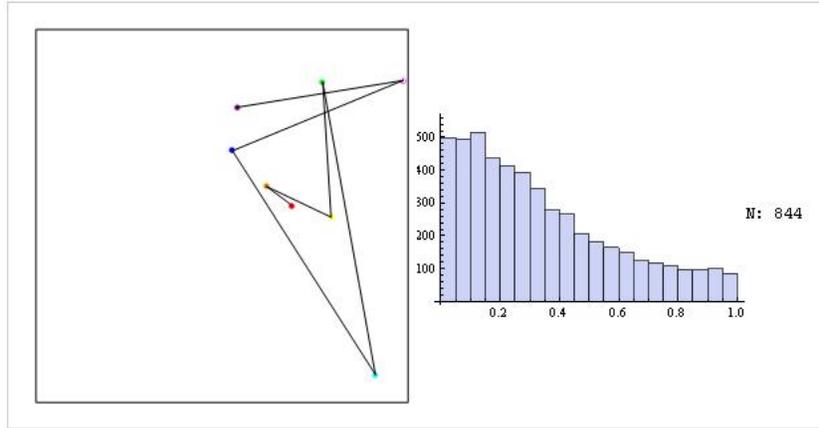

**Figure 95 2D**

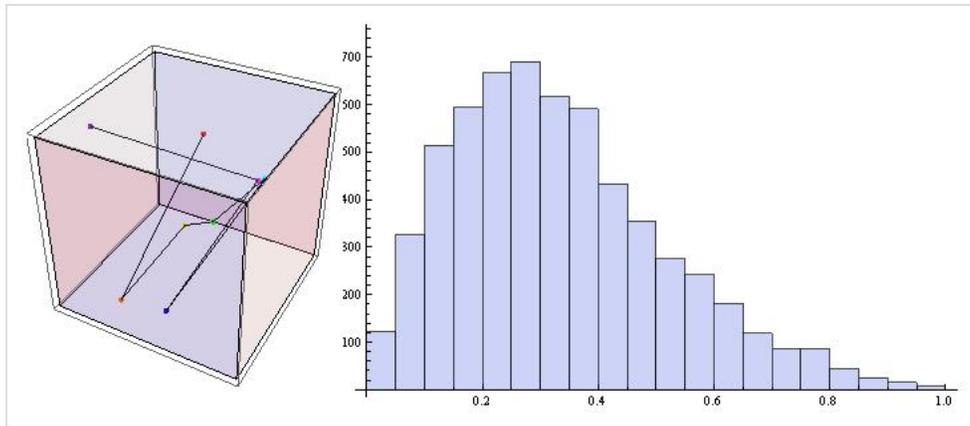

**Figure 96 3D**

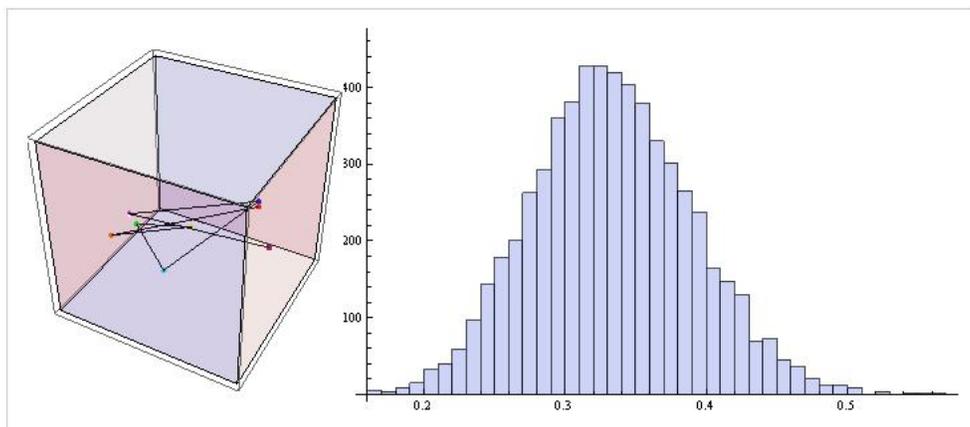

**Figure 97 20D**

### Theorem 1 Three random points in a circle

**Theorem 1** *Let $P_1, P_2, P_3$ be uniformly distributed random points on the unit circle. Then the probability density function of the angle $x\pi = \angle P_1 P_3 P_2$, $x \in [0, 1]$ is*

$$p(x) = 2(1 - x)$$

*with the according cumulative probability function and expectation*

$$\mathbb{P}(\angle P_1 P_3 P_2 \leq x\pi) = x(2 - x), \quad (0 \leq x \leq 1), \qquad \mathbb{E}(\angle P_1 P_3 P_2) = \frac{\pi}{3}.$$

PROOF: First note that the angle $\alpha = \angle P_1 P_3 P_2$ by the Inscribed Angle Theorem (click: http://en.wikipedia.org/wiki/Inscribed_angle#Theorem) depends mainly on the central angle $\gamma = \angle P_1 O P_2$: either $P_3$ is in the longer arc, then $\alpha = \gamma/2$ or $P_3$ is in the shorter arc, then $\alpha = \pi - \gamma/2$. Next, we can without loss of generality assume that $P_1 = (1, 0)$ since we can always bring any configuration into this situation by rotation and/or mirroring.

By the above comment we can reduce the probability $\mathbb{P}(\alpha \leq x\pi)$ to the 'simpler' probability $\mathbb{P}(\gamma \leq \tilde{x}\pi)$, where $\tilde{x}$ depends on $x$. We have to distinguish two cases.

- $\boxed{0 \leq \alpha \leq \pi/2:}$ Here $P_2$ lies somewhere on the arc (with angles in) $[0, \pi]$ and $P_3$ anywhere on the arc $[\gamma, 2\pi]$, where $\gamma = 2\alpha$. Any smaller angle $\alpha$ has a smaller angle $\gamma$. Therefore, for all $0 \leq x \leq 1/2$,

$$\mathbb{P}(\alpha \leq x\pi) = \mathbb{P}(P_2 \in [0, 2x\pi] \wedge P_3 \in [\gamma, 2\pi])$$

$$= \frac{1}{(2\pi)^2} \int_0^{2x\pi} \int_\gamma^{2\pi} 1 \, d\alpha \, d\gamma = x - \frac{x^2}{2}.$$

(Division by $2\pi$ before the integral is due to normalization, i.e., favorable cases (angles) over possible cases.)

- $\boxed{\pi/2 \leq \alpha \leq \pi:}$ Now $P_2$ again is somewhere on the arc $[0, \pi]$, and the point $P_3$ may take any place on the arc $[0, \gamma]$, where $\gamma = 2(\pi - \alpha)$. Any smaller angle $\alpha$ has a *larger* angle $\gamma$. So, for $1/2 \leq x \leq 1$,

$$\mathbb{P}(\alpha \leq x\pi) = \mathbb{P}(\alpha \leq \pi/2) + \mathbb{P}(\pi/2 \leq \alpha \leq \pi)$$

$$= \frac{3}{8} + \mathbb{P}(P_2 \in [2\pi - 2x\pi, \pi] \wedge P_3 \in [0, \gamma])$$

$$= \frac{3}{8} + \frac{1}{(2\pi)^2} \int_{2\pi(1-x)}^\pi \int_0^\gamma 1 \, d\alpha \, d\gamma$$

$$= \frac{3}{8} + \frac{-3}{8} + x - \frac{x^2}{2} = x - \frac{x^2}{2}$$

again.

So altogether, for arbitrary $x \in [0, 1]$,

$$\mathbb{P}(\alpha \leq x\pi) = \mathbb{P}(\alpha \leq x\pi \wedge x \in [0, \frac{1}{2}]) + \mathbb{P}(\alpha \leq x\pi \wedge x \in [\frac{1}{2}, 1])$$
$$= x(1 - x)$$

is the cumulative distribution function for $x \in [0, 1]$. From this we derive the probability density function $p(x) = 2(1 - x)$ and the expectation

$$\mathbb{E}(\alpha) = \int_0^1 x\pi \; p(x\pi) \, dx = 2\pi \int_0^1 x(1 - x) \, dx = \frac{\pi}{3}.$$

This concludes the proof. □

For the corresponding 3-dimensional case, i.e., three random points on a sphere, we cannot employ the Inscribed Angle Theorem since the three points generally will describe circles, whose midpoint is not $O$. So we have to resort to the general formula. We can however reduce the cases a little by assuming

$$P_1 = \begin{pmatrix} 1 \\ 0 \\ 0 \end{pmatrix}, \quad P_2 = \begin{pmatrix} \cos(\alpha) \\ 0 \\ \sin(\alpha) \end{pmatrix}, \quad P_3 = \begin{pmatrix} \cos(\beta)\cos(\gamma) \\ \cos(\beta)\sin(\gamma) \\ \sin(\beta) \end{pmatrix},$$

with $\alpha, \beta \in [0, \pi]$, $\gamma \in [0, 2\pi]$. That is, $P_1$ is a fixed point on the sphere, $P_2$ ranges along an arc and $P_3$ is free on the sphere. Any other configuration can be transformed to this situation.

A somewhat analogous approach to the two-dimensional case would be to calculate first, for each $P_2$ as above the area of all possible $P_3$ such that the angle $\angle P_1 P_3 P_2$ is at most a given $\alpha$. This amounts to determine the area within the intersection curve of a sphere and a cone – in general such curves are rather complicated to deal with, though (click: http://mathworld.wolfram.com/Cone-SphereIntersection.html).

– However, it seems these results are well-known in the literature, the respective (harder) problems for angles in disk and ball are mentioned in this unpublished note of Steven Finch, 2010: http://algo.inria.fr/csolve/rtg3.pdf, with the expected angle of $\pi/3$ in both cases.

# 6. Discussion

Sequentially linked GRB's create patterns that allow for the origin of 2 to 8 CPPC to emanate from every GRB location. Randomly generated noggs create patterns that allow for the origin of 2 to 3 CPPC to emanate from every point. The number of CPPC for GRB's is consistently higher than the number of CPPC for noggs. Sequentially linked GRB's present properties that nullify randomness.

How do we account for the multitude of intrinsic relationships that exist between all points in the GRB sequence? Does the behavior of any known system recapitulate the interconnectivity displayed by sequenced GRB's? If we are to believe GRB's are the final blasts of massive stars then can we expect to find an unknown underlying structural template that links the timing and location of each exploding star?

Because none of the current models used to describe the origin of GRB's consistently explains the variety of behaviors attributed to them it is difficult to concur that any one of them is valid. Therefore a new hypothesis is required, one that could explain the matrix of interrelated patterns revealed in this paper.

Gamma-ray bursts are a form of communication; the explosive "dots and dashes" are analogous to a supremely advanced geometric version of 3D Morse code. The tuned sequences of bursts represent the methodology of a cosmically ancient, extremely advanced civilization, transmitting a message that reveals the nature of their extraordinary technological achievements. The ultimate form of skywriting blasted from the past towards the galaxies of the future.

In 1950 Enrico Fermi said "where are they" He was puzzled as to why there was no sign of extraterrestrial intelligence. "The Fermi Paradox, Self-Replicating Probes" suggests that self replicating probes from advanced civilizations began to enter our galaxy millions of years ago. (Wiley 2010).

The Drake equation, $N = (R^* \times f_p \times n_e \times f_l \times f_i \times f_c) \times L$ shows the probability for intelligent life in our galaxy is high. In 1960 Francis Drake the father of SETI (Search for Extra Terrestrial Intelligence) began Project Ozma, directing a radio telescope to listen for messages from two stars just 10 light years away Tau Ceti and Epsilon Eridani. SETI continues to find the means and methods for designing the technology and digital parameters for an interstellar message. (Glade et al 1998), (A.P.V. Siemion et.al 2011), (Busch and Reddick 2009).

Nikolai Kardashev describes three levels of civilizations according to the amount of energy they are able to generate. (N.S. Kardeshev 1964)
Type I ~$4 \times 10^{19}$ erg/s,
Type II ~$4 \times 10^{33}$ erg/s,
Type III. ~$4 \times 10^{44}$ erg/s.

The Dyson sphere is a stellar engine from a Type II civilization, a spherical grid of connected satellites to collect all the energy from a star and transmit it to an orbiting planet. An interstellar archaeologist observing Dyson technologies altered spectral emissions would recognize the push to infrared as an indication of a Type II civilization (Search for Artificial Stellar Sources of Infrared Radiation, Dyson 1960). Makoto Inoue and Hiromitsu Yokoo refer to a highly advanced Type III civilization able to generate, collect and distribute energy using a Type III Dyson sphere that would collect radiation from the accretion disk of a super massive black hole.

When comparing the sight lines of quasars and GRB's Prochter, Prochaska et al.2006 discovered by examining the absorption lines in the spectra that GRB's were 4 times more likely to have passed through another galaxy than quasars were. Jason X. Prochaska, associate professor of astronomy and astrophysics at the University of California, Santa Cruz said "The result contradicts our basic concepts of cosmology, and we are struggling to explain it," **

Perhaps there is an explanation that also supports to the findings of this study. If GRB's are indeed the individual components of a trans-universal communication system it seems logical that those transmissions would be directed to galaxies because scattered throughout the galaxies are the planets where life resides.

Mathematica Graphics and Theorem 1- Gotlieb Prisic

References:

- Balazs et al. 1999, A&AS., 138, 417
- Barton, Fix (1963), *Random points in a circle and the analysis of*
- *chromosome patterns,* Biometrika, 50, pp. 23-30
- Belcynski et al. 2002, ApJ, 572, 407
- Busch and Reddick 2009), arXiv:0911.3976v3
- Cline et al. 1992, ApJ, 401, L57
- Davies 1995, MNRAS, 276, 887
- Della Valle et al. 2006, ApJ, 642, p. L103
- Dermer 1992, Phys. Rev. 68(12),1799,
- Di Porto et.al 2010 http://arxiv.org/abs/1008.1878
- Dyson, Search for Artificial Stellar Sources of Infrared Radiation, Freeman John Dyson, Science, Vol. 131, 1960
- Eichler, et al. 1989, Nature, 340, p. 126
- Fenimore et al.1993, Nature, 366, p. 40
- Fox D.B., et al., 2005, Nature, 437, 845
- Fynbo et al.2006, Nature 444, 1047
- Gal-Yam et al. 2006, Nature 444, 1053
- Glade et al 1998), arXiv:1112.1506v1
- Grindlay et al. 2006, Nature Physics 2, 116,
- Gurzadyan and R. Penrose 2010 arXiv:1011.3706v1 [astro-ph.CO]
- Hjorth et al.2003, Nature, 423, p. 847
- Horvath 1998, ApJ, 508, p. 757
- Inoue and Hiromitsu Yokoo Journal of British Interplanetary Society, Vol. 64, pp.58-62, 2011
- Ioka et al. 2007, Ap. J. 670, L77

- Kardashev, N. S. "Transmission of Information by Extraterrestrial Civilizations," *Soviet Astronomy*, **8**, 217 (1964)
- Klebesadel et al. 1973, ApJ, 182, p. L85
- Kluzniak & Ruderman 1998, Ap. J. 505, L113
- Kouveliotou et al. 1993, ApJ, 413, p. L101
- Lazzati et al. 2004, Astron. Astrophys. 422, 121
- Lee and A.Wynveen 2008 arXiv:physics/0511009v1
- Litvin et al.2001, Astron. Lett. 27(7), 416,
- Lyutikov et al. 2003, Ap. J. 597,998
- Mao & Paczynski 1992, Ap. J. Lett. 388, L45
- Maxham ,A. 2010,  http://www.newtonsseashore.com/solutions/pokerchips.pdf
- Meegan  et al. 1992, Nature, 355, p. 143
- Meszaros & Stocek 2003, A&A, 403, 443
- Meszaros et al. 2006, A&A, 455, p. 785
- Meszaros et al.2000, A&A, 403, 443
- Meszaros P., 2006, Rep. Prog. Phys., 69, 2259
- Mukherjee et al. 1998, ApJ, 508, p. 314
- Narayan, R., Paczy´nski, B., Piran, T. 1992, ApJ, 395, p. L83
- Nemiroff 1993, AIP Conf. Proc. 307, pp. 730-734
- Newman et al. 1980,ApJ, 242, 319
- O.V.Verkhodanov et al. 2010, Spec. Astrophys. Obs., 65, 238–249
- Paczynski & Xu 1994,  Ap. J. 427,708
- Paczynski 1991, Acta Astronomica 41, 257,
- Piran 1992, Ap. J. Lett. 389, L45
- Piran 2005, Rvmp, 76, 1143
- Prochter et al. 2006, ApJ, 648, L93
- Rosswog et al. 2003, AIP Conf. Proc. 662
- Schlovskii 1974 ,SovAstron, 18, 390
- Shaviv and Dar 1995,  ApJ, 447, 86
- Siemion et.al 2011 arXiv:1109.1136v1
- Solomon (Geometric Probability, 1978)
- Spruit et al. 2001, A&A, 369, 69
- Stanek et al. 2003, ApJ, 591, p. L17
- Toma et al. 2005, Ap. J. 635, 481
- Vavrek et al.2008, MNRAS, 391, p. 1741
- Wiley 2010, arXiv:1111.6131v1
- Woods & Loeb 1994,  Ap. J. 425, L63
- Woosley, & Bloom, J. S. 2006, Annual Review of Astronomy & Astrophysics
- Zhang and Meszaros 2002, ApJ, 581, p. 1236
- Zhang B., Meszaros P. 2004, IJMPA, 19, 2385
- Zhang B., Meszaros P., 2004, IJMPA, 19, 2385
- Zhang et al. 2004, ApJ, 608, p. 365
- Zhang et al. 2007, Adv. Space Res., 40, 1186
  **Jason X. Prochaska
  University of California Santa Cruz, University News July 31, 2006